 \documentclass[final,3p,times,preprint]{elsarticle}
\usepackage{comment}
\usepackage{rotating}
\usepackage{caption}
\usepackage{subcaption}
\usepackage{tablefootnote}
\usepackage{amssymb}
\usepackage{amsmath}
\usepackage{longtable}
\usepackage{fontspec}
\usepackage{minted} %\begin{minted}[fontsize=\small, frame=single]{octave}
\usepackage{enumitem}

\usepackage[ruled]{algorithm2e}
\SetKwInput{KwData}{Input}
\SetKwInput{KwResult}{Updated}

\SetCommentSty{mycommfont}
\SetArgSty{textup}

\usepackage{hyperref}
\hypersetup
{
    colorlinks=true,
    linkcolor=blue,
    filecolor=magenta,      
    urlcolor=cyan,
}
\usepackage{multirow}
\usepackage{hhline}
\usepackage{tikz}
\usepackage{soul}
\usepackage[dvipsnames,table]{xcolor}

\usetikzlibrary{fadings}
\usetikzlibrary{patterns}
\usetikzlibrary{shadows.blur}
\usetikzlibrary{shapes}

\usepackage{framed} % Framing content
\usepackage{multicol} % Multiple columns environment
\usepackage{nomencl} % Nomenclature package
\renewcommand*\nompreamble{\begin{multicols}{2}}
\renewcommand*\nompostamble{\end{multicols}}

\makeatletter
\def\smallunderbrace#1{\mathop{\vtop{\m@th\ialign{##\crcr
   $\hfil\displaystyle{#1}\hfil$\crcr
   \noalign{\kern3\p@\nointerlineskip}%
   \tiny\upbracefill\crcr\noalign{\kern3\p@}}}}\limits}
\makeatother

\journal{Computer Physics Communications}

\begin{document}

\begin{frontmatter}



%\title{A GPU-based Adaptive Mesh Refinement Scheme for the Lattice Boltzmann Method with Complex Geometries}
\title{GPU-native Embedding of Complex Geometries in Adaptive Octree Grids Applied to the Lattice Boltzmann Method}


\author[inst1]{Khodr Jaber\corref{email1}}
%\cortext[email1]{Corresponding author. Tel.: +1 (416) 834-4861}
\cortext[email1]{Corresponding author}
\ead{khodr.jaber@mail.utoronto.ca}

% \ead[inst1]{jaberjbr2@gmail.com}
\affiliation[inst1]{organization={Department of Mechanical and Industrial Engineering, University of Toronto},
addressline={5 King's College Rd}, 
            city={Toronto},
            postcode={M5S 3G8}, 
            state={ON},
       country={Canada}
            }
,%Department and Organization
\affiliation[inst2]{organization={Department of Mechanical, Industrial and Aerospace Engineering,  Concordia University  },
addressline={1515 Ste-Catherine St. W.}, 
            city={Montreal},
            postcode={H3G 1M8}, 
            state={QC},
            country={Canada}
            }            

\author[inst2]{E. E. Essel}
\author[inst1]{Pierre E. Sullivan}

%\author[inst1,inst2]{Author Three}

%\affiliation[inst2]{organization={Department Two},%Department and Organization
%           {addressline={Address Two}, 
%            city={City Two},
%            postcode={22222}, 
%            state={State Two},
%            country={Country Two}}

\begin{abstract}
%A major limitation of structured-grid computational fluid dynamics (CFD) methods like the lattice Boltzmann method (LBM), which are traditionally restricted to uniform grids.
%

%\input{Abstract_original}
Adaptive mesh refinement (AMR) reduces computational costs in CFD by concentrating resolution where needed, but efficiently embedding complex, non-aligned geometries on GPUs remains challenging. We present a GPU-native algorithm for incorporating stationary triangle-mesh geometries into block-structured forest-of-octrees grids, performing both solid voxelization and automated near-wall refinement entirely on the device. The method employs local ray casting accelerated by a hierarchy of spatial bins, leveraging efficient grid-block traversal to eliminate the need for index orderings and hash tables commonly used in CPU pipelines, and enabling coalesced memory access without CPU–GPU synchronization. A flattened lookup table of cut-link distances between fluid and solid cells is constructed to support accurate interpolated bounce-back boundary conditions for the lattice Boltzmann method (LBM).
We implement this approach as an extension of the AGAL framework for GPU-based AMR and benchmark the geometry module using the Stanford Bunny (112K triangles) and XYZ RGB Dragon (7.2M triangles) models from the Stanford 3D Scanning Repository. The extended solver is validated for external flows past a circular/square cylinder (2D, $Re = 100$), and a sphere (3D, $\text{Re}\in\{10,15,20\}$). Results demonstrate that geometry handling and interpolation impose modest overhead while delivering accurate force predictions and stable near-wall resolution on adaptive Cartesian grids. The approach is general and applicable to other explicit solvers requiring GPU-resident geometry embedding.

\end{abstract}


%\item Open source solver for CTF coefficients and response factors
%\item Series expansion estimation of hyperbolic conduction transfer function
%\item Performance under various time steps and wall weights
%\item Comparison with ASHRAE Handbook and Frequency-Domain Regression


\end{frontmatter}



\section{Introoduction} \label{sec:intro}

Researchers and industry professionals are increasingly using GPUs to accelerate computational fluid dynamics simulations (CFD) in heterogeneous computing systems. GPUs are capable of executing numerous concurrent threads; this makes the architecture highly suitable for explicit solvers such as the lattice Boltzmann method that are mainly composed of local computations and localized data exchanges between neighboring grid elements. Hardware acceleration of CFD solvers is maximized when the grid elements are processed contiguously and when data is exchanged with high spatial locality. This is naturally achieved with uniform (or structured) grids, but their memory and storage requirements make them infeasible in applications where high resolution is only needed in sparse regions of the domain (such as in turbulent flows or flows in complex geometries). This has motivated research on GPU-acceleration of CFD on non-uniform grids (such as unstructured and/or locally refined grids).

Grid refinement according to user-specified criteria (e.g., proximity of cells to a boundary) is referred to as adaptive mesh refinement (AMR) \cite{Berger1984,Berger1988,Berger1989}. This is static when the grid is processed once before the simulation starts, and dynamic when the grid is repeatedly refined at runtime based on the evolving numerical solution. AMR approaches can be categorized as block-structured (SAMR) or tree-based; we refer the reader to the works of Dubey et al. \cite{Dubey2014,Dubey2021} for further detail. The SAMR approach clusters patches of cells in logically rectangular grids without a strict parent-child relationship between the grids (which allows for flexible overlap in different grid resolutions). In contrast, tree-based AMR uses tree data structures (with a clear hierarchy and well-studied traversal and load-balancing patterns) to partition the domain. In some recent literature, a categorization of patch-, block- and cell-based AMR is made where block-based AMR is the special case of SAMR with regular-sized patches of cells, and cell-based AMR is simply cell-granular (and usually organized with a tree). AMR software packages are typically CPU-based (e.g., p4est \cite{p4est}, t8code \cite{t8code}) or hybrid (\cite{Daino}, GAMER \cite{GAMER2}) in that the mesh is organized and adapted on the CPU while the numerical solution is offloaded to the GPU for temporal integration.

Researchers have recently introduced GPU-native AMR where the mesh is moved to the GPU to completely eliminate CPU-GPU data transfers. This includes both unstructured AMR (\cite{Luo2016, Giuliani2019}) and cell-based tree AMR (\cite{Menshov2020, Pavlukhin2024}) with application to gas hydrodynamics with the Galerkin or finite volume methods. We recently developed a GPU-native AMR technique that was applied to simulate weakly-compressible flow with the LBM \cite{Jaber2025}. However, our approach was limited to domains where the surfaces aligned exactly with the faces of the grid cells. This work extends our original method to incorporate embeddings of complex geometries while retaining efficient dynamic adaptation during temporal integration. 

The embedding of complex geometries in axis-aligned Cartesian computational grids is closely related to the concept of voxelization in graphics processing \cite{Mierke2020}. Grids organized with tree data structures naturally lead to the concept of the sparse voxel octree \cite{Laine2011}, where fine-grained representation of the geometry is preserved on the finest levels of the tree while the coarser levels hide the interior in order to accelerate processes such as rendering and illumination. Analogously, the tree-based grid can be used to accurately resolve fluid flow near the geometry surface while retaining a coarser resolution of the solid interior. This saves memory and reduces the number of traversed elements, thereby accelerating temporal integration and the imposition of boundary conditions. However, tree data structures that represent grids for explicit solvers require additional features, such as 2:1 balancing \cite{Sundar2008} (where the relative size of two adjacent grid elements cannot exceed two) and metadata that enables proper synchronization at refinement interfaces and domain boundaries.
\setlength{\fboxrule}{0.4pt} % border thickness
\setlength{\fboxsep}{0pt}    % remove padding
\definecolor{csolidcell1}{HTML}{CCCCCC}
\definecolor{cgeom1}{HTML}{FF3399}
\definecolor{cboundarycell1}{HTML}{990000}
\definecolor{cfluidcell1}{HTML}{007FFF}
\definecolor{clinks1}{HTML}{D6B656}
\definecolor{cstairs1}{HTML}{CCFFFF}
\begin{figure}[t]
    \centering
    \includegraphics[width=0.8\textwidth]{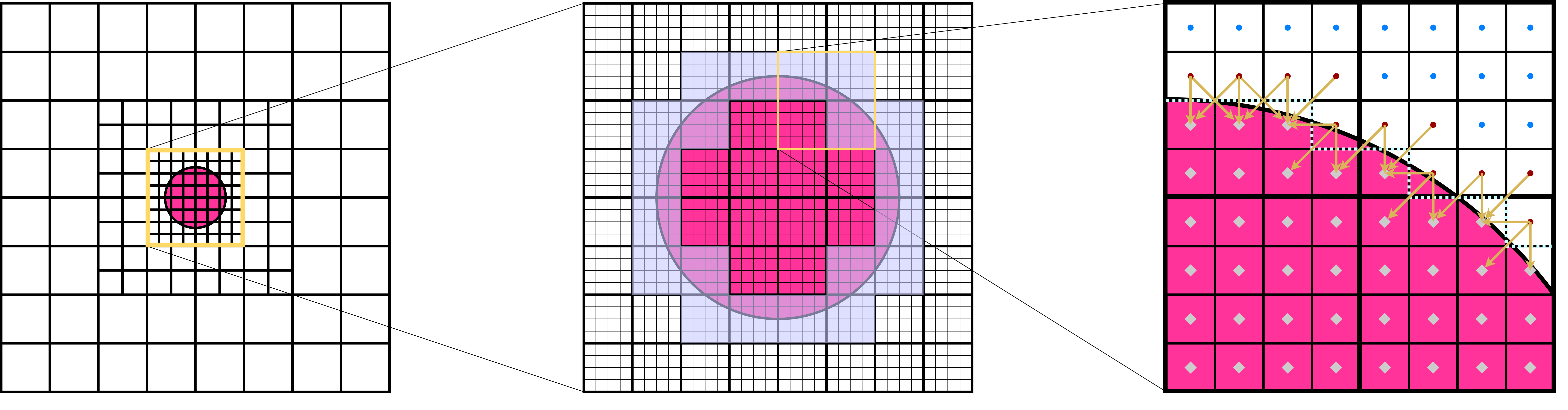}
    \caption{Embedding of a 2D circular cylinder (\textcolor{cgeom1}{\rule{1ex}{1ex}}) within an axis-aligned adaptive computational grid. Each square represents a $4\times 4$ block of cells. Cells are classified as fluid (\textcolor{black}{\rule{1ex}{1ex}}), boundary (\textcolor{cboundarycell1}{\rule{1ex}{1ex}}), and solid (\textcolor{csolidcell1}{\rule{1ex}{1ex}}). The links between the boundary and solid nodes are illustrated with arrows (\textcolor{clinks1}{\rule{1ex}{1ex}}). The staircase approximation of the cylinder imposed by voxelization is shown with the dashed line (\fbox{\textcolor{cstairs1}{\rule{1ex}{1ex}}}).}
    %Left: grid refinement near a 2D circular cylinder. Each square represents a $4\times 4$ block of cells. Middle: Blocks in the vicinity of the cylinder are marked as solid, adjacent to the solid, or fluid. Right: Nodes (cell-centers) are classified as fluid (\textcolor{black}{\rule{1ex}{1ex}}), boundary (\textcolor{red}{\rule{1ex}{1ex}}: boundary cells), and solid (\textcolor{gray}{\rule{1ex}{1ex}}: solid cells). The links between the boundary and solid nodes are illustrated with yellow arrows, and the staircase approximation of the cylinder imposed by the simple bounce-back boundary condition is shown in blue.}
    \label{fig:lit_grid_def}
\end{figure}

We summarize our contributions as follows:
\begin{itemize}[leftmargin=*,noitemsep]
    \item We present an efficient GPU-based embedding of complex geometries within an adaptive forest-of-octrees grid \cite{p4est} where the cells are aligned with Cartesian axes. Our approach feeds a block-based, cell-granular solid voxelization into a refinement and tree balancing pipeline that simultaneously builds the forest and embeds the geometry level by level to produce a mesh suitable for explicit computational fluid dynamics solvers. No CPU-GPU communications are involved at any stage of this procedure. The algorithm is implemented in C++/CUDA.
    \item The solid voxelization is based on a local ray cast procedure that first flags cells in the immediate vicinity of the geometry surface. These flags are then propagated to fill the interior. Ray cast acceleration is achieved with filtered hierarchical spatial binning of the geometry faces. The geometry and AMR grid are both hosted and processed entirely on the GPU. Our voxelization procedure is tested on standard computer graphics models from the Stanford 3D Scanning Repository \cite{Stanford3D}.
    \item We generate a flattened lookup table of lengths of solid-fluid cell links that cut through the geometry surface (Figure \ref{fig:lit_grid_def}), enabling accurate stencil computations in cells adjacent to the boundary. We also demonstrate how this lookup table can be used when imposing boundary conditions and computing forces at runtime. An application to external flow simulation with the lattice Boltzmann method validates the implementation.
    %We show that the overhead in accessing the cut-link lengths and incorporating interpolation at the boundary is minimal when compared with a typical staircase discretization of curved surfaces.
    \item Our proposed voxelization approach is similar to other approaches encountered in graphics processing for sparse voxel octrees \cite{Schwarz2010}; however, it can be distinguished in a few significant ways. The tree construction procedure is top-down and does not rely on a specific block ordering (such as space-filling curves). Instead, we use a custom index list representation for efficient data-parallel refinement and coarsening. Our approach also accounts for the features required for explicit computational fluid dynamics solvers such as the enforcement of a 2:1 balance for synchronized temporal integration.
\end{itemize} 

\section{Background} \label{sec:back}

The methodology introduced in this paper intersects several topics: voxelization in axis-aligned grids organized with tree data structures (which has historically involved a computer graphics pipeline), methods and GPU-parallelization of boundary treatments in the LBM in unaligned geometries, and GPU-accelerated adaptive mesh refinement. It also builds on our previous work \cite{Jaber2025}, where we developed an algorithm to facilitate the adaptation of a mesh organized as a block-based forest-of-octrees entirely on the GPU. However, the boundaries of this mesh were strictly axis-aligned, and only the simple bounce-back approach could be implemented for tests involving external flow. To our knowledge, the literature on GPU frameworks for voxelization into hierarchical grids suitable for CFD (such as the forest-of-octrees) is limited. This section will discuss relevant related work to clarify research gaps and contrast our proposed approach for incorporating complex geometries when the mesh metadata is entirely GPU resident.
%We first present an overview of our proposed methodology and then discuss related work by topic. Afterwards, we provide a brief summary of the GPU-based mesh adaptation procedure of \cite{Jaber2025} and introduce the main aspects of the framework that will be expanded on in Section \ref{sec:methodology}.

\subsection{GPU-based Computational Fluid Dynamics in Non-uniform Grids}

Non-uniform computational grids are commonly used to simulate fluid flow in complex geometries. They enable accurate resolution of flow features in the vicinity of the wall and around coherent structures that develop at higher Reynolds numbers. Non-uniformity is introduced either by relaxing the structure of grid elements (so that different shapes and topologies can be constructed to accommodate or approximate curved surfaces), or by introducing multi-resolution hierarchies where grid cells become finer near targeted features. These two options are generally referred to as unstructured grids and grid refinement, respectively. Most mature CFD software packages used for research and industry (e.g. OpenFOAM \cite{OpenFOAM}, Ansys, STAR-CCM+, Comsol, PowerFLOW) and some research codes (e.g., p4est \cite{p4est}/t8code \cite{Holke2018,t8code}, waLBerla \cite{Schornbaum2018,waLBerla}) support either or both of these features for meshing.

GPU-acceleration of CFD codes has attracted a great deal of attention \cite{Luo2016,Giuliani2019,Menshov2020,Pavlukhin2024,Wang2024,Wang2025} as a consequence of hardware improvements and the increased accessibility of heterogeneous compute clusters (hosted at universities or over the cloud). The degree of speedup that can be obtained from GPU-acceleration largely depends on how well CFD meshes can be organized in memory with respect to spatial locality. The mesh should therefore be chosen to expose data-parallelism with contiguous data arrangements that promote coalesced access and minimize branch divergence.

Unstructured grids are defined by an irregular connectivity that must be stored explicitly. This enables a more flexible partition of the domain that can adapt to complex geometries (e.g. via tessellation of curved boundaries) but comes at the cost of noncontiguous grid element data arrangements. In block-structured AMR, each individual block of cells is axis-aligned and logically rectangular, which promotes spatial locality. However, these grids cannot conform to arbitrary shapes with the same flexibility as unstructured grids. One option is to introduce arbitrarily-shaped blocks of cells (e.g., tetra-/hexahedral, wedge), but this results in a hierarchical unstructured grid whose cells are not processed as straightforwardly as axis-aligned blocks. Quad/octrees support a natural block-based decomposition of the domain where the blocks are equal-sized. This work therefore targets axis-aligned hierarchical grids. In particular, we focus our attention on the forest-of-octrees concept \cite{p4est}, where each element of an initial root grid partitions its corresponding subdomain with its own octree. In such grids, geometries are typically embedded in (or immersed within) via voxelization, where cells are flagged as fluid, solid, or on the boundary (in the fluid but adjacent to the solid to aid in imposing boundary conditions).

\begin{figure}[t]
    \centering
    \includegraphics[width=0.3\textwidth]{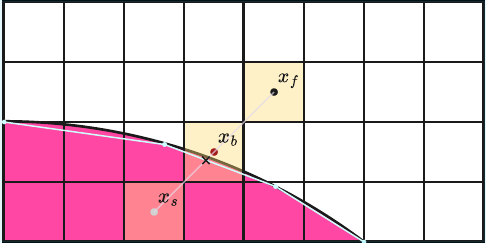}
    \caption{Stencil for the interpolated bounce-back condition along a link that intersects a tesselation of a 2D circular cylinder (shown in blue). The distance to the point of intersection between the link and a geometry edge is stored explicitly. A boundary node and one fluid node are required for the computation, and these may lie in different blocks as shown here.}
    \label{fig:lit_stencil}
\end{figure}

\subsection{Stencil Computations in lattice Boltzmann Simulations}

%The standard derivation of the lattice Boltzmann method from the Boltzmann equation discretized in particle-velocity space involves integration along its characteristics, which results in a linear relationship between the spatial and temporal step sizes via the discrete particle velocities. An exact streaming of the discrete density distributions effectively restricts the method to uniform grids. Non-uniformity has been introduced in the past by either formulating the method in a way that is suitable for unstructured grids (e.g. the finite volume method), or by employing piecewise uniform grids in a local grid refinement framework.

The standard lattice Boltzmann method is typically restricted to uniform grids with equidistant nodes to facilitate exact streaming of the discrete density distributions (DDFs) between nodes (i.e., no interpolation is required since DDFs never `land' between neighboring nodes). Non-uniformity has been introduced in the past by either reformulating the method for unstructured or body-fitted grids (e.g. the finite volume method \cite{Wang2020}, curvilinear coordinates \cite{Reyes2020}), or by employing piecewise-uniform grids in a local grid refinement framework \cite{Dupuis2003,Rohde2006,Lagrava2012}. Though the former approaches are more capable of adapting to the complexity of a given geometry, the performance penalties in GPU-acceleration that result from operating on non-structured grids hold, as mentioned above.

When a geometry is voxelized into a piecewise-uniform grid, boundary condition imposition is facilitated on nodes adjacent to the geometry surface by inspection of the flags of their neighbors. Boundary conditions are imposed when neighbors in a particular direction are marked as solid nodes. The simplest method in the LBM is the simple bounce-back (SBB) rule, which states that a DDF that streams along a link towards a wall is reflected back along the same link. This approach formally attains second-order accuracy when the wall is exactly halfway between boundary and solid nodes, and degrades to first-order accuracy with nonalignment of the geometry. Interpolation and extrapolation (or ghost) approaches have been introduced previously to account for the exact distance from the wall to recover accuracy; we refer the reader to Chapter 11 of Kr{\"u}ger et al. \cite{Kruger2017} for a recent review of approaches for fluid-structure interaction. Filippova and H{\"{a}}nel \cite{Filippova1998} introduced an extrapolation method which reconstructs the equilibrium state in a solid node downstream the boundary node to determine the boundary condition. Bouzidi et al. \cite{Bouzidi2001} introduced interpolated bounce-back (IBB) methods using the DDFs stored on a given boundary node and up to two DDFs distributed over two upstream fluid nodes. Ginzburg and d'Humi\`{e}res \cite{Ginzburg2003} developed a family of generalized bounce-back methods, denoted as the multireflection conditions, which are based on the Chapman-Enskog analysis. These can use up to five DDFs distributed over two upstream nodes, in addition to the boundary node. A common feature in all of these methods is the presence of interpolation weights that depend on the distance of a given boundary node from the wall along a given link. We refer to this quantity as a link-length for the remainder of this paper. Figure \ref{fig:lit_stencil} shows the case of a stencil involving one fluid and one solid node adjacent to a boundary node.

This paper introduces a routine for computing these link-lengths after the geometry has been embedded, which are stored in a lookup table that can be efficiently accessed in CUDA kernels. This enables the implementation of more advanced stencil computations near the geometry surface. Though we have motivated this lookup table with boundary conditions for the LBM, it could also be used for general finite-difference computations near the geometry (such as the determination of velocity gradients when implementing turbulence models).

We implement the linear IBB method of Bouzidi et al. \cite{Bouzidi2001} in the current work and compare its accuracy to the SBB method in simulations of external flow. The linear version requires data from only one fluid node for each boundary node in a given direction. We compute the time-averaged drag coefficient, the amplitude of the lift coefficient, and the Strouhal number of these simulations and compare with previous results in the literature.

\subsection{GPU-accelerated Adaptive Mesh Refinement}

Mesh adaptation is challenging to parallelize on the GPU since modification of the underlying data structure is difficult to achieve with data-parallelism (while still observing consistent coalesced global memory access and minimal branch divergence). For example, in tree-based AMR grids represented with space-filling curves, identifying neighboring nodes requires a search through a sorted index array. In a data-parallel approach, where one thread is assigned to each node in the tree, each thread would need to perform an individual search of logarithmic computational complexity. Due to the limited amount of memory available even on modern GPUs, it is common to employ multiple GPUs in distributed memory clusters to enable larger simulations. This necessarily requires host-side inter-node communication and synchronization patterns implemented, for example, with the Message Passing Interface (MPI) standard. As a result, software packages offering GPU-accelerated dynamic AMR are commonly hybrid \cite{Daino,GAMER2,waLBerla}, with the CPU managing the mesh and its distribution to various processors, and the GPU integrating the numerical solution.

The complex logic of static mesh adaptation can be handled once on the CPU before the grid data is offloaded to the GPU, where it is retained for temporal integration. However, in dynamic AMR there is repeated communication of metadata with the CPU with each adaptation, which shifts data in memory and necessitates auxiliary operations to restore ideal memory access patterns. These operations contribute to a bottleneck in large-scale simulations. In modern GPU-accelerated frameworks, the mesh data structure is typically hosted and adapted on CPUs due to the complex logic involved. This enables the communication of re-partitioned data between processors in distributed memory compute clusters with careful synchronization and load balancing. Data is then arranged so that computations can be suitably offloaded to the GPUs where they are efficiently processed in parallel. It is common to employ a GPU-resident approach \cite{Saetra2015,Beckingsale2015} where the data is permanently retained on the GPU and only communicated back to the CPU for host-side operations such as I/O. Software packages that utilize this hybrid approach include AMReX \cite{AMReX}, Daino \cite{Daino}, GAMER-2 \cite{GAMER2}, and waLBerla \cite{waLBerla}.

Communication between the CPU and GPU can be eliminated entirely in GPU-native AMR where the mesh adaptation takes place entirely on the latter. This has been introduced for triangular meshes that are suitable for Galerkin methods \cite{Luo2016,Giuliani2019,Wang2024}, and in octree grids suitable for the finite volume method \cite{Pavlukhin2019,Menshov2020}. Data parallelism is introduced to the refinement and coarsening on a carefully-defined list representation of the grid. Each element is assigned a unique index that corresponds to the location of its data in memory. Index recycling and defragmentation (the elimination of gaps in memory that form when coarsened elements are removed) are utilized to adapt the grid within a fixed memory allocation while preventing premature exhaustion after repeated coarsening.
Pavlukhin and Menshov \cite{Pavlukhin2024} recently extended their previous GPU-native cell-based octree AMR method \cite{Pavlukhin2019,Menshov2020} to moving rigid body geometries. They introduce nested 'cloud' boundaries in the form of virtual bodies that correspond to the rigid bodies to accelerate the search for cells that require intersection calculations for potential updates to their classification as the geometry is translated and rotated while maintaining a 2:1 balance in the octree during refinement and coarsening.

In previous work, we introduced our own block-based forest-of-octrees approach to AMR that was applied to simulate fluid flow with the lattice Boltzmann method. This method employs a list representation similar to previous unstructured grids (in contrast with the usual space-filling curve organization that is commonly used in tree-based AMR \cite{p4est,t8code,Schornbaum2018,Menshov2020,Wang2024}). This representation imposes specific memory access patterns for updating cell and cell-block metadata which will also be carefully reused here. Our adaptation algorithm incorporates an automatic reversion of flags for coarsening that enforces a 2:1 balance of the whole forest-of-octrees at all time. This will be exploited in our design of the geometry embedding algorithm.

\subsection{Voxelization in Hierarchical Data Structures}

Many early GPU-based voxelization approaches utilized the hardware rasterizer to efficiently process the geometries \cite{Fang2000, Li2005, Eisemann2006, Eisemann2008, Crassin2012} (we refer the reader to Aleksandrov et al. \cite{Aleksandrov2021} for a review). Using the graphics rasterization pipeline in CFD grid generation requires integration of the underlying data with texture-based formats (which are optimized for spatial data locality in uniform grids) and fixed-function rendering stages. This was done, for example, by Li et al. \cite{Li2005} in their GPU implementation of the LBM in moving geometries where the fixed-function rasterizer performs the voxelization in preparation for the imposition of second-order accurate boundary conditions.

Utilization of this pipeline restricts how the required data structures are chosen and how the corresponding algorithms are designed for more complex grid operations (e.g. imposing 2:1 balancing \cite{Sundar2008} of tree data structures for grid synchronization during explicit temporal integration). Compute-based approaches to voxelization have been introduced in the last two decades that now omit the fixed-function rasterizer in the grid generation pipeline. Schwarz and Seidel \cite{Schwarz2010} present several data-parallel techniques for surface and solid voxelization. Their surface voxelization is based on an improved version of the triangle-box overlap test of Akenine-Moller \cite{Moller2001}, which considers the overlap of two-dimensional projections of the triangles into the axis-aligned planes. They also introduce a sparse octree-based solid voxelization procedure. The octree is organized with a Z-order Morton space-filling curve that is constructed bottom-up. They begin by performing surface voxelization of a uniform dense grid with the same resolution as the second-finest layer, and subsequently extract the active nodes and construct the tree. The voxelization is then achieved by voxelizing the geometry into the finest level and then subsequently propagating inside-outside parity data in a hierarchical fashion.

The octree approach to solid voxelization continues to be used in CFD grid generation on both the CPU and the GPU. Park and Shin \cite{Park2012} introduced a GPU-based octree adaptive grid generation where cells intersecting the irregular geometry are clipped to produce a cut-cell computational mesh. Hasbestan and Senocak \cite{Hasbestan2018} employ a red-black tree organized as a Morton Z-order space-filling curve to efficiently adapt a computational grid with 2:1 balancing in the presence of an immersed geometry on a CPU. In block-based adaptive grids, a combination of hierarchical refinement and uniform grid voxelization \cite{Jansen2015,Ma2020,Shukla2022,Kumar2024} is required to achieve cell-granular embeddings.

Our proposed voxelization approach is designed with cell-granular data-parallelism within a block-structured grid in the absence of a particular ordering (e.g. space-filling curve). By modifying mesh metadata in a specific order of operations, we employ our earlier tree construction and balancing algorithm in an extended pipeline to build the hierarchical grid level-by-level. We produce a solid voxelization of the geometry into each level, and prepare subsequent ones for voxelization while maintaining a 2:1 balance at each step.

\subsection{Solid Voxelization Approaches}

In uniform grids, one can classify a given point as inside or outside the volume of a given geometry with the ray cast parity test. A ray is cast from the point in a given direction, and the odd-even parity of the number of intersections with the geometry determines the classification. However, if the geometry is known to be closed, watertight, and oriented, then it is sufficient to compute a signed distance between the point and the first geometry face encountered via intersection by a ray cast in any direction.

For high-resolution geometries, the total number of ray casts required for voxelization increases dramatically.  Ray cast acceleration, which restricts the set of faces that need to be traversed by each point in the mesh, can be achieved with complementary data structures such as bins (or buckets), bounding volume hierarchies, $k-$d trees, and quad-/octrees that spatially partition the triangles of the geometry to reduce the total number of ray casts and intersection tests. This is especially important when high resolution is specified for the geometry (where individual ray casts may potentially intersect numerous triangles in a relatively small neighborhood) or the computational grid (where ray casts might only be necessary in a sparse subset of voxels). Researchers have studied the construction of such data structures on GPUs in the last couple of decades \cite{Zhou2008,Kalojanov2009,Lauterbach2009,Pantaleoni2010,Schwarz2010,Garanzha2011,Kalojanov2011,Karras2012,Cornerstone}, and applications range from computer graphics (namely, scene rendering and illumination) to computational physics (such as particle simulations).

Though tree data structures can be constructed efficiently on GPUs, their use in accelerating voxelization is relatively limiting on GPUs if querying requires searches. An access cost of $\mathcal{O}(1)$ can be recovered in the case of octrees if combined with a hash table, but this increases the complexity of the overall algorithm and its implementation. This paper presents a ray cast acceleration based on a hierarchy of spatial bins (uniformly distributed on each level). The sizes of these bins are matched with those of the cell-blocks on each corresponding level.
%We will demonstrate that this added step accelerates voxelization into the hierarchy with minimal overhead.
Though we address only stationary geometries in this paper, our GPU-based spatial binning framework serves as the basis of a future expansion for handling moving geometries.

% \subsection{Summary}

% Our work 

%The methodology differs from what exists in the literature in several ways. We employ a hierarchy of bins that is constructed so that a set of uniform bins corresponds to each grid level in the AMR grid hierarchy. Each bin level is then filtered of faces that are guaranteed to never play a role in voxelization with parallel stream compaction. In other words, ray casts from voxels in the vicinity of the faces towards all neighboring cells in a surrounding halo will never intersect the face. The number of binned faces can be lower than 1\% of the total number of faces in the geometry, especially when the computational grid is coarse.

\section{Methodology} \label{sec:methodology}

\begin{figure}[t]
    \centering
    \begin{subfigure}[b]{0.45\textwidth}
        \centering
        \includegraphics[height=0.75\linewidth]{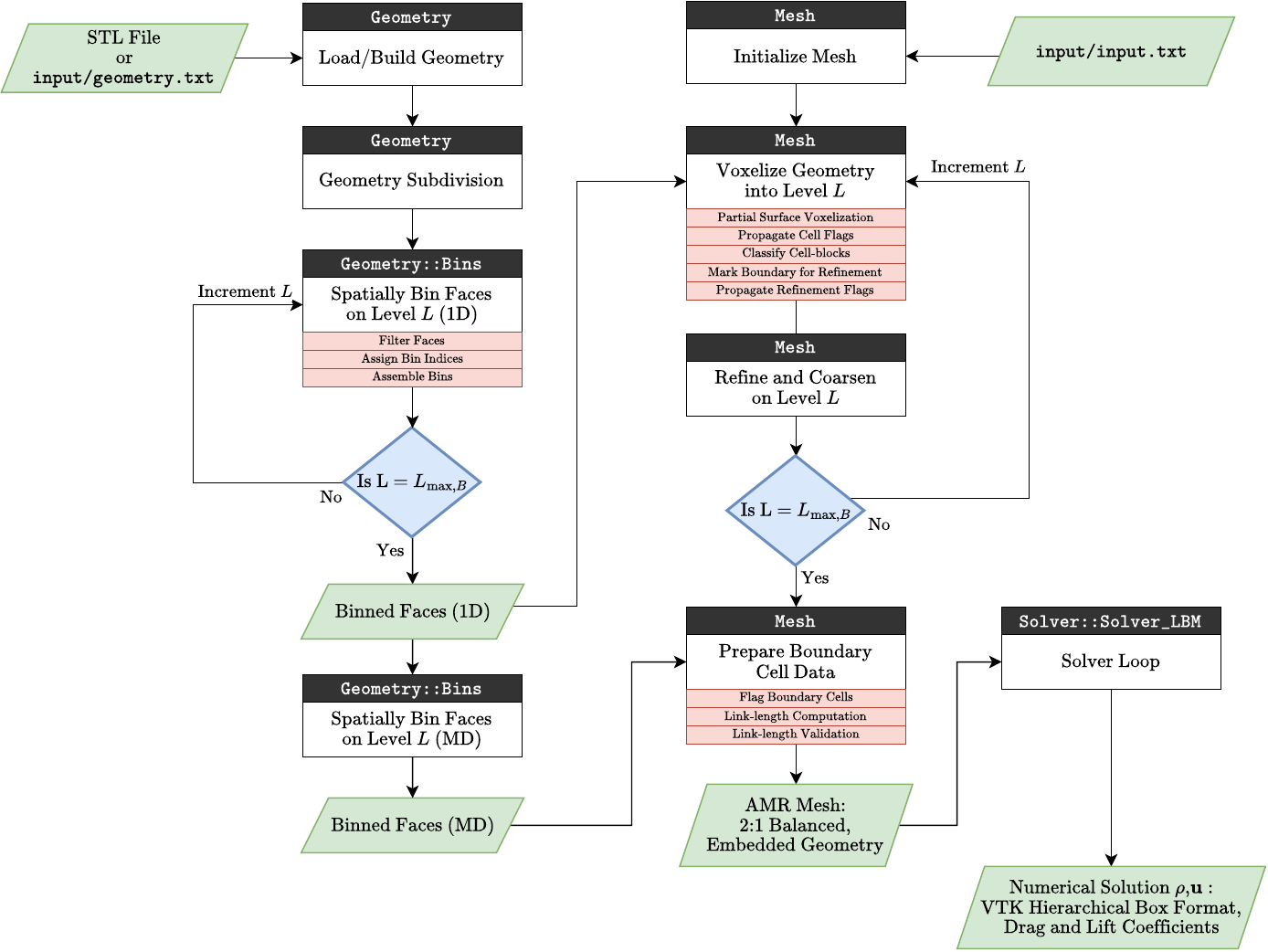}
        \caption{}
        \label{fig:overview_flowchart}
    \end{subfigure}
    \hfill
    \begin{subfigure}[b]{0.45\textwidth}
        \centering
        \includegraphics[height=0.75\linewidth]{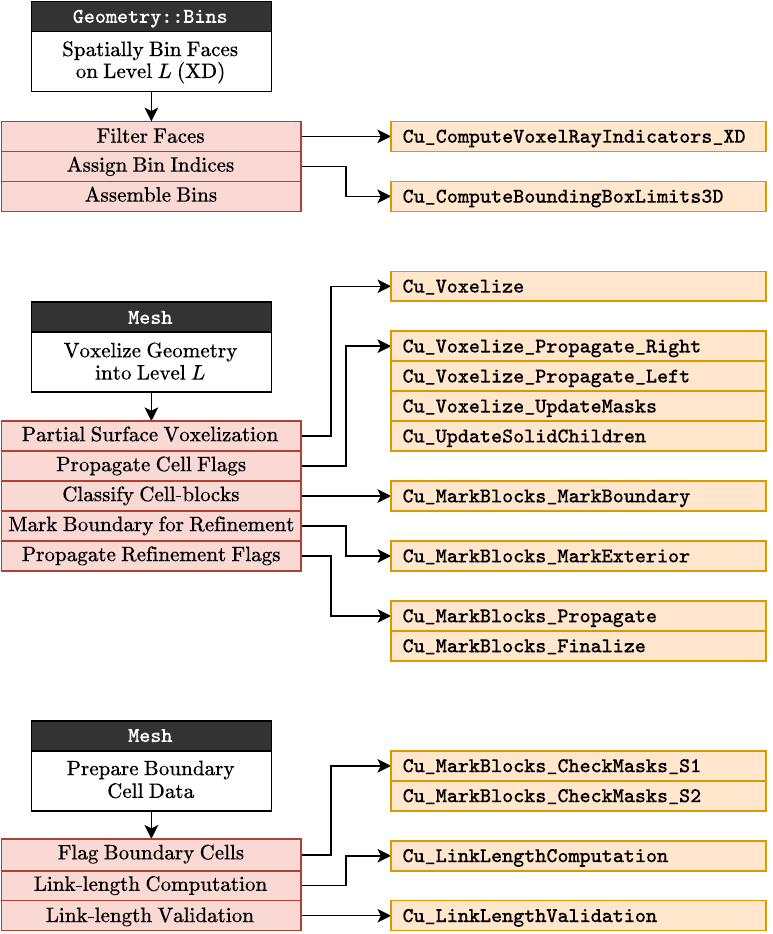}
        \caption{}
        \label{fig:overview_kernels}
    \end{subfigure}
    \caption{Left: Flow chart of proposed voxelization procedure. Right: Summary of our CUDA kernels for each subroutine.}
    \label{fig:overview}
\end{figure}

This section details our GPU-native procedure for embedding a complex geometry within a Cartesian grid organized as a block-based forest-of-octrees (Figure \ref{fig:overview}). We first present an overview of the procedure and then briefly describe how our mesh is organized to achieve data-parallel refinement.
%(including the concepts that will be reused and expanded to facilitate geometry embedding).
Next, we detail the construction of the hierarchy of spatial bins, and its subsequent application in a level-by-level solid voxelization and tree construction. Finally, we demonstrate how our setup recovers the link-length lookup table with an example application to lattice Boltzmann simulations.

%Since we demonstrate applicability of the resulting mesh in computational fluid dynamics with lattice Boltzmann simulations, we first summarize the LBM and present a simplified overview of the mesh adaptation procedure and memory access patterns. Then, we present voxelization and near-wall refinement schemes, and the procedure for computing an interpolated bounce-back condition in parallel. The former two require spatial binning procedures to reduce the search-space considered by the individual blocks of cells. We also provide schemes for parallel evaluation of the forces exerted by the fluid onto the geometry (under the assumption of full immersion) with the momentum exchange algorithm and the discrete analogy to the Reynolds transport theorem for momentum.

    %\subsection{The Lattice Boltzmann Method}

    %\input{p1_methodology_lbm}

    \subsection{Overview}

\begin{figure}[t]
    \centering
    % \begin{subfigure}[b]{0.4\textwidth}
    %     \centering
    %     \includegraphics[width=0.5\textwidth]{figs/setup/orient_edge-1.pdf}
    %     \caption{Edge.}
    %     \label{fig:method_orient_edge}
    % \end{subfigure}
    % \begin{subfigure}[b]{0.6\textwidth}
    %     \centering
    %     \includegraphics[width=0.5\textwidth]{figs/setup/orient_tri.pdf}
    %     \caption{Triangle.}
    %     \label{fig:method_orient_edge}
    % \end{subfigure}
    \includegraphics[width=0.4\textwidth]{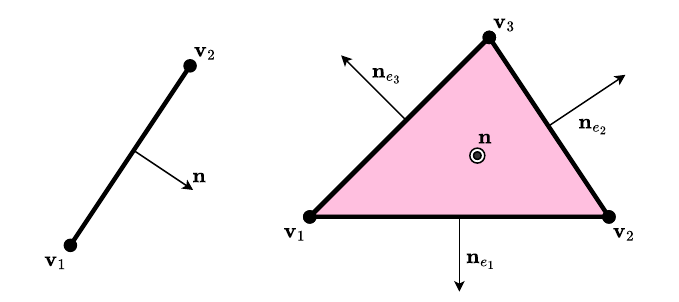}
    \caption{Labeling and orientation of the geometry elements. Left: Edge in 2D. Right: Triangle in 3D.}
    \label{fig:method_orient}
\end{figure}

A closed, oriented triangle mesh (referred to simply as the "geometry" from this point forward) comprises the domain boundary (Figure \ref{fig:method_orient}). It is initialized either by constructing primitive objects (e.g., spheres, prisms) described in an input text file on the fly, or by reading an STL file (currently, only ASCII format is supported). The geometry is complex in that 1) it can be locally convex or concave, 2) its faces may vary widely in size, and 3) its faces do not necessarily align with the cells of the multi-resolution lattice grid implicitly defined by the forest-of-octrees. We utilize two data structures to represent the geometry: an index list format where the faces store the indices of their vertices within a coordinate list, and a coordinate list format where each face stores its vertex coordinates separately. The benefit of the former is that connectivity of the mesh is preserved and that it is generally possible to describe the mesh without duplicating vertex data. The latter format loses both advantages but allows the face data to be arranged in strict structure of arrays (SoA) or arrays of structures (AoS) formats. This is ideal for coalesced global memory load transactions on GPUs in triangle-parallel kernels, which is why we adopt it to represent the geometry in some of our embedding operations. However, we will also show that the array of structures can be employed in kernels where the parallelism targets the cells of the computational mesh instead. We therefore store the data in both formats in the current implementation. These lists are first assembled using C++ vector containers and then loaded into arrays allocated on the GPU. A coordinate list is employed immediately when the user supplies an STL input to the program, whereas the index-list representation is used internally when generating primitive objects, after which it is converted to a coordinate-list format.

Embedding of the complex geometry is a two-fold process. We first perform a top-down tree construction based on solid voxelization starting from an initial coarse grid. Cell-blocks are refined in the vicinity of the geometry according to a user-specified near-wall distance refinement criterion. The geometry is voxelized with an initial partial surface voxelization, which determines how cells near the geometry surface should be flagged (either inside or outside). This is followed by flag propagation throughout the mesh. We note that this surface voxelization differs from the usual surface voxelization methods encountered in graphics processing contexts wherein a given voxel's inclusion is considered by triangle-AABB overlap. In contrast, we perform ray cast computations and exploit the known orientation of the geometry to locally determine if the voxel's center lies inside or outside of the geometry without traversing a larger set of faces to perform odd-even parity testing. Flag values are stored in the same arrays as cell masks, so the two terms will be used interchangeably in the remainder of this section. This is essentially an automated refinement of the mesh around the geometry that performs enough subdivision around its surface to enable maintaining a 2:1 balance across all levels while allowing for a sufficient number of temporal integrations on finer grids before synchronization along the refinement interface. After this construction, we identify a set of cell layers around the geometry surface on a specified grid level and compute the necessary cut-link lengths, storing them in a SoA format that is suitable for lookup during computation.
%This process is GPU-native in the sense that as soon as the geometry is uploaded to the GPU, no further communication of mesh or geometry metadata is required with the CPU to facilitate the embedding. 

The overall approach is essentially a manipulation of cell and cell-block flags. We pass this data into our previous refinement and coarsening pipeline \cite{Jaber2025} to construct the forest-of-octrees and to enforce 2:1 balancing.
%implemented in $\texttt{mesh\_amr.cuh}$ where mesh adaptation proceeds according to the original process of \cite{Jaber2025} with some modifications to reduce the costs of setting up intermediate arrays (\ref{sec:refcoarse}).
Voxelization and refinement are sequentially carried out on the individual grid levels of the hierarchy to construct the octrees, terminating when the specified near-wall grid level is reached. Once the mesh has been adapted to the geometry, the total number of boundary cells adjacent to the geometry are counted, and memory is allocated on the GPU for storage of the face-cell linkage lengths. The final result, after computation of the link-lengths, is a minimal bookkeeping approach for implementing interpolated boundary conditions in the vicinity of the geometry. Incorporation of the geometry in this way does not depend on any specific solver -- the output could potentially be used in finite difference solvers or for the calculation of interpolated quantities in post-processing. However, our motivation is to extend an LBM solver that was previously limited to an axis-aligned Cartesian mesh.

    \subsection{GPU-based Forest-of-Octrees} \label{sec:amr_review}

\begin{figure}[t]
    \centering
    \begin{subfigure}[b]{0.32\textwidth}
        \centering
        \includegraphics[height=0.5\linewidth]{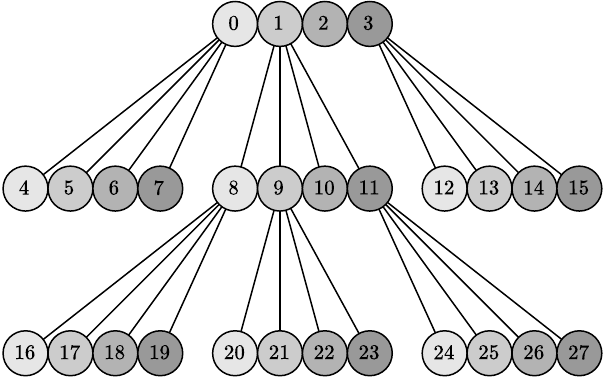}
        \caption{}
        \label{fig:lit_mesh_octree}
    \end{subfigure}
    \begin{subfigure}[b]{0.32\textwidth}
        \centering
        \includegraphics[height=0.5\linewidth]{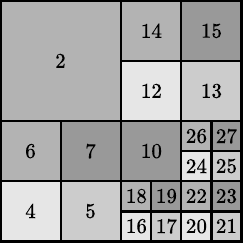}
        \caption{}
        \label{fig:lit_mesh_grid}
    \end{subfigure}
    \begin{subfigure}[b]{0.32\textwidth}
        \centering
        \includegraphics[height=0.5\linewidth]{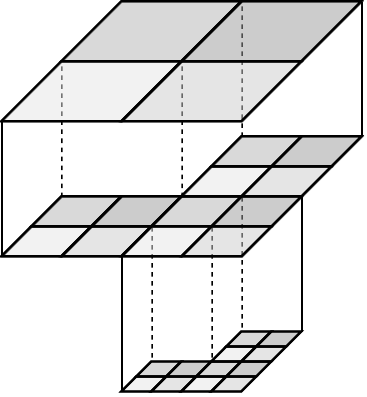}
        \caption{}
        \label{fig:lit_mesh_hierarchical}
    \end{subfigure}
    \caption{Representations of the hierarchical grid: a) octree, b) flattened grid, c) hierarchical view of overlap.}
    \label{fig:lit_mesh}
\end{figure}

Our current methodology depends on an earlier framework \cite{Jaber2025} for data-parallel construction and adaptation of forest-of-octrees meshes (Figure \ref{fig:lit_mesh}) on GPUs. We will briefly summarize how the mesh is organized in memory and adapted dynamically so that the procedures outlined in this paper are made clearer.

Although we refer to octrees for the remainder of this paper, our discussion is also valid for quadtrees in 2D. An octree here is a tree data structure where each node can be subdivided into exactly eight child nodes. A subdivided node (or internal node) is denoted as the parent node to its children, and a node that has not been subdivided is denoted as a leaf. The root node is the unique node without a parent. The depth of a node is the number of parent-child links between it and the root node. The nodes correspond to cell-blocks in the computational grid each with size $4^D$. The forest-of-octrees is therefore referred to as the grid hierarchy, and the set of nodes with depth $L$ across the forest is referred to as the grid level $L$. Metadata stored explicitly include the spatial coordinates of the blocks, their level (the depth in the tree), neighbor-links, and mask IDs that indicate participation in different subroutines of the grid refinement scheme. We use Id lists to store active block Ids by grid level, and a complementary list of available Ids that can be used to store data in allocated memory on the GPU.
%It is common for octree-based AMR to utilize space filling curves \cite{Weinzierl2011,p4est,Schornbaum2018,Menshov2020,Wang2024}, however, we depart from this representation and instead use Id lists...

Mesh connectivity is described with two arrays: neighbor indices that point to the location of metadata for neighboring blocks, and `neighbor-children' which are the indices of the first child of each neighboring block. Cell-blocks store $N_{Q,\text{max}}$ values of each index so that they are connected with a full halo of neighbors. Other metadata include refinement Ids indicating cell-block status during the refinement and coarsening procedure, grid level, spatial coordinates (measured from the lower left corner), an integer mask value indicating whether at least one group of cells in the block participates in inter-grid communication, and an integer value indicating whether the block is in the vicinity of the domain boundaries.

Grid communication is intra-level when blocks on the same grid level exchange data, and inter-level when data is exchanged along the refinement interface. Cells and blocks are assigned integer masks that indicate how they participate during communication and mesh adaptation. Cells are classified as ghost cells if they acquire data via interpolation on the fine grid for temporal integration, interface cells if they participate in averaging from the fine grid to the coarse grid, interior cells if they do not participate in inter-level grid communication but lie within the fluid region, solid cells if the cell center lies within the solid region of the geometry, or boundary cells if they are in the interior but adjacent to at least one solid cell. Blocks can be classified as communicating if among their child blocks there are interface and/or ghost cells, solid if all cells in the block lie within the solid region, solid-boundary if at least one cell in the block is adjacent to a cell in the solid region, or regular otherwise. A 2:1 balance \cite{Sundar2008} of the forest-of-octrees is enforced at all times, meaning that the difference in depth between adjacent nodes never exceeds 1 at any point.
\begin{figure}[t]
    \centering
    \begin{subfigure}[b]{0.6\textwidth}
        \centering
        \includegraphics[height=3.5cm]{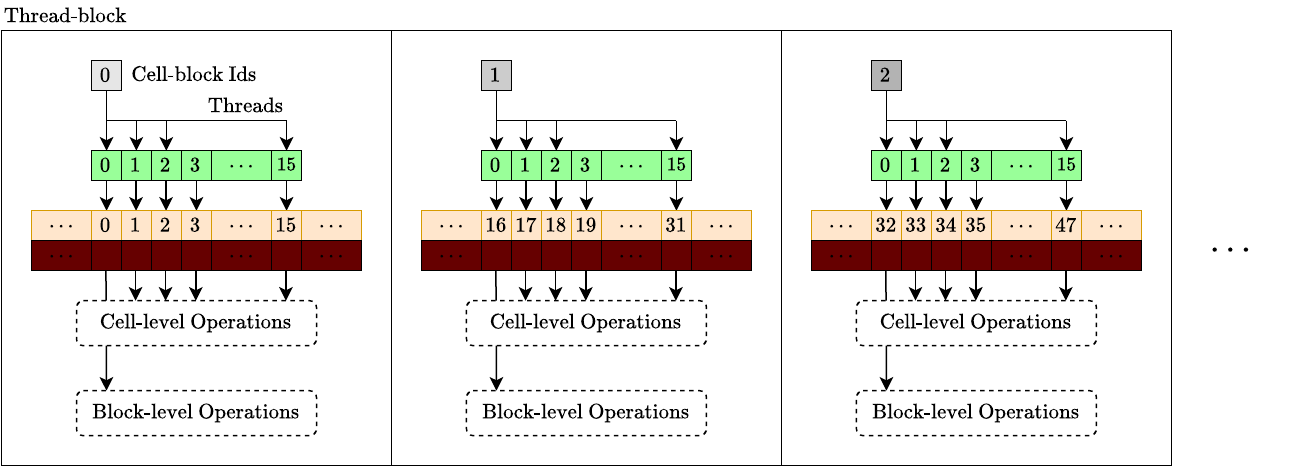}
        \caption{Primary mode.}
        \label{fig:lit_access_primary}
    \end{subfigure}
    \begin{subfigure}[b]{0.35\textwidth}
        \centering
        \includegraphics[height=3.5cm]{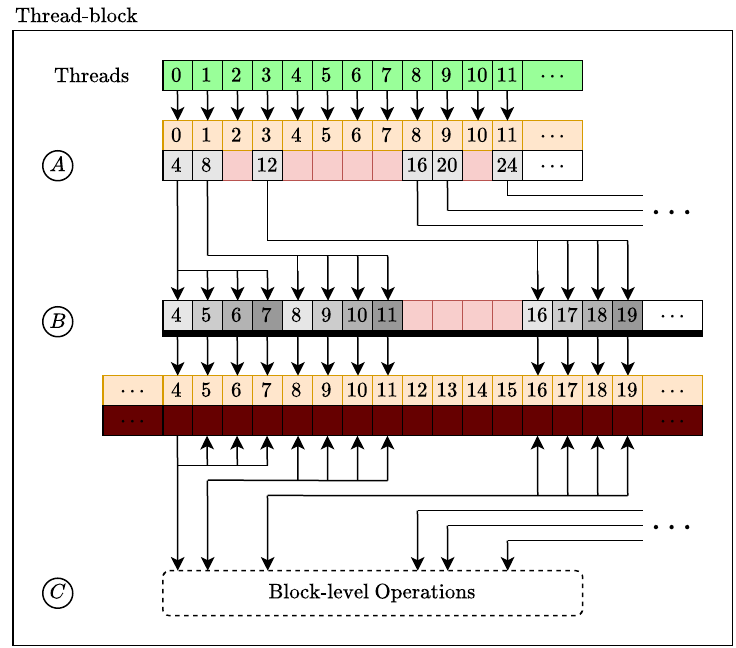}
        \caption{Secondary mode.}
        \label{fig:lit_access_secondary}
    \end{subfigure}
    \caption{Memory access patterns used to process cell and cell-block data.}
    \label{fig:lit_access}
\end{figure}

The mesh is manipulated entirely on the GPU in an eight-step procedure, in addition to a preparation step prior to invocation of the routine. For preparation, data is averaged from the finest grid up to the root grid to ensure that coarse cells are up to date. Interpolation then proceeds starting from the root grid down to the finest level to ensure that ghost cells are also up to date. The specified refinement criterion is then computed, and marks for refinement or coarsening are applied to the appropriate blocks. The procedure is as follows:
\begin{enumerate}[noitemsep]
    \item The set of blocks participating in the update is identified (these are either marked for refinement/coarsening or adjacent to marked blocks),
    \item New child Ids are assigned to blocks marked for refinement,
    \item Blocks violating 2:1 balancing are prevented from being coarsened,
    \item Child block metadata is constructed from their parents' data,
    \item Removed blocks are eliminated from the Id sets and added to the gap set,
    \item New blocks are added to the Id sets and decremented from the gap set,
    \item Branch-leaf relationships and communicator blocks are identified,
    \item New blocks calculate the indices of their neighbors for explicit storage and cell masks are updated.
\end{enumerate}

To implement the methodology proposed in this paper, we introduce post-voxelization metadata (such as the computed link-lengths in cells adjacent to the geometry surface). Memory for these metadata is allocated after voxelization and grid hierarchy construction are completed so that only the amount required is utilized. Since the geometry is stationary in this work, the allocation is performed once and metadata is reused in all iterations of the fluid solver. We use the number of cell-blocks adjacent to a solid and multiply it by the current block size to get the size of the memory allocation, which ensures proper alignment during thread-cell assignment. This new memory allocation does not follow the usual convention for cell metadata where the cell-block indices enumerate the arrays. Instead, we introduce an auxiliary set of indices that map cell-block indices to the correct locations of the post-voxelization metadata.

We employ two templates for memory access of cell-block data at the cell and block levels, denoted respectively as the primary and secondary modes of access (Figure \ref{fig:lit_access}). The primary mode is cell-parallel, and is structured so that one thread-block is assigned to each cell-block and one thread maps exactly to each cell. Kernels are each provided an Id set, along with its length. The resulting location of the cell-block data is mapped via $\kappa M_b + t = \texttt{id\_set}[\texttt{blockIdx.x}] \ M_b + \texttt{threadIdx.x}$ where $M_b=M_t$ is the cell-block size (which is always set equal to the thread-block size), $\kappa$ is the global cell-block index, and $t=\texttt{threadIdx.x}$ is the thread index. The secondary mode is cell-block-parallel based on a fixed, user-specified thread-block size $\overline{M_t}$ (not necessarily equal to $M_t$). This mode of access minimizes the number of transactions needed when making decisions based on neighbor or child data. Each cell-block loads its neighbor's (or child) cell-block indices into shared memory in such a way so as to recover order in the indices (based on the idea that cell-blocks are inserted during refinement in groups of 4 and 8 in 2D and 3D, respectively). These indices are then traversed within the kernel, and data is subsequently loaded and processed. Afterwards, the indices can be replaced by the results since they are no longer needed. A final traversal of the original cell-block indices now enables decision-making based on the data accessed in an uncoalesced way but in shared memory rather than global memory.

%Both modes of access are employed explicitly to enable embedding of complex geometries natively on the GPU.

%We will address how cell and blocks are modified to incorporate the geometry in a manner that is consistent with this procedure.

%The data type for this quantity matches the type chosen for the Geometry class during instantiation.
%One may either recompute link-lengths during each imposition of boundary conditions, or compute it once and store it permanently in an array. The latter requires up to 26 lengths per boundary node, however, the former is computationally expensive. Assuming the geometry is stationary, we can simply allocate memory strictly for boundary nodes once during voxelization so that the cost of storing the numerous lengths are minimized. 

%We are also interested in storing force contributions on a cell-block basis. The force calculation procedure is parallelized by having the cells compute their contribution and then reducing this result at the end per block. If only the MEA is being used for force calculation, this memory allocation can follow the same convention as above since only boundary nodes contribute to the computation. However, if a control volume approach is being used, then the total momentum in the interior of the volume also needs to be accounted for. For simplicity of implementation, we allow this array to be sized in the original way according the maximum number of cell-blocks that is calculated after surveying free memory during initialization of the Mesh object.

    \subsection{Spatial Binning}

In naive solid voxelization, each cell in the computational grid traverses the full set of faces to perform an orientation test, which determines whether or not the cell is inside the fluid or the solid. In a cell-parallel GPU implementation of this approach, one thread would be assigned to each cell to perform the required computations (e.g. based on ray casts). The numerous global memory accesses required to access vertex data for relatively simple orientation tests would result in a memory-bound program with low arithmetic intensity. However, cells directly adjacent to high-resolution geometries only need to search within a relatively small subset of the faces to determine adjacency. We therefore utilize spatial binning to reduce the search space for the small cell-blocks of the current framework. Each bin contains a reduced set of faces, which enables efficient cell-parallel execution suitable for the current block-based grid. We employ a hierarchical set of spatial bins where the sizes of the subsets of faces are proportional to the resolution on each grid level. Each level of this hierarchy contains a uniform grid of bins.

\subsubsection{Hierarchical Binning}

Spatial binning is defined in the \texttt{Geometry::Bin} class and takes as input the dimensions of the computational grid $(l_x,l_y,l_z)$, its root-grid resolution $(N_x)$ along $x$, a `bin density' $B$ that defines the resolution of the grid of bins, a maximum number of bin levels $L_{\text{max},B}$, and a value $N_{\text{spec.}}$ that imposes a maximum number of bins that a single face can overlap. The binning routines are called prior to voxelization and near-wall refinement. We introduce separate $(D-1)$- and $D$-dimensional sets of spatial bins for partial surface voxelization in one-step and two-step approaches (i.e., involving one and two kernels, simultaneously). With the former, a larger set of faces is exposed by extending the bin across $x$. This extended set is then used to flag cells in the whole grid simultaneously within a single kernel. A standard intersection parity test is used as the criterion for flagging. The latter set of bins employs the two-step voxelization procedure outlined earlier.
The $(D-1)$-dimensional bins are discretized along $y$ in 2D (or $(y,z)$ in 3D) and span the domain along $x$. In preliminary numerical experiments, we found that the first approach was highly memory-intensive even after reducing the total search space to a small window. The measured execution times were an order of magnitude higher than those obtained from the second approach due to the larger volume of considered faces per cell-block, and the increased total number of cell-blocks processed in the kernel. The remainder of the methodology and the numerical tests that follow will therefore rely only on the $D$-dimensional bins. We have retained the $(D-1)$-dimensional routines for completeness.

The output of the module is a set of three integer arrays for each bin level $0 \leq L < L_{\text{max},B}$, storing 1) the indices of binned faces in AoS format ($\texttt{bins\_face\_ids\_3D}$), 2) the total number of faces in each bin ($\texttt{bins\_face\_ids\_n\_3D}$), and 3) the offsets (or starting index) in the binned index array for each bin ($\texttt{bins\_face\_ids\_N\_3D}$). These are passed into the partial surface voxelization kernel, and define which faces cells need to consider when determining adjacency to the geometry.

The following procedure generates spatial bins of a given dimensionality $D'\in\{D-1,D\}$ on grid level $L$:
\begin{enumerate}[noitemsep]
    \item Filter the faces of the geometry that do not participate in voxelization or link-length computation.
    \item Compute the face axis-aligned bounding boxes (AABBs), and determine which bins these faces overlap.
    \item Perform stream compaction to remove unused elements in the face-bin index arrays.
    \item Sort the face indices using the bin indices as keys.
    \item Extract the unique bins and their total count.
    \item Scatter the bin sizes to their respective locations by index.
    \item Identify where the bin index changes in the sorted array.
    \item Transform the values at these positions into starting indices.
    %These will be used to retrieve the face indices of each bin.
    \item Scatter these starting indices by bin index.
\end{enumerate}

The first step is designed around the observation that most faces eventually lie in between cells without the possibility of an interaction with ray casts. The faces of a high-resolution geometry can therefore be heavily filtered for processing by coarse grids. After filtering, the faces are mapped to the appropriate bins using their bounding boxes. This is possible since the bins are arranged in a uniform grid. AABB overlap tests are used to filter out bins that lie in the bounding box but do not actually overlap the actual faces. The remaining steps are performed in a precise sequence to generate the output arrays based primarily on standard parallel algorithms implemented in the Thrust library.

In voxelization, rays only need to be cast along $x$ since partial surface voxelization relies on ray casts in this single direction. However, when computing the lengths of links connected to boundary cells that cut into the geometry, the full set of rays defined by the velocity set needs to be considered. Consequently, we define two versions of this kernel: 1) \texttt{Cu\_ComputeVoxelRayIndicators\_1D}, which is utilized by the partial surface voxelization kernel, and 2) \texttt{Cu\_ComputeVoxelRayIndicators\_MD}, which is utilized only during link-length computation.

In principle, ray cast intersections are complemented with point-in-line/triangle tests which can be implemented as a sequence of orientation tests. However, round-off errors may result in missed faces. In the absence of exact floating-point arithmetic, we instead utilize a triangle-AABB overlap test (Algorithm \ref{alg:triangleAABB}) in the 1D variation of the kernel. The AABB is defined with a width and height of machine epsilon, and a length that spans the domain along $x$. This represents a more inclusive test which slightly increases the total number of faces considered but ensures that the faces required for voxelization are all considered.

\subsubsection{Implementation}

For the first step, we compute integer `ray cast indicators' in a custom kernel \texttt{Cu\_ComputeVoxelRayIndicators} (Algorithm \ref{alg:ray_indicator}), which mark the faces that participate in voxelization. The kernel employs a face-parallel traversal, where each thread casts rays in the vicinity of each face to test for intersections. If at least one intersection is detected, an indicator is assigned to the face to mark it for consideration for binning (which results in filtering out the remainder). Indicator computation is followed by stream compaction and scatter operations, which provide these faces with indices that map to a more compact subset. The corresponding reduction in the total set of faces improves efficiency in subsequent steps.
\begin{figure}[b]
    \centering
    \begin{subfigure}[b]{0.31\textwidth}
        \centering
        \includegraphics[height=3cm]{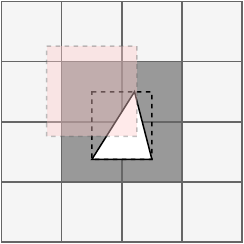}
        \caption{$N_{\text{spec.}}=2$}
        \label{fig:method_nspec_1}
    \end{subfigure}
    \hfill
    \begin{subfigure}[b]{0.31\textwidth}
        \centering
        \includegraphics[height=3cm]{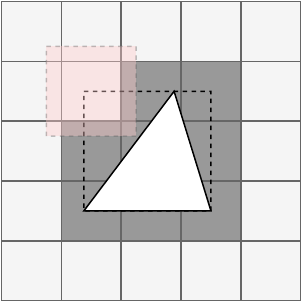}
        \caption{$N_{\text{spec.}}=3$}
        \label{fig:method_nspec_2}
    \end{subfigure}
    \hfill
    \begin{subfigure}[b]{0.31\textwidth}
        \centering
        \includegraphics[height=3cm]{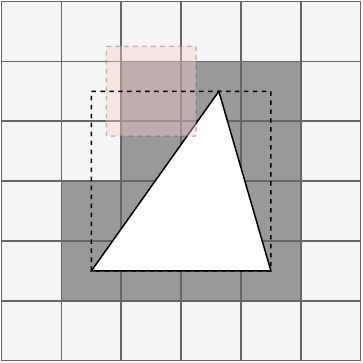}
        \caption{$N_{\text{spec.}}=4$}
        \label{fig:method_nspec_3}
    \end{subfigure}
    \caption{The variable $N_{\text{spec.}}$ enforces the potential range of bins that a face can occupy by restricting the maximum length of the edges. The actual set of bins considered at runtime is determined by face-AABB overlap tests, where the AABB is defined by the corners of each bin with an offset of cell-spacing $\Delta x_L$ on grid level $L$. (\textcolor{gray}{\rule{1ex}{1ex}}: potential bin, \textcolor{darkgray}{\rule{1ex}{1ex}}: overlapped bin, \textcolor{red}{\rule{1ex}{1ex}}: overlap region)}
    \label{fig:method_nspec}
\end{figure}

\begin{algorithm}[H]
\small
\caption{\texttt{Cu\_ComputeVoxelRayIndicators\_MD}: Ray Indicator Computation}
\label{alg:ray_indicator}
\SetAlgoLined
\DontPrintSemicolon

\tcp{Preparation}
Load vertex data $\textbf{v}_1,\textbf{v}_2,\textbf{v}_3$ of triangle $\mathcal{T}_t$\;
Compute global grid index $\textbf{I}_{\min,t} = \left(\lfloor \min_k v_{k,d} \rfloor \right)_{0\leq d < D}^T$\;
Compute global grid index $\textbf{I}_{\max,t} = \left(\lfloor \max_k v_{k,d} \rfloor \right)_{0\leq d < D}^T$\;
Initialize intersection indicator $s \leftarrow 0$\;
\vspace{0.5em}

\tcp{Check different directions}
\For{$p \gets 0$ \KwTo $N_{Q,\max} - 1$}{
    \If{$p = 26$ \textbf{or} $(\text{mod}(p - 1, 2) = 0$ \textbf{and} $p < 25)$}{
        \ForEach{$\textbf{I}_t$ such that $\textbf{I}_{\min,t} \leq \textbf{I}_t \leq \textbf{I}_{\max,t}$}{
            Compute lattice node coordinates $\textbf{v}_t = (\Delta x_L / 2)\textbf{1} + \Delta x \, \textbf{I}_t$\;
            Compute distance $d_p = ((\textbf{v}_1 - \textbf{v}_t) \cdot \textbf{n}) / (\textbf{c}_p \cdot \textbf{n})$\;
            Compute intersection point $\textbf{v}_i = \textbf{v}_t + d_p \, \textbf{c}_p$\;
            \If{$\textbf{v}_i \in \mathcal{T}_t$}{
                Set $s \leftarrow 1$\;
                Set $\textbf{I}_{\max,t} \leftarrow \textbf{I}_{\min,t} - \textbf{1}$\;
            }
        }
    }
    \If{$s = 1$}{
        \textbf{break}\;
    }
}
\vspace{0.5em}

\tcp{Store result}
\If{$s = 1$}{
    Store indicator global face index in ray indicator array\;
}
\end{algorithm}

% Cu_ComputeVoxelRayIndicators_MD

% 1- Load vertex data $\textbf{v}_1,\textbf{v}_2,\textbf{v}_3$ of triangle $\mathcal{T}_t$.
% 2- Compute global grid index $\textbf{I}_{\text{min},t} = \lfloor \min_{k} \textbf{v}_k  \rfloor$ of the minimum bounding box corner.
% 3- Compute global grid index $\textbf{I}_{\text{max},t} = \lfloor \max_{k} \textbf{v}_k  \rfloor$ of the maximum bounding box corner.
% 4- Initialize intersection indicator $s = 0$.
% 5- For $0 \leq p < N_{Q,\text{max}}$:
%    a) If ($p==26$ or ($\mod((p-1),2) = 0$ and $p < 25$)):
%       For ($\textbf{I}_{\text{min},t} \leq \textbf{I}_t \leq \textbf{I}_{\text{max},t}$):
%           1) Compute lattice node coordinates $\textbf{v}_t = (\Delta x_L/2)\textbf{1} + \Delta x \textbf{I}_t$
%           2) Compute distance $d_p = (\textbf{v}_1-\textbf{v}_t)\cdot \textbf{n} / \textbf{c}_p \cdot \textbf{n} $.
%           3) Compute intersection point $\textbf{v}_i = \textbf{v}_t + d \textbf{c}_p$.
%           4) If ($\textbf{v}_i \in \mathcal{T}_t$), then:
%              a) Set $s = 1$.
%              b) Set $\textbf{I}_{\text{max},t} = \textbf{I}_{\text{min},t} - \textbf{1}$.
%    b) If $s = 1$, then break from for loop.    
% 6- If $s = 1$, then store an indicator global face index in ray indicator array.

In Step 2, the faces are traversed again in the custom kernel \texttt{Cu\_ComputeBoundingBoxLimits3D} (Algorithm \ref{alg:bounding_box}). We identify all pairs $(f,b)$ of face index $f$ and bin index $b$ that satisfy a face-bin overlap test. These are stored using two arrays $\texttt{c\_bounding\_box\_limits}$ and $\texttt{c\_bounding\_box\_index\_limits}$ arranged in a SoA format. This operation requires the user-specification of a minimum bin cover (Figure \ref{fig:method_nspec}) $N_{\text{lim.}} = (2+N_{\text{spec.}})^{D'}$ for each face. In $D$-dimensional binning, the volumes of the considered bins are extended in all directions by $\Delta x_L$ so that cells can access faces that might lie outside of their respective bins while still in their vicinity. Face subdivision is required to enforce the minimum bin cover, as the implementation accepts arbitrary geometries. We implement a recursive geometry mesh refinement routine $\texttt{RecursiveRefineFace}$ based on midpoint subdivision. This routine is executed on the host side prior to binning, and forces the maximum length of any face to be less than the corresponding specified length
\begin{align}
    l_{\text{spec.}} = \min_{0 \leq i < D} l_{x_i} \times \left( 0.95 N_{\text{spec.}} \right) \big/ \left( 2^{L_{\text{max}}-1} N_B  \right).
\end{align}

%\noindent that is executed on the host side prior to spatial binning to accommodate $N_{\text{spec.}}$.
Memory for the two arrays is allocated based on the number of filtered faces $N_{\text{filtered}}$ and $N_{\text{spec.}}$ for a total of $N_{\text{filtered}} N_{\text{lim.}}$ integers each. The values of the former array are reset to $N_{\text{bins},L}$, and the latter to $-1$ to indicate an unused index for stream compaction in the next step. Each face first computes the bounding box from its vertices and snaps its corners based on the specified bin density. The snapped corners recover the loop limits for the bin index range covered by the face. Each thread then loops over these indices and performs a triangle-AABB intersection test to determine whether to include the current bin among the pairs. If so, the bin and face indices are written to $\texttt{c\_bounding\_box\_limits}$ and $\texttt{c\_bounding\_box\_index\_limits}$, respectively.

\begin{algorithm}[H]
\small
\caption{\texttt{Cu\_ComputeBoundingBoxLimits3D}: Bounding Box Triangle Binning on the GPU}
\label{alg:bounding_box}
\SetAlgoLined
\DontPrintSemicolon

\tcp{Definitions}
\textbf{Define:} $\texttt{LINEAR}(\mathbf{i}, N) = i_x + N i_y + N^2 i_z$\;
\vspace{0.5em}

Initialize map index $m \gets t$\;

\If{using ray indicator mapping}{
    Update $m$ with value from ray indicator array\;
}

\If{$m \geq 0$}{
    \tcp{Prepare triangle data}
    Load vertex data $\textbf{v}_1, \textbf{v}_2, \textbf{v}_3$ of triangle $\mathcal{T}_t$\;

    Compute lower triangle bounding box corner $\textbf{v}_{\mathcal{T},m} = \left( \min_k \textbf{v}_k \right)_{0 \leq d < D}$\;
    Compute upper triangle bounding box corner $\textbf{v}_{\mathcal{T},M} = \left( \max_k \textbf{v}_k \right)_{0 \leq d < D}$\;

    Initialize boolean $C \gets \texttt{true}$\;

    \If{$\mathcal{T}_t$ is outside domain}{
        $C \gets \texttt{false}$\;
    }
    \vspace{0.5em}

    \tcp{Loop over potential bins}
    \If{$C$}{
        Compute $\textbf{I}_{\min,t} = \left( \lfloor B(\textbf{v}_{\mathcal{T},m})_d \rfloor \right)_{0 \leq d < D} - \textbf{1}$\;
        Compute $\textbf{I}_{\max,t} = \left( \lfloor B(\textbf{v}_{\mathcal{T},M})_d \rfloor \right)_{0 \leq d < D} + \textbf{1}$\;

        Initialize counter $v \gets 0$\;

        \ForEach{$\textbf{I}_t$ such that $\textbf{I}_{\min,t} \leq \textbf{I}_t \leq \textbf{I}_{\max,t}$}{
            Compute $\textbf{v}_m = (\textbf{I}_t \circ \textbf{l}) - \Delta x_L \cdot \textbf{1}$\;
            Compute $\textbf{v}_M = ((\textbf{I}_t + \textbf{1}) \circ \textbf{l}) + \Delta x_L \cdot \textbf{1}$\;

            Update $C \gets \texttt{TriangleBinOverlap3D}(\textbf{v}_m, \textbf{v}_M, \textbf{v}_1, \textbf{v}_2, \textbf{v}_3)$\;

            \If{$C = \texttt{true}$ \textbf{and} $v < (2 + N_{\text{spec.}})^D$}{
                Compute global bin index: $I_B \gets \texttt{LINEAR}(\textbf{I}_t, 4)$\;
                Write $I_B$ to bounding box limits array\;
                Write $t$ to bounding box index limits array\;
                $v \gets v + 1$\;
            }
        }
    }
}
\end{algorithm}

\begin{figure}[t]
    \centering
    \includegraphics[width=0.35\linewidth]{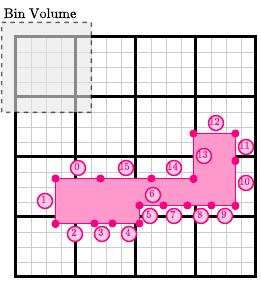}
    \caption{A sample 2D geometry with 16 faces and $N_{\text{spec.}}=1$ that we used to illustrate the construction of bins on a given grid level. Shown also is the bin overlap used to include nearby faces that do not necessarily intersect the bin directly but still require inclusion to account for ray casts from cells at the edges.}
    \label{fig:binning_sample_geom}
\end{figure}
\begin{figure}[b]
    \centering
    \begin{subfigure}[b]{0.48\textwidth}
        \centering
        \includegraphics[width=1\linewidth]{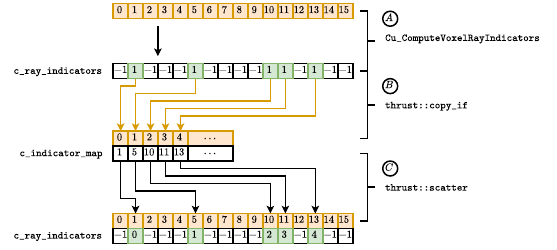}
        \caption{Step 1}
    \end{subfigure}
    \begin{subfigure}[b]{0.48\textwidth}
        \centering
        \includegraphics[width=1\linewidth]{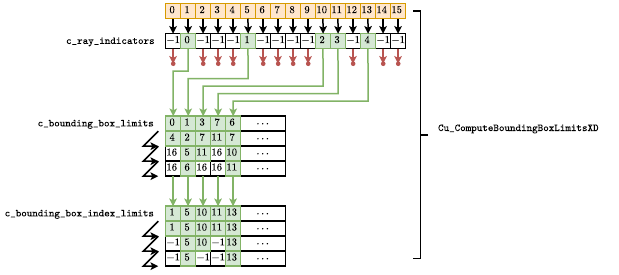}
        \caption{Step 2}
    \end{subfigure}
    \caption{Application of the GPU-based uniform spatial binning algorithm to the sample set of faces from Figure \ref{fig:binning_sample_geom}.}
    \label{fig:binning_sample_geom_applied_S12}
\end{figure}

Step 3 is a stream compaction (via Thrust's \texttt{remove\_if}) of the two arrays zipped together. The remaining values of $N_{\text{bins},L}$ in $\texttt{c\_bounding\_box\_limits}$ are filtered out so that only valid index pairs remain. With $\texttt{thrust::sort\_by\_key}$, Step 4 now sorts the face indices in $\texttt{c\_bounding\_box\_index\_limits}$ using the values of $\texttt{c\_bounding\_box\_limits}$ as keys to get contiguous sets of face indices that belong to the same bins.

It now remains to extract the number of bins on level $L$ and to identify their starting indices. In Step 5, $\texttt{thrust::reduce\_by\_key}$ is called on a constant iterator with a value of unity. The sorted bin indices are used as keys to obtain both the unique set of bins and their sizes. Step 6 scatters the bin sizes to $\texttt{bins\_face\_ids\_n\_3D}$ using the unique bin index set as the map, placing them in the correct locations. Step 7 calls $\texttt{thrust::adjacent\_difference}$ on the sorted $\texttt{c\_bounding\_box\_limits}$ to identify the indices where two consecutive indices correspond to differing bins. Step 8 begins with a transformation (with $\texttt{thrust::transform}$) that replaces positive difference values with their locations in memory. A stream compaction operation follows that produces a contiguous set of these starting indices. A final scattering to the locations of their corresponding bins (with the unique bins still stored in an intermediate array) in $\texttt{bins\_face\_ids\_N\_3D}$ forms Step 9, which completes the construction of the bins on this grid level.

The construction is repeated separately for $(D-1)$- and $D$-dimensional bins and for each grid level, but memory allocations for intermediate arrays are retained for constructions on the same level. The whole procedure requires six intermediate arrays to store the ray indicator values, the indices that map filtered faces to a compacted region, bin and face index pairs, unique bins indices, and the results of reduction/transformation. The first two require an allocation of $N_{\text{faces}}$ integers since potentially all faces participate in the binning process. As mentioned above, the bin-face index pair arrays are allocated an amount $N_{\text{filtered}} N_{\text{lim.}}$ after filtering is completed to reduce memory consumption. This can otherwise quickly become enormous for sufficiently large $N_{\text{faces}}$ combined with even relatively small values of $N_{\text{spec.}}$. The last two arrays require only $N_{\text{bins},L}$ integers.

\begin{figure}[t]
    \centering
    \includegraphics[width=0.75\linewidth]{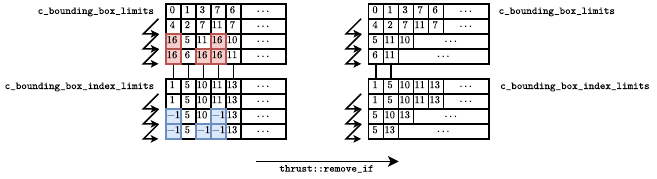}
    \caption{Application of the GPU-based uniform spatial binning algorithm to the sample set of faces from Figure \ref{fig:binning_sample_geom}.}
    \label{fig:binning_sample_geom_applied_S3}
\end{figure}

\begin{figure}[b]
    \centering
    \begin{subfigure}[b]{0.48\textwidth}
        \centering
        \includegraphics[width=1\linewidth]{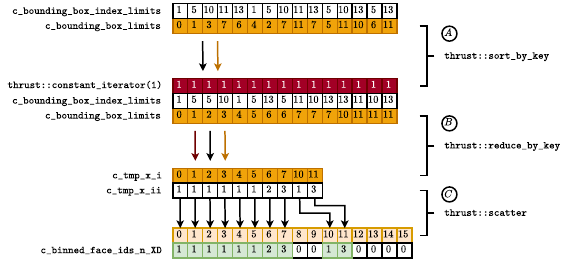}
        \caption{Step 4}
    \end{subfigure}
    \begin{subfigure}[b]{0.48\textwidth}
        \centering
        \includegraphics[width=1\linewidth]{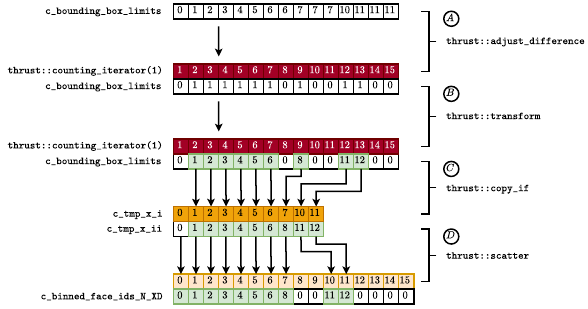}
        \caption{Step 5}
    \end{subfigure}
    \caption{Application of the GPU-based uniform spatial binning algorithm to the sample set of faces from Figure \ref{fig:binning_sample_geom}.}
    \label{fig:binning_sample_geom_applied_S45}
\end{figure}

A 2D example is visualized in Figure \ref{fig:binning_sample_geom}. The geometry is composed of 16 faces that are overlaid on a domain that has been discretized using a $4\times 4$ grid of bins. The volume extension applied during Step 2 is also shown. The binning procedure detailed above is illustrated for this example in Figures \ref{fig:binning_sample_geom_applied_S12}, \ref{fig:binning_sample_geom_applied_S3}, and \ref{fig:binning_sample_geom_applied_S45}.

    \subsection{Voxelization and Near-Wall Refinement}

We divide the voxelizer procedure into four subroutines. Solid voxelization is performed level-by-level, and the grid is refined near the surface accordingly, resulting in a top-down construction of the forest-of-octrees. Boundary cells are identified after all solids cells on a given level have been flagged. Memory is then allocated for the link-length lookup table based on the total number of boundary cells, and for auxiliary arrays for performing parallel force calculations. The procedure ends with a computation of the link-lengths on all boundary cells. We further subdivide the solid voxelization subroutine into six steps to carefully manage cell and cell-block flagging in a thread-safe manner. These steps also ensure that a sufficient number of cell-blocks are placed on subsequently finer grids to adequately achieve 2:1 balancing.
\begin{figure}[t]
    \centering
    \includegraphics[width=0.35\linewidth]{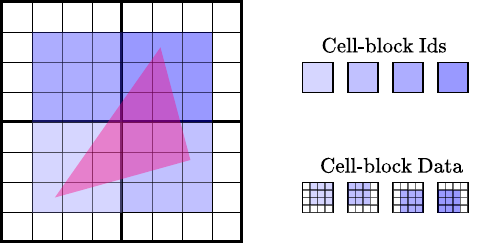}
    \caption{In a triangle-parallel approach to the current computational grid, threads may cross over several cell-blocks, requiring uncoalesced global memory loads.}
    \label{fig:voxelizer_coords_mapping}
\end{figure}

Since the cell-blocks are not arranged in any particular ordering in the current AMR grid (such as a SFC ordering or a traversal or a hash table), it is not necessarily desirable to assign individual threads to triangles (a typical data-parallel approach to voxelization in the literature). If the range of cells in the triangle bounding boxes covers several cell-blocks, the mapping of the Ids strictly from the coordinate limits of the range (Figure \ref{fig:voxelizer_coords_mapping}) to the locations of data in memory is non-trivial. Such data structures could be implemented in principle to enable this assignment. An SFC ordering of block data enables the computation of neighboring block Id encodings, but their location in memory typically requires a supplementary search in sparse storage (which is common on CPUs). A hash table can be assembled instead to enable $\mathcal{O}(1)$ access to the block Id locations. However, this requires care when implemented on GPUs to ensure high performance in the absence of data locality. Another concern is that a thread traversing cells across several cell-blocks incurs several global memory reads of the block Ids and writes to the corresponding data blocks, even with the current AMR grid setup. In addition to being generally uncoalesced without an initial ordering of the triangles (which would ensure that contiguous threads are writing to nearby cell-blocks), this approach may require atomic operations for thread safety in case the bounding box cell ranges intersect. We avoid constructing these data structures for these reasons, even though a combination of SFC and hash are encountered in the literature \cite{Hasbestan2018}.

We therefore depart from the usual rasterization algorithms and employ point-plane distance calculations via ray cast combined with point-in-face tests. Our partial surface voxelization procedure relies on the following assumptions about the input geometry: 1) it is watertight so that ray casts always intersect the geometry at some location, 2) it is not self-intersecting and there are no dangling edges. These assumptions allow us to use only local ray casts to determine if a given cell is inside the geometry.

\subsubsection{Partial Surface Voxelization}

Cells compute their respective bin indices to retrieve the indices $p$ of the faces $\mathcal{T}_p$ to be considered. Vertex data is retrieved from global memory one face at a time in a loop over $p$. For each face, a ray cast is performed from the center of each cell $\textbf{v}$ along $x$ as part of a point-plane distance calculation, where the plane is defined by the face's first vertex $(\textbf{v}_0)_p$ and its normal $\textbf{n}_p$. This results in a signed distance $d$ between $\textbf{v}$ and $\mathcal{T}_p$ that is computed with
\begin{align}
    d= \frac{((\textbf{v}_0)_p-\textbf{v})\cdot\textbf{n}_p}{\textbf{e}_x\cdot \textbf{n}_p}.
\end{align}

\noindent The point of intersection between the ray and the plane $\textbf{v}_i =\textbf{v} + d\textbf{e}_x$ is also computed. A point-in-face test is conducted for each intersection point with the face vertices. If this test passes, the shortest distance $d_{\text{min}}$ is retained. The point-in-face test is usually a sequence of half-plane tests defined, in the case of a triangle, by the edges. However, round-off errors resulted in false negatives in high-resolution computational grids due to a lack of exact floating-point arithmetic features in our implementation. Instead, a triangle-AABB test serves as the point-in-face test in this work, where the AABB is defined as a thin cuboid that spans the domain along $x$ with dimensions defined by floating-point epsilon. We found that this reduced false negatives in most grids. However, they were not totally eliminated when the grids were high-resolution.

After traversing the faces, each cell uses the sign of its shortest ray-plane links and the orientation of the corresponding face $\mathcal{T}_{p_\text{min}}$ to assign its type. If $n_{p_\text{min},x} d_\text{min} > 0$, the cell-center lies inside the geometry and the cell is flagged as a solid cell. However, if $n_{p_\text{min},x} d_\text{min} < 0$, the cell is flagged as a `guard' cell that will be used to manage the propagation of the solid flags as the interior of the geometry is filled.

\begin{figure}[b]
    \centering
    \begin{subfigure}[b]{0.3\textwidth}
        \centering
        \includegraphics[width=0.5\linewidth]{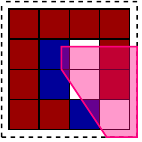}
        \caption{}
        \label{fig:method_prop_internal_1}
    \end{subfigure}
    \hfill
    \begin{subfigure}[b]{0.3\textwidth}
        \centering
        \includegraphics[width=0.5\linewidth]{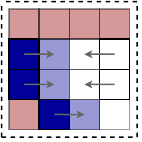}
        \caption{}
        \label{fig:method_prop_internal_2}
    \end{subfigure}
    \hfill
    \begin{subfigure}[b]{0.3\textwidth}
        \centering
        \includegraphics[width=0.5\linewidth]{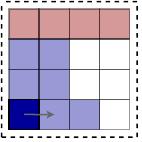}
        \caption{}
        \label{fig:method_prop_internal_3}
    \end{subfigure}
    \caption{(a) Cell classification is performed in the direct vicinity of the geometry. (b) Internal propagation begins; fluid cells look to their left and right and take on the values of solid and guard cells. (c) At the end, solid/guard cell masks are visible at the $\pm x$ edges. (\fbox{\textcolor{white}{\rule{1ex}{1ex}}}: solid cells, \textcolor{blue}{\rule{1ex}{1ex}}: guard cells, \textcolor{red}{\rule{1ex}{1ex}}: fluid cells)}
    \label{fig:method_prop_internal}
\end{figure}

Now that cells adjacent to the geometry surface are marked, those in its interior remain to be flagged. We avoid cell-granular propagation throughout the whole interior as this would necessarily involve thread-level data exchanges that further complicate the implementation. Instead, we perform an internal propagation of the flags (Figure \ref{fig:method_prop_internal}) to ensure that all cells in a scan line along $x$ within each cell-block are assigned a flag. When these flags are propagated to neighboring cell-blocks, a single comparison is evaluated by all cells in parallel to determine how their cell flags should be updated.
\setlength{\fboxrule}{0.4pt} % border thickness
\setlength{\fboxsep}{0pt}    % remove padding

Cells inspect the masks of their neighbors to the left and right in three iterations (which is sufficient for propagation from one side of a $4^D$ cell-block to the other). If a cell is currently a fluid cell and its neighbor is a guard, then it is switched to a guard. If it is not a guard cell, but its neighbor is a solid cell, then it is also marked solid. In cell-blocks where at least one cell was adjacent to the solid, new cell-masks are written into global memory.

The base implementation of this procedure (Algorithm \ref{alg:voxelize_block}) employs the primary mode of access as described in Section \ref{sec:amr_review}, where threads are assigned to individual cells. When cells within a cell-block lie in a single bin, face vertex data retrieval is achieved with a broadcast from global memory. Distance calculations, point-in-face tests, and internal propagation are all executed in parallel. Shared memory is used to communicate cell flags with their neighbors, and to perform the reduction of the indicators that determine if these flags need to be written.

\begin{algorithm*}
\small
\caption{\texttt{Cu\_Voxelize\_V1}: Partial Surface Voxelization}
\label{alg:voxelize_block}
\SetAlgoLined
\DontPrintSemicolon

%\KwData{Cell-block coordinates, bin mapping, face data}
%\KwResult{Updated cell mask values in global memory}

\tcp{Definitions}
\textbf{Define:} 
$V_S = \texttt{MaskCell::Solid}$\;
$V_G = \texttt{MaskCell::Guard}$\;
$V_{GH} = \texttt{MaskCell::Ghost}$\;
$V_I = \texttt{MaskCell::Interface}$\;
$\texttt{LINEAR}(\mathbf{i}, N) = i_x + N i_y + N^2 i_z$\;

\vspace{0.5em}

\tcp{Preparation}
Compute local cell indices $\mathbf{I}$ such that thread $t = \texttt{LINEAR}(\textbf{I},4)$\;
Compute cell coordinates: $\mathbf{v}_{t} = \mathbf{v}_b + (\mathbf{I} + (1/2)\textbf{1})\Delta x$\;
Compute global bin index: $\texttt{LINEAR}(\lfloor B\mathbf{x} \rfloor, B\mathbf{1})$\;
Retrieve number of bin faces $n_f$ and the starting index $N_f$\;
Retrieve original cell mask $H_{to}$ and store it\;
Initialize current cell mask $H_t$\;
Set $p_{\min} \gets -1$, $d_{\min} \gets 1$\;

\vspace{0.5em}

\tcp{Binned face traversal and cell mask identification}
\For{$p \gets 0$ \KwTo $n_f - 1$}{
    Load face vertex data $\mathbf{v}_1, \mathbf{v}_2, \mathbf{v}_3$ of face $\mathcal{T}_p$\;
    Compute norm $\textbf{n} = (\textbf{v}_2 - \textbf{v}_1)\times (\textbf{v}_3 - \textbf{v}_1) / || (\textbf{v}_2 - \textbf{v}_1)\times (\textbf{v}_3 - \textbf{v}_1) || $\;
    Compute $d = ({(\mathbf{v}_1 - \mathbf{v}_t) \cdot \mathbf{e}_x})/{n_x}$\;
    Compute intersection point: $\mathbf{v}_i = \mathbf{v}_t + d \mathbf{e}_x$\;
    
    \If{$|d| < d_{\min}$ \textbf{and} $\mathbf{v}_i \in \mathcal{T}_p$}{
        $p_{\min} \gets p$;\quad $d_{\min} \gets |d|$\;
        \eIf{$n_x \cdot d < 0$}{
            $H_t \gets V_S$\;
        }{
            $H_t \gets V_G$\;
        }
    }
}

\vspace{0.5em}

Store indicator $\eta = (p_{\min} \neq -1)$ in shared memory\;

Perform blockwise reduction over $\eta$ to determine if internal propagation is needed\;

\tcp{Internal propagation and global store}
\If{$\sum_t \eta_t > 0$}{
    \For{$l\gets 0$ \KwTo $3$}{
        \tcp{Consider neighbors $t'=\texttt{LINEAR}(\textbf{I}\pm \textbf{e}_x,4)$}
        Load cell masks into shared memory\;
        \If{$H_{t'} = V_G$ \textbf{and} $H_t = V_I$}{
            Set $H_t \gets V_G$\;
        }
        \If{$H_{t'} = V_S$ \textbf{and} $H_t \neq V_G$}{
            Set $H_t \gets V_S$\;
        }
    }
    \If{$H_t = V_S$ \textbf{or} $H_{to} \notin \{V_{GH}, V_I\}$}{
        Write $H_t$ to global memory\;
    }
}
\end{algorithm*}

\begin{figure}
    \centering
    \begin{subfigure}[t]{0.31\textwidth}
        \centering
        \includegraphics[height=4cm]{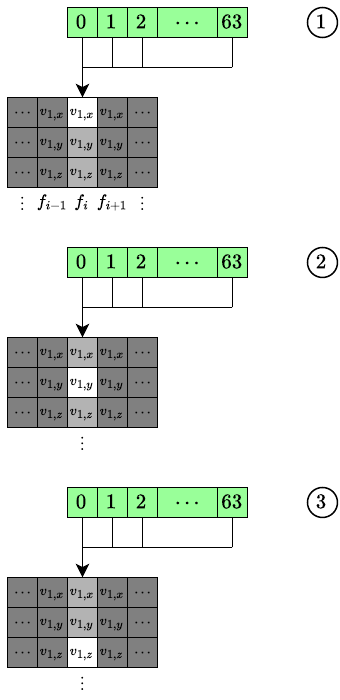}
        \caption{Variant 1 (SoA)}
        \label{fig:method_soavsaos_v1_soa}
    \end{subfigure}
    \begin{subfigure}[t]{0.31\textwidth}
        \centering
        \includegraphics[height=4cm]{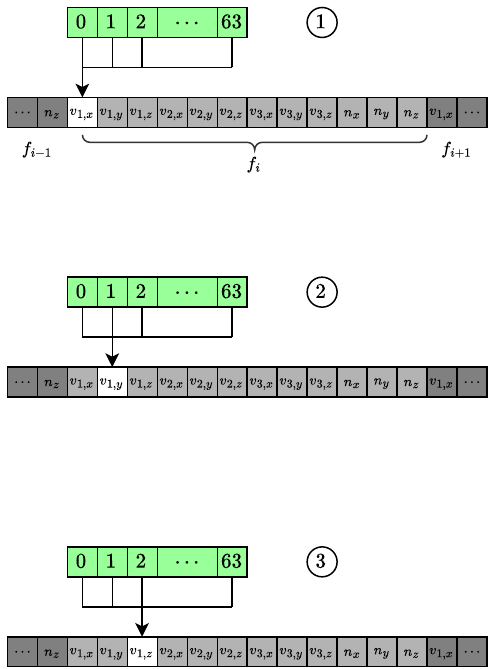}
        \caption{Variant 1 (AoS)}
        \label{fig:method_soavsaos_v1_aos}
    \end{subfigure}
    \begin{subfigure}[t]{0.31\textwidth}
        \centering
        \includegraphics[height=4cm]{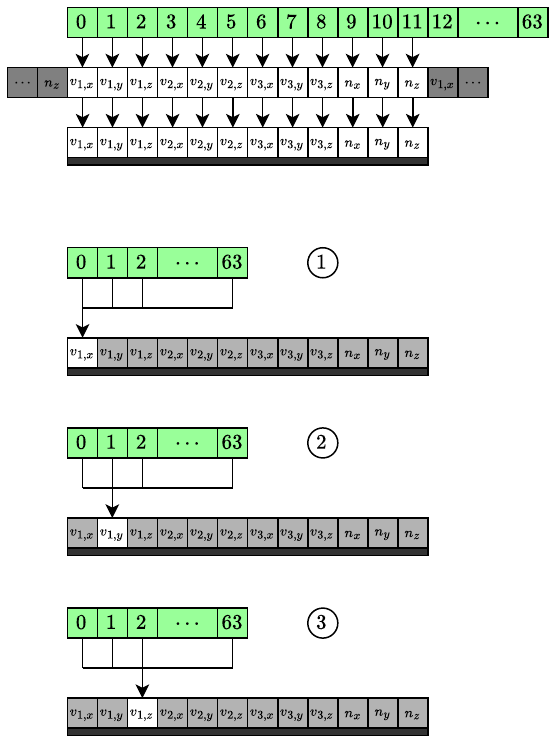}
        \caption{Variant 2}
        \label{fig:method_soavsaos_v2}
    \end{subfigure}
    \caption{Parallel accesses of the vertices of the first face in the different variants of the partial surface voxelization kernel.}
    \label{fig:soavsaos}
\end{figure}

Since vertex data is broadcast to individual threads, there are no repeated uncoalesced accesses within the warps. However, retrieval of this data becomes memory-intensive as the number of faces increases (even after spatial binning). The impact of data arrangement is not obvious when memory bandwidth becomes saturated as a result of the total volume of reads. This can potentially diminish the benefits of utilizing the cache hierarchy. The earlier spatial binning phase loaded vertex data arranged in a SoA format, where contiguity was established with respect to the faces. However, retrieval may result in cache hits if the data is arranged as an AoS instead. In this case, the individual addresses are part of a contiguous section of memory that was serviced during the read of the first vertex component. Retaining both formats of the vertex data doubles the geometry's GPU memory consumption, so it is worth investigating whether an array of structures format is necessary.

\begin{figure}
    \centering
    \begin{subfigure}[t]{0.31\textwidth}
        \centering
        \includegraphics[height=4cm]{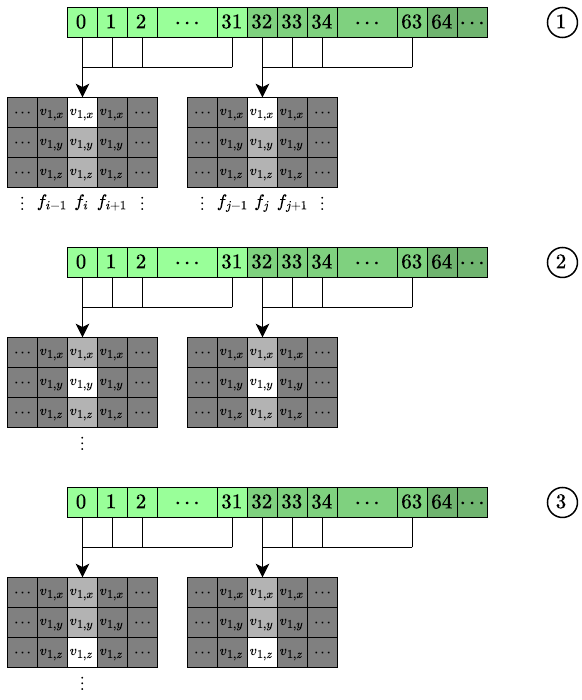}
        \caption{Variant 1 (SoA)}
        \label{fig:method_soavsaos_v1_soa_warp}
    \end{subfigure}
    \begin{subfigure}[t]{0.31\textwidth}
        \centering
        \includegraphics[height=4cm]{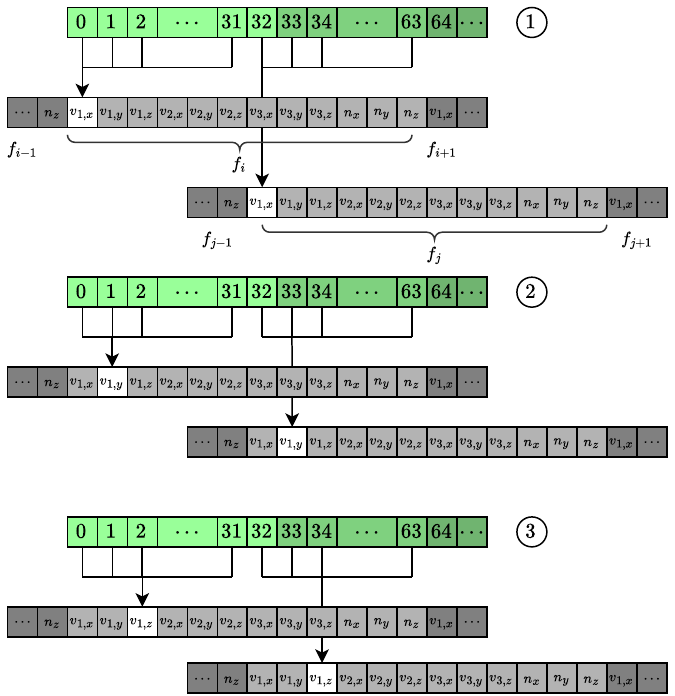}
        \caption{Variant 1 (AoS)}
        \label{fig:method_soavsaos_v1_aos_warp}
    \end{subfigure}
    \begin{subfigure}[t]{0.31\textwidth}
        \centering
        \includegraphics[height=4cm]{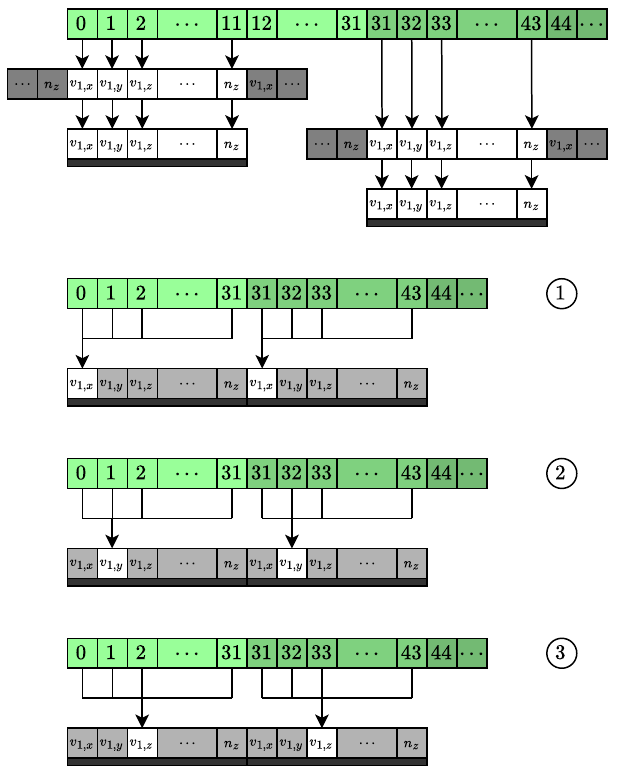}
        \caption{Variant 2}
        \label{fig:method_soavsaos_v2_warp}
    \end{subfigure}
    \caption{Warp-level approaches to the variants of the partial surface voxelization kernel (Figure \ref{fig:soavsaos}).}
    \label{fig:soavsaos_warp}
\end{figure}

We therefore developed a second variation (Algorithm \ref{alg:voxelize_warp}, Figure \ref{fig:soavsaos}) of this kernel to investigate the effects of the vertex data arrangement and parallel reads of the bin and face data on overall execution time. This variant still employs the primary mode of access. However, shared memory is used to store the data that is read from global memory. This is then broadcast to threads prior to computations for cell flag assignment. We also developed warp-level variants of this and the base kernel (Figure \ref{fig:soavsaos_warp}). These are launched so that individual warps process a full cell-block in 3D (in contrast with the primary mode of access, where two warps are needed). The latter kernels employ warp-level primitives to reduce the cell-flag indicators, and to broadcast vertex data without using shared memory.

\begin{algorithm}[H]
\small
\caption{\texttt{Cu\_Voxelize\_V2}: Partial Surface Voxelization with Parallel Face Reads}
\label{alg:voxelize_warp}
\SetAlgoLined
\DontPrintSemicolon

\tcp{Preparation}

Initialize shared memory arrays $s_{fI}, s_{fD}$ to store binned face indices and data\;
\tcp{Remaining preparation as per \texttt{Cu\_Voxelize\_V1} ($n_f, N_f$)}

\vspace{0.5em}

\tcp{Partial surface voxelization}
\For{$0 \leq j < \left\lfloor n_f / M_t \right\rfloor + 1$}{
    Set $p_{\text{lim}} \gets M_t$\;
    
    \If{$(j+1)M_t \geq n_f$}{
        Set $p_{\text{lim}} \gets M_t - ((j+1)M_t - n_f)$\;
    }

    \If{$t < p_{\text{lim}}$}{
        Parallel read $s_{fI}[t] \gets$ global index of face $\mathcal{T}_{jM_t + t}$\;
    }

    \For{$0 \leq p < p_{\text{lim}}$}{
        Parallel read vertex data $\textbf{v}_1,\textbf{v}_2,\textbf{v}_3$ for face $\mathcal{T}_{jM_t + p}$ using index $s_{fI}[p]$, store in $s_{fD}$\;
        Each cell loads data for calculations component by component from $s_{fD}$\;
        \tcp{Do partial surface voxelization computations as per \texttt{Cu\_Voxelize\_V1}}
    }
}

\tcp{Blockwise reduction, internal propagation, and global write as per \texttt{Cu\_Voxelize\_V1}}

\end{algorithm}

% Cu_Voxelize_V2
%
% 1. Initialize shared memory array $s_{fI,t} \gets -1 \quad \forall t$.
% 2. Compute local cell indices $\mathbf{I}$.
% 3. Compute cell coordinates: $\mathbf{v}_p = \mathbf{v}_b + (\mathbf{I} + \tfrac{1}{2})\Delta x$.
% 4. Compute the global bin index: LINEAR$(\lfloor B\mathbf{x} \rfloor, B\mathbf{1})$.
% 5. Get the bin data: number of faces $n_f$ and starting index $N_f$.
% 6. Retrieve the current cell mask $V_{to}$ and store for later. Initialize a current cell mask $V_t$.
%
%
% 1. Initialize shared memory arrays $s_{fI}, s_{fD}$ to store binned face indices and data.
% 2. Remaining preparation as per \texttt{Cu\_Voxelize\_V1} ($n_f, N_f$).
% 3. For $0 \leq j < n_f/M_t + 1$:
%    a) Set $p_{\text{lim}} = M_t$.
%    b) If $(j+1)M_t \geq n_f$ then set $p_{\text{lim}} = M_t - ((j+1)M_t - n_f)$.
%    c) If $t < p_{\text{lim}}$, then set $s_{fI}[t]$ to global index of face $\mathcal{T}_{jM_t + t}$.
%    d) For $0 \leq p < p_{\text{lim}}$:
%       i) Get vertex data $\mathbf{v}_{1 \leq k \leq 3}$ for face $\mathcal{T}_{jM_t + p}$ using index $s_{fI}[p]$.
%       ii) Do partial surface voxelization computations as per \texttt{Cu\_Voxelize\_V1}.
%
% 4. Blockwise reduction, internal propagation, and global write as per \texttt{Cu\_Voxelize\_V1}.
%
%
% Notes:
% 1. Since this kernel reads the face data in parallel, there is an implicit assumption
%    that the bin density is fixed equal to grid resolution divided by 4, otherwise
%    some threads might be reading the data of the wrong face.

\subsubsection{External Cell Flag Propagation}

Flags externally propagate from cell-blocks in the vicinity of the geometry surface to their neighbors to complete the solid voxelization. Multiple layers of solid cells are required to ensure that a sufficient number of cell-blocks are marked for refinement in the interior of the geometry. The boundary of the geometry is resolved with the resulting children during top-down construction of the hierarchical grid.

Separate kernels propagate the flags along $\pm x$. Propagation starts from cell-blocks whose neighbors in the opposite direction do not exist (i.e., blocks on the domain boundary or along the refinement interface). Propagation in both directions is necessary on grid levels $L > 0$ to cover blocks that are inaccessible due to holes that correspond to solid cell-blocks on coarser levels that were not refined. We omit propagation along $-x$ on the root grid since no such holes exist. We will describe the kernel for propagation along $+x$ without loss of generality (Algorithm \ref{alg:voxelize_propagate}).

\begin{figure}[]
    \centering
    \begin{subfigure}[b]{0.85\textwidth}
        \centering
        \includegraphics[width=0.85\linewidth]{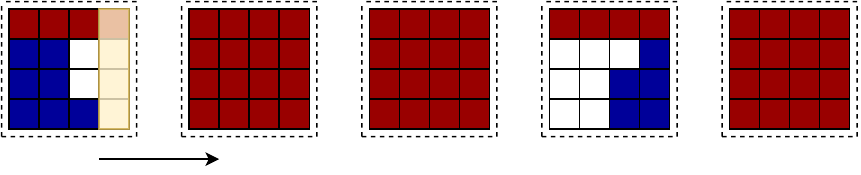}
        \caption{}
        \label{fig:method_prop_external_1}
    \end{subfigure}
    \begin{subfigure}[b]{0.85\textwidth}
        \centering
        \includegraphics[width=0.85\linewidth]{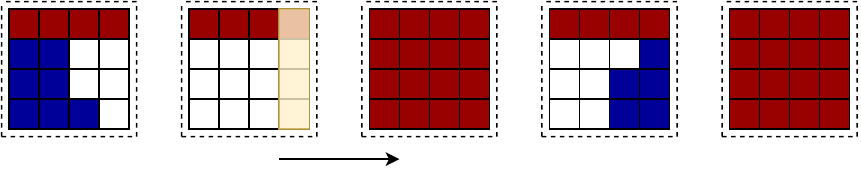}
        \caption{}
        \label{fig:method_prop_external_2}
    \end{subfigure}
    \begin{subfigure}[b]{0.85\textwidth}
        \centering
        \includegraphics[width=0.85\linewidth]{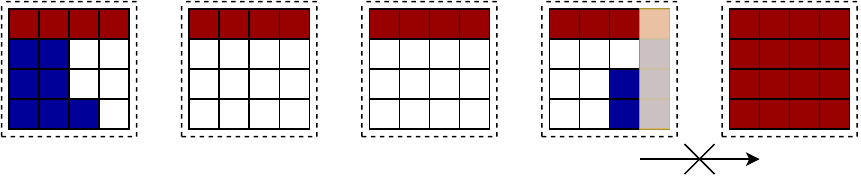}
        \caption{}
        \label{fig:method_prop_external_3}
    \end{subfigure}
    \caption{(a) Statuses are identified after cell masks have been loaded in the starting block. (b) Cell masks in the neighbor block are modified simultaneously. All cells in a scanline along $+x$ are adjusted based on the status flag in the starting block. (c) Propagation ceases once the statuses indicate only interior or guard cells. It can restart again if statuses in future neighbor blocks are set back to solid. (\fbox{\textcolor{white}{\rule{1ex}{1ex}}}: solid cells, \textcolor{blue}{\rule{1ex}{1ex}}: guard cells, \textcolor{red}{\rule{1ex}{1ex}}: fluid cells)}
    \label{fig:method_prop_external}
\end{figure}

Cell flags in the starting block are loaded into shared memory. The first neighbor cell-block index is subsequently loaded to initiate propagation. Neighbor cells in a scan line along $x$ check the stored flag of the cells along the $+x$ face in the starting block. These flags, referred to as the cell `status', indicate how the cells in the neighbor need to be updated. The cell update can be performed simultaneously in the current cell-block under consideration. This is possible since internal propagation has already filled out all necessary guards, which prevents cells from being marked as solid beyond the geometry boundary. If the status is `solid' but the current cell is a guard cell, then no change is made. If the cell is in the fluid, then it is flagged as solid. After this evaluation, the neighbor block becomes the current one, and the next neighbor is identified. This is repeated in a while loop until an invalid neighbor index is reached. This procedure is visualized in Figure \ref{fig:method_prop_external} for propagation along $+x$ in 2D.

%\fcolorbox{blue}{lightgray}{\textcolor{red}{This is colored text in a bordered box.}}

\begin{algorithm}[H]
\small
\caption{\texttt{Cu\_Voxelize\_Propagate\_Right}: External Propagation (Rightward)}
\label{alg:voxelize_propagate}
\SetAlgoLined
\DontPrintSemicolon

\tcp{Using definitions from \texttt{Cu\_Voxelize\_V1}}

Read left neighbor cell-block index $\kappa_{l,\text{start}}$\;

\If{$\kappa_{l,\text{start}} < 0$}{
    Compute local cell indices $\mathbf{I} = (I, J, K)^T$ such that thread $t = \texttt{LINEAR}(\mathbf{I}, 4)$\;
    
    Load cell masks $H_t$ of the current block into shared memory $s_V$\;

    Compute status indicator $s \gets s_V[\texttt{LINEAR}((3, J, K)^T, 4)]$\;
    
    \If{$\kappa_{l,\text{start}} = N_{\text{solid nbr}}$ \textbf{and} $s \neq V_G$}{ \tcp{$N_{\text{solid nbr}}$ indicates that propagation begins within the solid.}
        Set $s \gets V_S$\;
    }

    Get right neighbor cell-block index $\kappa_r$\;

    \While{$\kappa_r \geq 0$}{
        Get neighbor cell masks $H_{t,r}$\;

        \If{$s = V_S$ \textbf{and} $H_{t,r} \neq V_G$}{
            Set $H_{t,r} \gets V_S$\;
        }

        Load $s_V[t] \gets H_{t,r}$\;

        Update $s \gets s_V[\texttt{LINEAR}((3, J, K)^T, 4)]$\;

        \If{working on root grid \textbf{and} $H_{t,r} = V_G$}{
            Set $H_{t,r} \gets V_I$\;
        }

        Store $H_{t,r}$ in cell-block $\kappa_r$\;

        Make right neighbor the current block and update new neighbor (i.e., $\kappa_r \gets$ next right neighbor)\;
    }
}
\end{algorithm}

% Cu_Voxelize_Propagate_Right

% 1- Read left neighbor cell-block index $\kappa_{l,\text{start}}$.
% 2- If $\kappa_{l,\text{start}} < 0$ (doesn't point to a real block):
%    a) Compute local cell indices $\mathbf{I}=(I,J,K)^T$ such that thread $t = \texttt{LINEAR}(\textbf{I},4)$\;
%    b) Load cell masks $V_t$ of current block into shared memory $s_V$.
%    c) Compute status indicaor $s = s_V[\texttt{LINEAR}((3,J,K)^T,4)]$.
%    d) If $\kappa_{l,\text{start}} = N_{\text{spec.}}$ and $s \neq V_G$, then $s = V_S$.
%    e) Get right neigbor cell-block index $\kappa_r$.
%    f) While ($\kappa_r \geq 0):
%       i) Get neighbor cell masks $V_{t,r}$.
%       ii) If $s = V_S$ and $V_{t,r} \neq V_G$, then $V_{t,r} = V_S$. 
%       iii) Load $V_{t,r}$ into $s_V[t]$.
%       iv) Update $s = s_V[\texttt{LINEAR}((3,J,K)^T,4)]$.
%       v) If working on the root grid and $V_{t,r}=V_G$, then set $V_{t,r} = V_I$.
%       vi) Store $V_{t,r} in cell-block $\kappa_r$.
%       vii) Make right neighbor the current block and update new neighbor $\kappa_r$.

The external propagation kernels also employ the primary mode of access. Invalid cell-block neighbor indices are indicated by a negative sign. Cells use their local coordinates to compute the indices in shared memory of their corresponding statuses. Along $+x$, the index of the status flag is found with $3+4J+16K$ in 3D. The statuses are then used in a parallel evaluation of the conditional statements to determine if flags need to be modified. By default, propagation assumes that the initial status of the starting block is fluid. This is the case for geometries embedded in open domains, where the starting point is at the inlet or outlet. A special case occurs during propagation on grid levels $L > 0$. If cell-blocks start from the interior of the solid near a hole left by an unrefined coarser cell-block, the status should be solid initially. To account for this, we assign a special negative neighbor index to the children of these coarse solid cell-blocks. This later informs cells in the starting block to switch to solid before initiating propagation.

After executing both propagation kernels, guard cells are switched back into fluid cells, and cell-blocks that possess at least one solid cell are marked as solid cell-blocks. This step is performed in a standalone kernel.
%$\texttt{Cu\_Voxelize\_UpdateMasks}$: Now solid voxelization at the cell level is complete, block masks are updated to mark blocks as solid if at least one cell is a solid cell.

\subsubsection{Refinement Flagging and Propagation}

Cell-blocks are marked for refinement now that they have been classified. The objective is to meet the user-specified near-wall distance refinement criterion $d_{\text{spec.}}$, and to incorporate enough refinement within the interior of the geometry to ensure that future refinement of the tree on finer grid levels remains possible. We first identify the solid cell-blocks that are adjacent to at least one fully fluid cell-block (i.e., all cells lie in the fluid). This marks the `solid boundary' of the grid. From this boundary, we refine two layers of blocks into the geometry, and a number of layers $N_{\text{prop.}}$ into the fluid according to
\begin{align}
    N_{\text{prop.}} = 1+ \left\lfloor \frac{1}{2^L}\left( \frac{d_{\text{spec.}}}{\sqrt{2} \Delta x_{b,L}}\right) \right\rfloor,
\end{align}

\noindent where $\Delta x_{b,L} = 4\Delta x_L$ is the cell-block length along one axis. Propagation of the refinement flags accomplishes this (Figure \ref{fig:method_voxel_steps}). Our numerical experiments indicated that it was sufficient to propagate inward twice to enable partial surface voxelization on subsequent finer grids. The pre-factors $\sqrt{2}$ and $1/2^L$ are used to normalize the distance by the diagonal cell-block length , and to restrict propagation by half on subsequently finer grids, respectively. The procedure is split into three kernels. Unlike cell flag propagation, cell-block metadata is only contiguous in groups of 4/8 in 2D/3D based on the insertion pattern during refinement. Consequently, all three kernels employ the secondary mode of access.

The first kernel identifies the solid boundary. Masks are loaded in a first traversal of the cell-blocks. If the blocks are solid, their neighbor block indices are copied into shared memory. These are arranged to expose a limited degree of contiguity in the neighboring blocks that arises from the insertion pattern. In a second traversal, the masks of the neighbor blocks are loaded. If a block is found to be a fluid block, then its index in shared memory is replaced to indicate that the original block lies on the boundary of the solid. Otherwise, the index is reset to a negative value. The original cell-blocks can now access the indicators assigned to its neighbors, with uncoalesced access in shared memory rather than global memory. The masks of solid blocks with at least one neighbor are updated and, if eligible, marked for refinement. A cell-block is ineligible if it lies on the refinement interface, since refining it would lead to a violation of the 2:1 balance of the forest-of-octrees.

The second kernel follows a similar memory access strategy. However, all cell-blocks in the grid are considered this time. The refinement criterion is that a cell-block must be adjacent to at least one solid-boundary block. If the cell-blocks under consideration were fluid blocks, then they are marked as solid-adjacent (meaning that they are directly adjacent to the solid boundary). This mask value is used within the mesh adaption routine to avoid an incorrect overwrite.

The last kernel is repeated $N_{\text{prop.}}$ times to propagate the refinement flags. The scope is restricted to fluid or, in just the first iteration, solid cell-blocks. This results in a total of $N_{\text{prop.}}$ propagation iterations into the fluid region, and two iterations into the solid region. The refinement criterion here is that a neighboring cell-block is already marked for refinement. The kernel employs an intermediate array to avoid a race condition where a neighboring refinement flag is modified while it is being checked. Consequently, there are two possible destination arrays: the original array storing the refinement flags and the intermediate array. The kernel selects the destination array based on the parity of the iteration index, and separate refinement flags are assigned to each.
\definecolor{csolidcell}{HTML}{7ECDFF}
\definecolor{csolidblock}{HTML}{000DA4}
\definecolor{csolidblockb}{HTML}{B6004C}
\definecolor{csolidblocka}{HTML}{DE00B8}
\definecolor{cmref}{HTML}{13FF9E}
\definecolor{cboundarycell}{HTML}{FFDA7E}
\begin{figure}[h]
    \centering
    \begin{subfigure}[b]{0.3\textwidth}
        \centering
        \includegraphics[width=\textwidth]{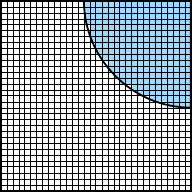}
        \caption{}
    \end{subfigure}
    \hfill
    \begin{subfigure}[b]{0.3\textwidth}
        \centering
        \includegraphics[width=\textwidth]{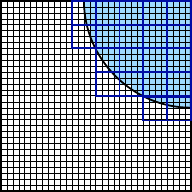}
        \caption{}
    \end{subfigure}
    \hfill
    \begin{subfigure}[b]{0.3\textwidth}
        \centering
        \includegraphics[width=\textwidth]{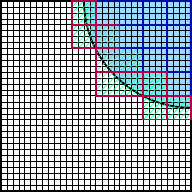}
        \caption{}
    \end{subfigure}
    \begin{subfigure}[b]{0.3\textwidth}
        \centering
        \includegraphics[width=\textwidth]{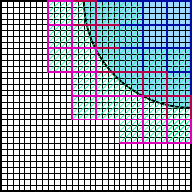}
        \caption{}
    \end{subfigure}
    \hfill
    \begin{subfigure}[b]{0.3\textwidth}
        \centering
        \includegraphics[width=\textwidth]{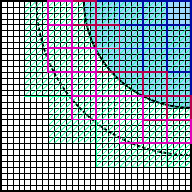}
        \caption{}
    \end{subfigure}
    \hfill
    \begin{subfigure}[b]{0.3\textwidth}
        \centering
        \includegraphics[width=\textwidth]{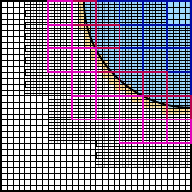}
        \caption{}
    \end{subfigure}
        \caption{Steps for voxelization into and refinement of a single grid level. a) Initial state after partial surface voxelization and cell flag propagation (with desired near-wall refinement distance shown). b) Blocks with at least one solid cell are marked as solid. c) Solid blocks neighboring regular blocks are marked as solid-boundary and marked for refinement. d) Blocks neighboring solid-boundary blocks are marked for refinement; regular blocks neighboring solid-boundary blocks are marked as solid-adjacent. e) Refinement flags are propagated in regular blocks (and one within the solid) until desired distance has been reached. f) Blocks are refined, and cells adjacent to solid cells are marked as boundary cells. (\textcolor{csolidcell!75}{\rule{1ex}{1ex}}: solid cell, \textcolor{cboundarycell!75}{\rule{1ex}{1ex}}: boundary cell, \textcolor{csolidblock!100}{\rule{1ex}{1ex}}: solid block, \textcolor{csolidblockb!100}{\rule{1ex}{1ex}}: solid-boundary block, \textcolor{csolidblocka!100}{\rule{1ex}{1ex}}: solid-adjacent block, \textcolor{cmref!100}{\rule{1ex}{1ex}}: marked for refinement)}
        %Steps for voxelization and refinement of a grid level $L$. a) Cells perform raycast computations to identify themselves as solid or fluid cells, effectively voxelizing the solid. b). Blocks marked as solid that are adjacent to fluid blocks are marked as solid-boundary blocks and marked for refinement. c) Blocks adjacent to solid-boundary blocks are marked with an intermediate flag that is propagated until the near-wall distance criterion is reached. d) Intermediate flags are finalized. e) Cells adjacent to solid cells are marked as boundary cells.      (\textcolor{blue}{\rule{1ex}{1ex}}: solid cell, \textcolor{red}{\rule{1ex}{1ex}}: boundary cell, \textcolor{cyan}{\rule{1ex}{1ex}}: solid block, \textcolor{magenta}{\rule{1ex}{1ex}}: solid-boundary/refined block, \textcolor{lime}{\rule{1ex}{1ex}}: interior block, \textcolor{green}{\rule{1ex}{1ex}}: propagated block)}
    \label{fig:method_voxel_steps}
\end{figure}

\subsubsection{Boundary Cell Identification}

A separate kernel finalizes the first voxelizer subroutine by updating refinement flags to a value that is accepted by the mesh adaptation routine. Boundary cells are now identified in the grid after constructing the whole grid hierarchy. A cell is defined to be on the boundary if it is adjacent to at least one solid cell in any of its surrounding directions.
\begin{figure}
    \centering
    \begin{subfigure}[b]{0.49\textwidth}
        \centering
        \includegraphics[width=0.75\linewidth]{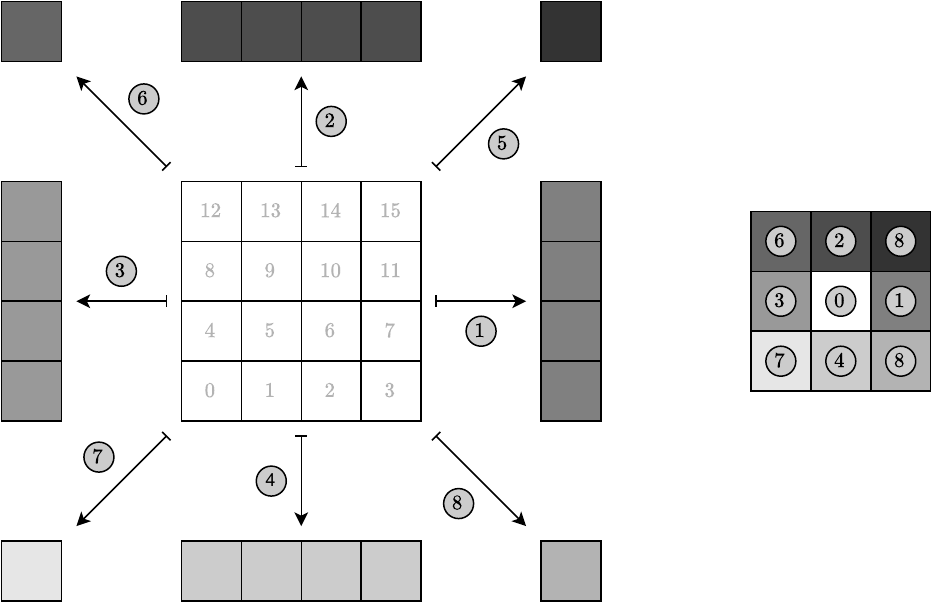}
        \caption{}
        \label{fig:method_lookup_A}
    \end{subfigure}
    \hfill
    \begin{subfigure}[b]{0.49\textwidth}
        \centering
        \includegraphics[width=1\linewidth]{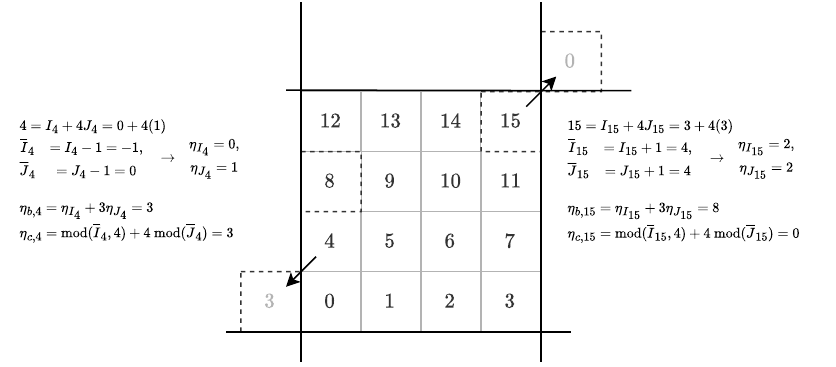}
        \caption{}
        \label{fig:method_lookup_B}
    \end{subfigure}
    \caption{(a) When cells need to sample data from neighbors, a lookup table is formed for the indices of the neighboring blocks. Local cell coordinates incremented along a direction are mapped to the lookup table, and the retrieved value is used to further retrieve the data in the neighboring cell. (b) Example lookup of addresses of neighbor cells $\eta_c$ and corresponding neighbor blocks $\eta_b$ for cells 4 and 15 in directions 7 ($\textbf{e}_7=[-1,-1]^T$) and 5 ($\textbf{e}_5=[1,1]^T$), respectively.}
    \label{fig:method_lookup}
\end{figure}

This step is split into two kernels to avoid a race condition in which cells check the flags of their neighbors while they are being updated. Both of these kernels employ the primary mode of access. The first kernel traverses all solid-boundary and solid-adjacent cell-blocks and evaluates the masks of cells in neighboring blocks. Cells are then flagged as boundary cells if at least one of their neighbors is solid, and the result is written to an intermediate array. If at least one cell is flagged, an indicator is stored in a separate intermediate array that marks the current cell-block as requiring a memory allocation for boundary metadata. The second kernel traverses these same blocks and finalizes the results by copying them into the proper cell mask array.

In the first kernel, there is little re-use since threads access the same region of global memory when retrieving neighbor cell masks within the same cell-block. Re-use can be achieved by storing the cell-masks in a shared memory array with a single-layer halo around the current cell-block. However, this comes at the cost of a non-trivial setup stage. We note that cache reuse may offset the penalty of redundant access since it is the same section of memory being retrieved repeatedly. We implemented both approaches to confirm which is faster.

The version employing re-use stores the current cell-block masks in the center of the shared memory block. Then, for each direction, threads are mapped to cells in the appropriate neighbor cell-block (Figure \ref{fig:method_lookup}). The masks are then retrieved and placed in the halo in shared memory. We use the following mapping of local cell coordinates $\textbf{I}_t$ to achieve this. We define a function $\text{LINEAR}(\textbf{I},N)=I_x + NI_y + N^2I_z$ that maps the coordinates back to the cell indices within the block. In 2D, $I_z$ is set to zero. For the current $4^D$ blocks in the AMR grid, $t$ relates to $\textbf{I}_t$ via $t=\text{LINEAR}(\textbf{I}_t,4)$. An increment to the coordinates in direction $p$ is given by $\textbf{I}_{t,p} = \textbf{I}_t + \textbf{c}_p$. This corresponds to the correct cell index $t_p$ when the neighbor lies in the current cell-block. For a shared memory array sized to fit the data of one cell-block with a single-layer halo, we define the indices $t_h$ that map cell-block data to the shared memory array; this is given by $t_h = \text{LINEAR}(\textbf{I}_t+\textbf{1},6)$.

When a value is retrieved from a neighboring cell-block, at least one of the components of $\textbf{I}_{t,p}$ will be outside of the range $\{0,...,3\}$. We use this violation to map threads to the correct neighbor indices $t_p$ to limit retrievals only from neighboring cell-blocks. This also limits writes strictly to the one-layer thick halo. The modified incremented cell coordinates $I_{t,p}'$ given by
\begin{align}
    I_{t,p,d}' = \text{mod}((4 + \text{mod}(I_{t,p,d}+c_{p,d},4)), 4),
\end{align}

\noindent where $0 \leq d < D$, map to the correct index in the neighboring cell-block. We check the following condition to determine whether the incremented coordinates $\textbf{I}_{t,p}$ have at least one violating component:
\begin{align}
    \bigwedge_{d=0}^{D-1}  \left( (I_{t,p,d}' \neq I_{t,p,d}) \vee (c_{p,d}=0) \right).
\end{align}

\noindent This condition ignores components where there was no increment (e.g. if incrementing along $+x$, only the first modified coordinate plays a role). If this condition returns true, the current neighbor cell is one cell-layer away from the current cell-block. Its mask is therefore retrieved and added to the halo. After setting up the shared memory array, cells loop over their stored neighbor masks to determine if they are boundary cells.

The other version of the first kernel relies on the same index mappings. However, the loop is structured in a way that allows for repeated retrievals of masks from the current cell-block. This results in less overhead for setting up access to neighbor cell masks, but increases the number of redundant global memory accesses. At the end of the flagging step in both kernels, a shared memory array stores unit indicators corresponding to flagged boundary cells. A block-wise reduction determines the total number of boundary cells in the cell-block. If this number is positive, it is stored in an intermediate array, and the cell-block is flagged as `on the domain boundary.' 

\subsubsection{Memory Allocations for Boundary Cells}

Memory is now allocated to store the link-lengths required by the boundary cells. We approach the allocation in a block-granular rather than cell-granular way, which ensures that cells can safely read data in parallel in a SoA format even if only a fraction are classified as boundary cells. The number of cell-blocks participating in the allocation $N_{b,\text{solid}}$ is obtained with Thrust's \texttt{count\_if}; the routine is applied to the intermediate array that stored the reduction results of the first boundary-cell flagging kernel.
%We also apply a reduction on this same array with \texttt{reduce} to get the total number of individual boundary cells as an optional step to compare the block-granular memory allocation with an ideal cell-granular one; we report these statistics in the test cases to follow.

We currently assume that a no-slip condition is applied on all geometry surfaces. Consequently, we do not explicitly identify which boundary conditions need to be applied on different faces in the current implementation. However, we have implemented an allocation for a lookup table of boundary condition Ids that correspond to the link-lengths. This table is accessed in the same way (SoA format). Future work will expand the present framework to accommodate a more general specification of conditions on the geometry boundary (such as space- and time-dependent conditions). We expect that a unique Id will be specified per face, and that this Id will be obtained from the data arrays during link-length computation.

Cell-block indices $\kappa$ are usually mapped to cell data $\kappa M_b + t$, where $0 \leq t < M_b$, and $M_b$ is the cell-block size and $t$ are cell indices. Since the lookup tables are allocated only for cells on the boundary, this no longer holds. We therefore prepare a mapping of boundary cell-block indices to the contracted set in advance. This array is dense in the overall set of cell-blocks, but its elements indicate when cell-blocks are on the boundary in the solver kernels. When the retrieved value is non-negative, the indices specify where to retrieve boundary cell data. Negative values indicate that the cell-blocks are not on the boundary.
% \begin{align}
%     \kappa\in \mathbb{N}: [0,N_{\text{max,b}}] \to \kappa'\in \mathbb{N}:\{-1\} \cup [0,N_{b,\text{solid}}]
% \end{align}

If there is a positive number of cell-blocks along the geometry boundary (i.e., cell-blocks with at least one boundary cell), then we allocate memory for the following arrays: 1) the link-length lookup table, requiring $N_{b,\text{solid}} N_{q,\text{max}}$ floats, 2) the boundary-index lookup table, requiring $N_{b,\text{solid}}N_{q,\text{max}}$ integers, and 3) the cell-block metadata array that maps cells from their usual indices in global memory to these lookup tables, requiring $N_{b,\text{solid}}$ integers. To construct the mapping, we apply Thrust's \texttt{copy\_if} on a counting iterator to collect the indices of cell-blocks with a positive number of boundary cells. We then apply \texttt{scatter} on the counting iterator to scatter its elements to the mapped index array using these collected indices as the stencil.

\subsubsection{Link-Length Computation}

The link-length computation kernel is called after memory has been allocated. The kernel utilizes the same structure and parallelization as the partial surface voxelization kernel (e.g., primary mode of access, spatial binning to reduce the total number of ray casts). However, all surrounding directions are now considered rather than just $\textbf{e}_x$. The kernel only applies the computation procedure in cell-blocks with a non-negative contraction mapping index.

Cells loop over faces $\mathcal{T}_p$, with vertices $(\textbf{v}_k)_p$ and normal $\textbf{n}_p$, in parallel. For each direction $q$, we compute the line-plane intersection distance $d_{q,p}$ from a ray that is cast from cell-center $\textbf{v}$, where the plane is defined by the first vertex of the face and its normal. Intersection points $\textbf{v}_{i,q,p} = \textbf{v} + d_{q,p} \textbf{c}_q$ that lie inside $\mathcal{T}_p$ are potential candidates for the final link-length along $q$. The shortest distance $d_{q,\text{min}} = \text{min}_p d_{q,p}$ is retained and stored in the lookup table after normalizing by $\Delta x_L$.

Since each link-length is taken as a minimum among a set of faces, we store them all explicitly and update them simultaneously. This avoids repeated traversals of the faces. When each new face is considered, the links in all directions are updated accordingly. To reduce register pressure and enable indexed access to these links when looping over the different directions, we store them in local memory rather than in register memory (i.e., $\texttt{REAL dQ[N\_Q\_max]}$ vs. $\texttt{REAL dQ\_0, dQ\_1}\dots$, where $\texttt{REAL}$ is the specified floating-point precision). All link-lengths are initialized with a value of $-1$. The value remains unchanged if there was no intersection with the geometry. A negative sign therefore indicates that the neighbor cell in the corresponding direction is a fluid cell.

\section{Test Cases} \label{sec:tests}

This section describes performance and validation tests for our proposed methodology. Spatial binning and voxelization kernels are tested with the Stanford bunny and XYZ RGB Dragon models from the Stanford 3D Scanning Repository\footnote{We obtained a binary STL variant of the original PLY model from Wikimedia Commons \cite{BunnyII}, which has a different triangle count.} \cite{Stanford3D} (Table \ref{tab:res_models_bb}). Two sets of tests are conducted for these models with $L_{\text{max}}=5,7$, denoted respectively as L5 and L7 (i.e., BunnyL5/DragonL5 and BunnyL7/DragonL7). Laptop GPUs are limited in memory, so the L5 tests enable execution time comparisons with the datacenter GPUs. The L7 tests provide large enough workloads to saturate the GPU, highlighting the performance-limiting kernels. We repeated simulations 100 times and extracted the average and a 95\% confidence interval to ensure that comparisons across different GPUs and kernel variants are not due to random fluctuations.

Although the models of the Stanford 3D Scanning Repository are not representative of typical CFD geometries in shape, their large triangle counts and complex shapes expose potential flaws in the presence of faces of greatly-varying size and local concavity/convexity.
%We have included performance tests for a sphere primitive and applied it in flow simulations for validation.
\begin{table}[]
    \centering
    \begin{tabular}{c|cc}
        \hline
        Model & Bounding Box & No. Triangles \\ 
        \hline
        %Sphere & $[0.5-1/(2D),0.5+1/(2D)]^3$ & 7,938 \\
        Stanford Bunny & $[0.170,0.843]\times[0.243,0.792]\times[0.150,0.825]$ & 112,394 \\
        XYZ RGB Dragon & $[0.115,0.844]\times[0.282,0.767]\times[0.335,0.662]$ & 7,218,906 \\
         \hline
    \end{tabular}
    \caption{Bounding boxes of the models in the reported test cases. The Stanford bunny and XYZ RGB dragon were rescaling to fit in a unit cube domain in the Blender software. $D$ is the diameter of the sphere.}
    \label{tab:res_models_bb}
\end{table}

Geometry embedding is assessed from three perspectives. We first compare performance with the sparse voxel octree method of Schwarz and Seidel \cite{Schwarz2010} (developed for graphics processing). The comparison is not exact, as Schwarz and Seidel perform the voxelization within a bounding box of the model, while we embed the model into a larger computational grid. However, it serves as a baseline case with an ideal minimum time taken for the models considered in our tests. It can also reveal the degree to which the block-based structure of the grid affects performance. This is contrasted with the approach of Schwarz and Seidel, where bottom-up octree construction is preceded by an initial embedding into a uniform grid. Second, we compare execution times for the various steps in the spatial binning and voxelization routines on four GPUs (Table \ref{tab:tests_device_specs}): 1) GTX 970M, 2) RTX4070 Mobile, 3) A100, and 4) H100, to study performance across different hardware bandwidths and computing power. This analysis also identifies the bottlenecks of the implementation that will be targeted for optimization in future work. Finally, we compare the variants of the partial surface voxelization and boundary cell identification kernels introduced in Section \ref{sec:methodology}. We execute the tests on the 970M, which saturates the hardware more quickly due to relatively limited bandwidth. This better highlights the differences in performance between the variants.
% We note that the comparison is not exact. We embed a geometry within an outer computational grid and build the trees top-down, while Schwarz and Seidel \cite{Schwarz2010} build their tree bottom-up (starting with an embedding into a uniform grid) and voxelize only within the bounding box of the model. Our effective voxelization resolution is therefore lower.
\begin{table}[]
    \centering
    \begin{tabular}{l|lll}
    \hline
        OS & g++ & nvcc/CUDA & Device \\
    \hline
        Debian 10.13 (Buster) & 8.3.0 & 9.2 & NVidia GTX970M (3GB) \\
        Windows 10, WSL 2 & 13.3.0 & 12.0 & NVidia RTX 4070 Mobile (8GB) \\
        Rocky Linux 8.10 (Green Obsidian) & 9.3.0 & 11.4 & NVidia A100 SXM4 (A100) \\
        Rocky Linux release 9.6 (Blue Onyx) & 12.3.1 & 12.6 & NVidia H100 SXM (80GB) \\
    \hline
    \end{tabular}
    \caption{Characteristics of software and hardware used in the current tests.}
    \label{tab:tests_device_specs}
\end{table}

An application to computational fluid dynamics simulations via the lattice Boltzmann method is also presented. We simulate external flow over a 2D cylinder (both square and circular) and a 3D sphere, extracting values of the drag coefficient, lift coefficient, and the Strouhal number. These values are compared with previous results in the literature to validate the implementation. In previous work \cite{Jaber2025}, we employed a 2D square cylinder whose edges were aligned to the edges of the cell-blocks in the finest grid to validate the AMR algorithm. The circular cylinder and sphere will now demonstrate the cell-granular resolution of the embedded geometries, and the accuracy of the solver in the presence of surface curvature. The solver is extended to incorporate the interpolated bounce-back condition of Bouzidi et al. \cite{Bouzidi2001} for comparison with the simple bounce-back method. Double precision is specified for the geometry to ensure that no links to the geometry are missed, and for the solver to ensure that force computations are accurate.

\subsection{Geometry Embedding}

\subsubsection{Comparison with Sparse Voxel Octree Construction}

The total execution times for the BunnyL5 and DragonL5 tests are tabulated in Table \ref{tab:tests_voxelization_comparison}. We distinguish the times required to build and balance the forest-of-octrees, generate the hierarchy of spatial bins, and perform voxelization in each grid level. Our values are compared with the sparse solid voxelization results of Schwarz and Seidel \cite{Schwarz2010} applied to the same models. When a root grid of $64^3$ is selected, refining four times results in an effective resolution near the geometry surface of $N_{\text{eff.}} = 1024$. We therefore extracted results from \cite{Schwarz2010} corresponding to sparse solid voxelization in a grid size of $1024^3$. However, the actual voxelization resolution defined with respect to the model bounding boxes within the unit cube domain. Values for the effective resolution at different levels (up to a total of seven) are shown in Table \ref{tab:tests_voxelization_effective}. In particular, we observe effective voxelization resolutions of $689\times 562 \times 691$ and $746 \times 497 \times 335$ at the finest levels of the BunnyL5 and DragonL5 tests, respectively. The results of \cite{Schwarz2010} were obtained from an NVidia GeForce GTX 285. To make the comparison as fair as possible, we tabulated our results on the NVidia GeForce GTX 970M (which was released approximately five years after the GTX 285) along with results on the H100.
\begin{table}[]
    \centering
    \begin{tabular}{c|rrr}
        \hline
        \multirow{2}{*}{Step} & \multicolumn{3}{c}{Execution Times (ms)} \\ \cline{2-4}
        & 970M & H100 & \cite{Schwarz2010} \\
        \hline
        \rowcolor{gray!40}
        & \multicolumn{3}{c}{Stanford Bunny} \\
        Refinement, Balancing & $69.62 \pm 0.44 \ (26.57\%)$ & $8.18 \pm 0.02 \ (27.52\%)$ & \multirow{3}{*}{-} \\
        Spatial Binning & $39.05 \pm 0.03 \ (14.90\%)$ & $8.53 \pm 0.01 \ (28.68\%)$ &  \\
        Voxelization & $153.40 \pm 0.36 \ (58.53\%)$ & $13.02 \pm 0.04 \ (43.80\%)$ &  \\
        \hline
        \rowcolor{gray!20}
        Total & \multicolumn{1}{c}{$262.07 \pm 0.57$} & \multicolumn{1}{c}{$29.73  \pm 0.05$} & 93.27 \\
        \hline
        \rowcolor{gray!40}
         & \multicolumn{3}{c}{XYZ RGB Dragon} \\
        Refinement, Balancing & $38.42 \pm 0.04 \  (17.33\%)$ & $7.76 \pm 0.01 \ (28.23\%)$ & \multirow{3}{*}{-} \\
        Spatial Binning & $108.41 \pm 0.04 \ (48.91\%)$ & $10.90 \pm 0.01 \ (39.65\%)$ &  \\
        Voxelization & $74.82 \pm 0.04 \ (33.76\%)$ & $8.83 \pm 0.04 \ (32.12\%)$ &  \\
        \hline
        \rowcolor{gray!20}
        Total & \multicolumn{1}{c}{$221.65 \pm 0.07$} & \multicolumn{1}{c}{$27.49  \pm 0.05$} & 178.4 \\
        \hline
    \end{tabular}
    \caption{Total execution times for the current forest-of-octrees tree construction and voxelization procedure compared with the computer graphics method of Schwarz and Seidel \cite{Schwarz2010}.}
    \label{tab:tests_voxelization_comparison}
\end{table}

Our 970M results require more time in total than those of Schwarz and Seidel \cite{Schwarz2010} by factors of approximately $3.1\times$ and $1.2\times$ for the bunny and dragon models, respectively. Tree refinement and balance times are operations that are roughly comparable to the combined determination of active nodes and octree construction steps in \cite{Schwarz2010}; these required approximately $69.62\text{ ms}$ versus $25.1 \text{ ms}$ for the bunny and $38.42 \text{ ms}$ versus $83.8 \text{ ms}$ for the dragon. The most expensive step in our refinement and balancing algorithm (Step 8 in \cite{Jaber2025}) involves a kernel that specifies whether cells in the vicinity of the refinement interface are ghosts, which is required for synchronization during temporal integration. If the step executing the cell mask update kernel is not counted in the timings, our values drop to $26.14 \text{ ms}$ and $17.50 \text{ ms}$ (i.e., by factors of approximately $2.5$ and $2.2$) for the bunny and dragon models, respectively, which are more comparable to \cite{Schwarz2010}. This step contributes to more than half of the tree construction process since it involves a full traversal of the grid for the cell mask update. The proportion of time taken for this step relative to the total increases rapidly in larger tests, up to $60\%$ when refining on the second-finest level of the DragonL7 test (Table \ref{tab:tests_octree_refinement}). It will be targeted for optimization in future work (e.g., with a permutation iterator defined across the indices of cell-blocks along the refinement interface).
\begin{table}[h]
    \centering
    \begin{tabular}{cc|ccc|ccc}
        \hline
        \multirow{2}{*}{Level} & \multirow{2}{*}{$N_{\text{eff.}}$} & \multicolumn{3}{c|}{Stanford Bunny} & \multicolumn{3}{c}{XYZ RGB Dragon} \\ \cline{3-8}
        & & $x$ & $y$ & $z$ & $x$ & $y$ & $z$ \\
        \hline
        0 & 64 & 43 & 35 & 43 & 47 & 31 & 21 \\
        1 & 128 & 86 & 70 & 86 & 93 & 62 & 42 \\
        2 & 256 & 172 & 141 & 173 & 187 & 124 & 84 \\
        3 & 512 & 345 & 281 & 346 & 373 & 248 & 167 \\
        \rowcolor{gray!20}
        4 & 1024 & 689 & 562 & 691 & 746 & 497 & 335 \\
        5 & 2048 & 1378 & 1124 & 1382 & 1493 & 993 & 670 \\
        \rowcolor{gray!20}
        6 & 4096 & 2757 & 2249 & 2765 & 2986 & 1987 & 1339 \\
        \hline
    \end{tabular}
    \caption{Voxelization resolutions of the Stanford bunny and XYZ RGB dragon models at each level in a grid hierarchy defined by a root grid resolution of $64^3$.}
    \label{tab:tests_voxelization_effective}
\end{table}

Spatial binning on the 970M requires the least amount of time among the three groups of steps for the BunnyL5 test ($14.90\%$ of the total), but the most for the DragonL5 test ($48.91\%$). On the H100, spatial binning requires approximately as much time as refinement and balancing for the BunnyL5 test and remains the most expensive step for the DragonL5 test. In fact, the three groups require a nearly equal proportion of time relative to the total, with the largest difference being about $5 \text{ ms}$ between voxelization and spatial binning in the BunnyL5 test. The larger set of faces in the dragon model requires more time for filtering and bin index assignment due to the increased memory traffic. However, higher bandwidth on the newer GPU minimizes the impact of traffic on throughput. This is also reflected in the partial surface voxelization kernel (Variant 1 is applied in all cases), where warps likely stall on the older GPU while they wait for vertex data to be retrieved from memory. 
\begin{table}[h]
    \centering
    \begin{tabular}{c|rrrrrr}
        \hline
        \multirow{2}{*}{Step} & \multicolumn{6}{c}{Execution Times ($\mu \text{s}$)} \\ \cline{2-7}
        & $L=0$ & $L=1$ & $L=2$ & $L=3$ & $L=4$ & $L=5$ \\
        \hline
        Pre & $92\pm1$ &$57\pm1$ &$55\pm1$ &$57\pm1$ &$57\pm1$ &$74\pm2$ \\
        S1 & $719\pm3$ &$424\pm2$ &$424\pm2$ &$487\pm2$ &$823\pm2$ &$2063\pm2$ \\
        S2 & $230\pm2$ &$126\pm1$ &$126\pm1$ &$135\pm1$ &$195\pm1$ &$423\pm1$ \\
        S4 & $24\pm1$ &$12\pm1$ &$13\pm1$ &$17\pm1$ &$44\pm1$ &$146\pm1$ \\
        S6 & $812\pm2$ &$735\pm2$ &$739\pm2$ &$773\pm2$ &$963\pm2$ &$1,794\pm2$ \\
        S7 & $141\pm1$ &$154\pm1$ &$192\pm1$ &$242\pm1$ &$334\pm1$ &$586\pm1$ \\
        S8 & $150\pm2$ &$105\pm1$ &$169\pm1$ &$528\pm1$ &$1,941\pm1$ &$7,869\pm1$ \\
        \hline
        Total & $2,168\pm24$ &$1,613\pm13$ &$1,718\pm13$ &$2,239\pm13$ &$4,357\pm13$ &$12,955\pm16$ \\
        \hline
    \end{tabular}
    \caption{Execution times for forest-of-octrees refinement for the DragonL7 test on the H100. Step 5 is omitted as it pertains to coarsening.}
    \label{tab:tests_octree_refinement}
\end{table}

Our DragonL5 tests require less time in total than the BunnyL5 tests on both GPUs, even though there are more faces in the former by an order of magnitude. On the H100, they are nearly equal with a difference of approximately $2 \text{ ms}$. This is due to our approach to voxelization, which is done with respect to the computational grid rather than the model itself. The ray cast filtering process that eliminates faces prior to voxelization reduces the overall workload by effectively matching the resolution of the model to that of the grid. Another factor is our top-down construction strategy, which means that an initial coarse root grid can be used for the starting point for voxelization (in contrast with the step for determining active nodes in \cite{Schwarz2010}, which is done starting from a relatively finer bottom level).

Overall, our method is capable of voxelizing complex geometries in execution times of comparable order of magnitude (albeit with lower effective resolutions on the finest levels). The face filtering step ensures that the dragon model requires almost as much time to embed into the grid as the bunny, even though the former has a number of faces that is two orders of magnitude larger. The required time for spatial binning naturally grows with the number of faces. We found that the most expensive steps where sortation and bounding box index assignment (which requires numerous triangle-AABB tests). The time taken to construct and balance and forest-of-octrees is also non-negligible, though there is room for optimization in steps such as cell classification along the refinement interface.

\subsubsection{Performance and Optimization of Spatial Binning and Voxelization}

\begin{figure}
    \centering
    \includegraphics[width=0.975\linewidth]{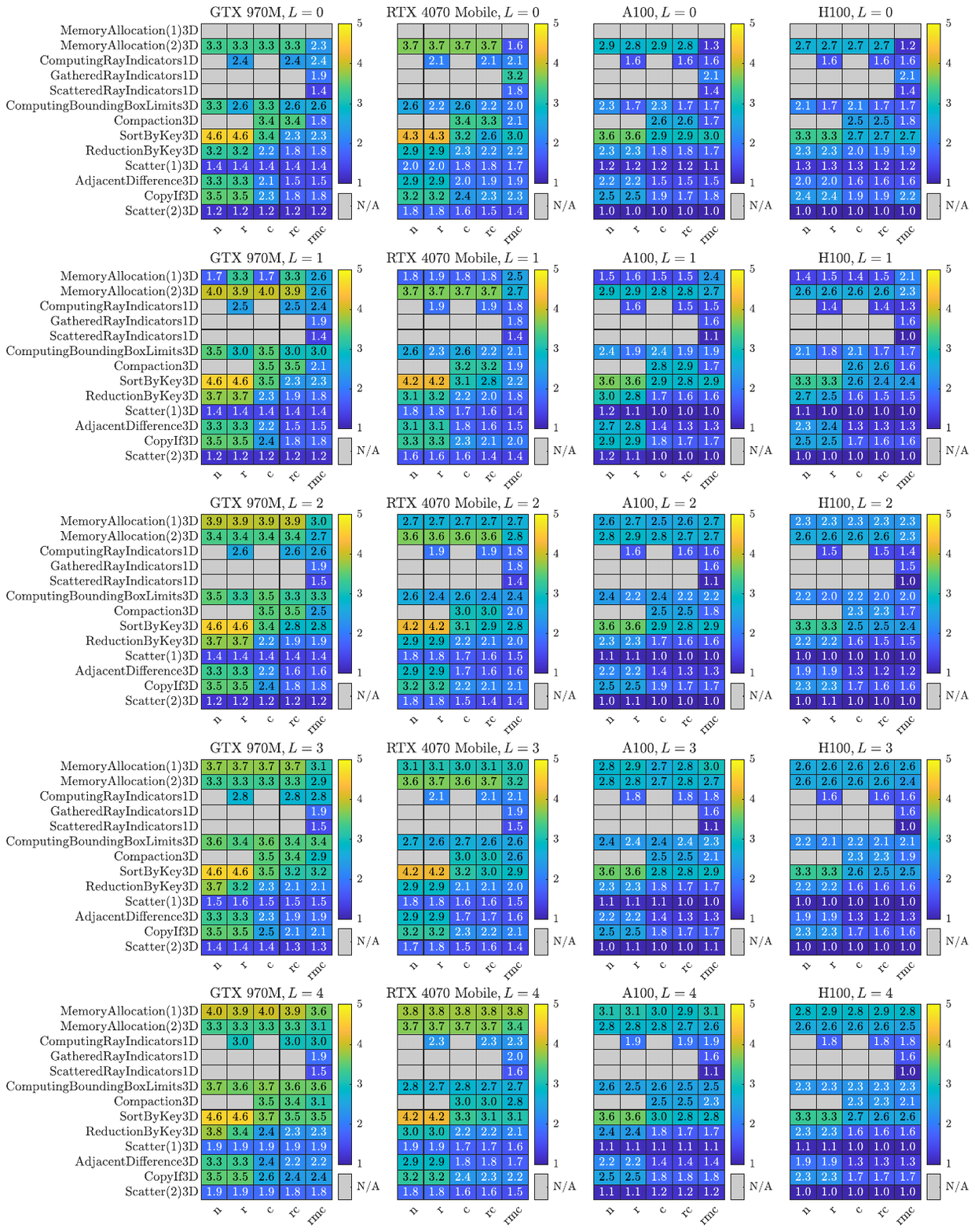}
     \caption{Execution times for the hierarchical spatial binning tests.}
    \label{fig:res_binning_timings}
\end{figure}

\begin{figure}
    \centering
    \includegraphics[width=0.975\linewidth]{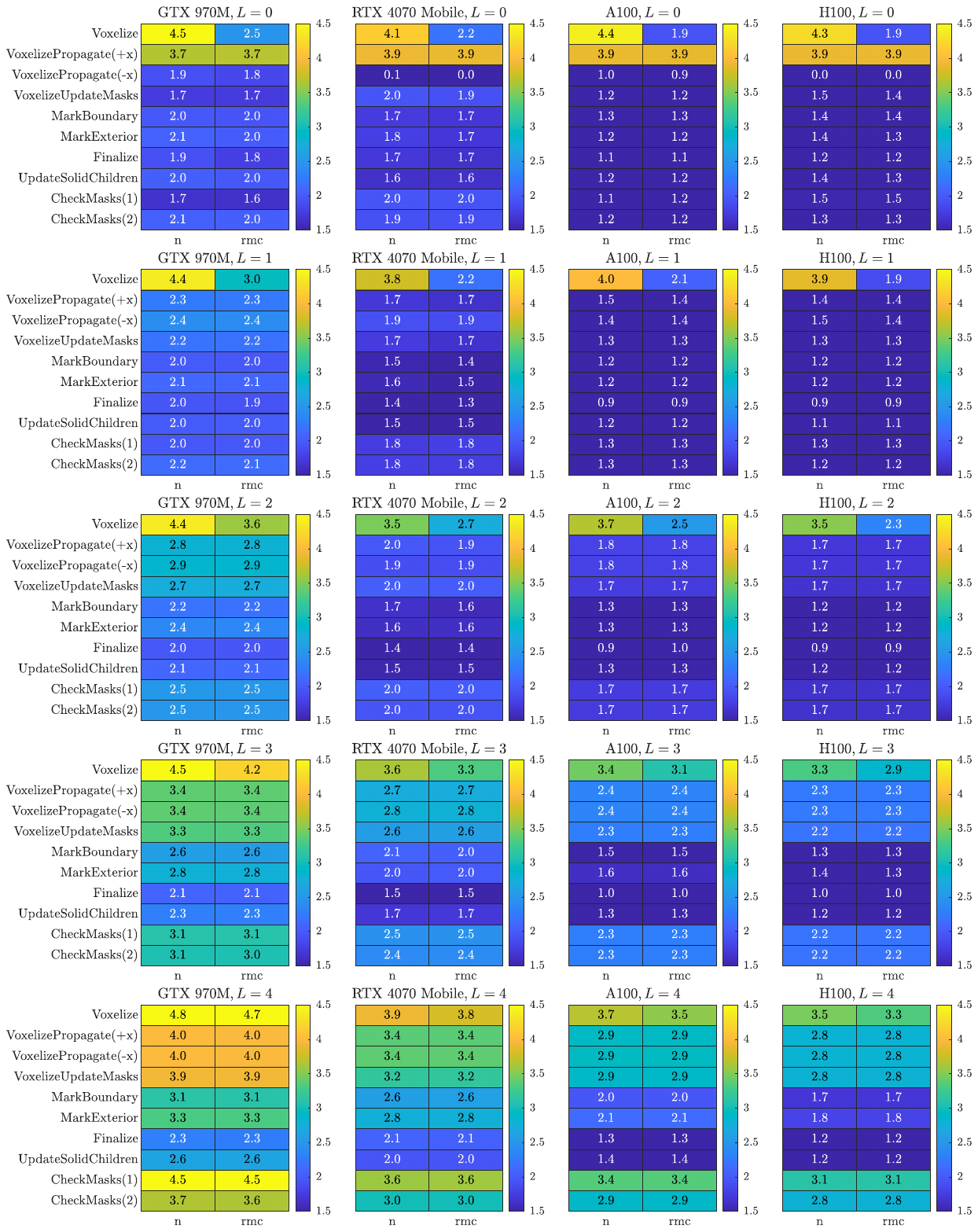}
    \caption{Execution times for the voxelizer tests.}
    \label{fig:res_voxelizer_timings}
\end{figure}

Spatial binning is replicated for the BunnyL5 test (single-precision) under five optimization configuration to demonstrate their benefits:
\begin{enumerate}[noitemsep]
    \item No optimizations (`n'): all faces are considered in the provided geometry, and a full sort and reduction are performed on the bounding box limit arrays.
    \item Filtering (`r'): faces are filtered prior to spatial binning and voxelization without compaction of the arrays.
    \item Compaction (`c'): arrays are compacted only prior to sorting and reduction without face filtering.
    \item Filtering and compaction (`rc'): filtering and compaction are both performed.
    \item Filtering, mapping, and compaction (`rmc'): compaction and filtering are both performed, in addition to a re-mapping of faces after filtering.
\end{enumerate}
We also illustrate how these execution times change at each grid level. Heat maps of the execution times in $\log_{10}$ $\mu \text{s}$ are displayed in Figures \ref{fig:res_binning_timings} and \ref{fig:res_voxelizer_timings} for the spatial binning and voxelizer subroutines, respectively. In all tests, the root grid is set to $64^3$ with a bin density that is matched at $16^3$. This ensures that all threads assigned to each cell-block in the partial surface voxelization kernel retrieve faces from the same bin. We set $N_{\text{spec.}} = 2$, which induces appreciable subdivision of the Stanford bunny for the L7 tests (but not the dragon). A minimal near-wall distance of 0.05 is also set.


The most expensive step in spatial binning is Step 4 (sortation by key), which ranges from $10^{3.3} \text{ $\mu$s} \approx 2 \text{ ms}$ on the H100 to $10^{4.6} \text{ $\mu$s} \approx 40 \text{ ms}$ on the 970M when no optimizations are used. This decreases by two orders of magnitude on the laptop GPUs and one order of magnitude on the datacenter GPUs when all optimizations are activated. The difference in reduction across these GPUs indicates an appreciable impact of increased memory bandwidth on performance. Steps that are not impacted by the optimizations include Step 2 (bounding box limit computation), scatter operations, the initial ray-indicator computation (which precedes all other computational operations). Filtering and compaction reduce the total amount of data processed by Thrust routines. This is reflected in the reduction in Steps 4, 5 (reduction by key), 7 (adjacent difference) and 8 (copy).

We found that compaction provides a bigger reduction in total execution time than filtering alone. For example, there is a decrease from $10^{4.6}$ to $\sim 10^{3.5} \text{ $\mu$s}$ ($\approx 3.2 \text{ ms}$) in Step 4, while filtering alone decreases the time taken to compute the bounding box limit data from $\sim 10^{3.5}$ to $10^{2.6} \text{ $\mu$s}$ ($\approx 0.4 \text{ ms}$) on the coarse grid (or $10^{3.3} \text{ $\mu $s}$ on the finest grid). This latter reduction is mainly observed in the tests on the older GPU.
Steps 5, 7, and 8 also experience a decrease in time taken by an order of magnitude when compaction is activated alone.

Combining filtering and compaction results in an overall reduction in time taken for sortation and reduction by key. The addition of index mapping after filtration further reduces the time taken for compaction. The additional costs in introducing filtering and compaction (e.g. $10^{2.4} \text{ $\mu$s}$ and $10^{3.4} \text{ $\mu$s}$ at the root grid on the 970M, respectively) are outweighed by the reduction provided to sortation and reduction by key (e.g. $10^{4.6} \text{ $\mu$s}$ and $10^{3.2}$ down to $10^{2.3} \text{ $\mu$s}$ and $10^{1.8} \text{ $\mu$s}$ for a net reduction of $\sim 10^4 \text{ $\mu$s}$). The costs of adding a gather and scatter for index mapping ($\sim 10^{1.9} \text{ $\mu$s}$ and $10^{1.4} \text{ $\mu$s}$) are, again, smaller than the reduction to compaction ($10^{3.4}$ down to $10^{1.8} \text{ $\mu$s}$) for a net reduction of $\sim 10^3 \text{ $\mu$s}$.

Memory allocations typically depend on the specified bin densities and the number of faces, so these only decrease when the choice of GPU differs. One exception is when index mapping is applied after filtering to further reduce the total size of the bounding box limit data arrays, which shortens the time taken for compaction. Since the maximum total sizes of these latter arrays can be predicted by the mapping, a smaller amount of memory can be allocated which reduces time taken in older GPUs or coarser grid levels (this is part of $\text{MemoryAllocations(2)3D}$). The time taken for the first memory allocation fluctuates and likely incorporates execution times in steps that come before the binning step is called. The time taken for the very first allocation on the root grid is omitted as its magnitude greatly exceeds that of the time taken for all of other steps and likely does not reflect on the actual binning algorithm.

On the A100 and H100, the only steps that significantly contribute to the total execution of spatial binning after activation of the optimizations are the memory allocations, bounding box limit computation, compaction, and sortation by key. These are generally in the range $\sim 10^2 \text{ $\mu$s} = 0.1 \text{ ms}$ to $\sim 10^3 \text{ $\mu$s} = 1 \text{ ms}$. The remaining steps are below the lower limit and negligible regardless of the grid level.

The partial surface voxelization kernel is the primary bottleneck in the voxelizer routine in all grid levels except the root grid, where external flag propagation must be performed in the whole grid, and the finest grid, where the identification of boundary cells requires processing the most cells. Since the faces of the geometry are only accessed by the former voxelization kernel, it is the only one that experiences the reduction in total time taken across the optimization variants. All other subroutines are negligible in the coarser grids $L \leq 2$ (at $\sim 10^2 \text{ $\mu$s}=0.1\text{ ms}$ in more recent GPUs). On finer grids $L \geq 3$, the execution times of partial surface voxelization, external flag propagation, and boundary cell classification are all comparable in order of magnitude, from $\sim 10^4 \text{ $\mu$s} = 10 \text{ ms}$ on the 970 to $\sim 10^3 \text{ $\mu$s} = 1 \text{ ms}$ on the H100. The kernels targeting cell-block metadata (such as MarkBoundary and MarkExterior) are naturally quicker to execute since there is less data to process overall. Although negligible when processed on datacenter GPUs for all $L$, they are still felt on the laptop GPUs for $L \geq 3$ and require $\sim 10^3 \text{ $\mu$s}$ to execute.

The filtration of faces produces significant reductions in execution time of partial surface voxelization by 1-2 orders of magnitude in coarser grids across all GPUs. This can be explained by the sparsity of the cell centers relative to the geometry mesh in these grids. Cell-centers safely ignore most of the faces (especially when the geometry is of relatively high resolution, as in the XYZ RGB dragon) and only need to interact with the faces that intersect rays cast along $\textbf{e}_x$. Execution times on the finest grids are nearly identical pre- and post-filtration.

In summary, the optimizations introduced in the spatial binning procedure demonstrably speed up both the binning of the faces and the subsequent partial surface voxelization on each grid level. Filtering, index mapping, and compaction all play a role in reducing the total workload required to process the faces when arranging their assignment to the bins. Only filtering affects voxelization, but its inclusion noticeably decreases the execution time of partial surface voxelization in the coarser grid levels.

\subsubsection{Performance of Kernel Variants}

Section \ref{sec:methodology} introduced several variants of the partial surface voxelization kernel that retrieve vertex data using different approaches or memory arrangements. Variant 1 reads the data directly from a global memory array in SoA format (V1-SoA) or AoS format (V1-AoS). Variant 2 is structured so that the vertex data is read from an AoS array in parallel to shared memory, and then broadcast to individual threads for subsequent calculations. Dual versions complement these variants that employ warp-level primitives for inter-block communication (e.g., block-wise reduction). We also introduced two variants of the boundary cell identification kernel. The first variant employs shared memory to organize cell masks read from neighboring cell-blocks. The second variant directly reads data from global memory, which potentially results in redundant reads.

The kernel variants are all executed on the 970M with the BunnyL5 test. The required time on grid levels $L=3,4$ are displayed in Table \ref{tab:tests_voxelizer_vars}. For the partial surface voxelization kernel, performance was similar across all variants, with differences in the range of execution times of $4.4 \text{ ms}$ on $L=3$ and $8.7 \text{ ms}$. The variant requiring the least time taken was V1-AoS. Warp-level variants were all more expensive, which indicates that restructuring the kernels to make use of warp-level primitives resulted in an overall penalty. The penalty is minimized in V2, where the differences between the duals were $2.2 \text{ ms}$ and $2.5 \text{ ms}$ on levels $3$ and $4$, respectively. However, V2 is slightly more expensive than both cases of V1 by up to $\sim 2 \text{ ms}$.

We found that the direct-access variant of the boundary cell identification variant required less time than the shared memory variant by $6 \text{ ms}$ on level $4$. Performance was identical to one decimal place on level $3$. This indicates that the overhead in setting up shared memory loads (i.e., thread mapping, conditional statements) requires more time than simply accessing the data even when some of the transactions are redundant.
\begin{table}[t]
    \small
    \centering
    \begin{tabular}{ll|cc}
        \hline
        \multirow{2}{*}{Kernel} & \multirow{2}{*}{Variants} & \multicolumn{2}{c}{Execution Times ($\text{ms}$)} \\ \cline{3-4}
         & & $L=3$ & $L=4$ \\
        \hline
        \multirow{6}{*}{Voxelize} & V1-SoA & $13.3 \pm 0.020$ & $43.1 \pm 0.018$ \\
        & V1-SoA (Warp) & $17.4 \pm 0.028$ & $51.3 \pm 0.014$ \\
        & V1-AoS & $13.0 \pm 0.022$ & $42.6 \pm 0.012$ \\
        & V1-AoS (Warp) & $17.3 \pm 0.034$ & $50.9 \pm 0.026$ \\
        & V2 & $13.4 \pm 0.019$ & $44.4 \pm 0.010$ \\
        & V2 (Warp) & $15.6 \pm 0.023$ & $46.9 \pm 0.015$ \\
        \hline
        \multirow{2}{*}{Boundary Cell Id.} & Shared Memory & $01.3 \pm 0.001$ & $40.4 \pm 0.010$\\
        & Direct & $01.3 \pm 0.001$ & $34.2 \pm 0.006$ \\
        \hline
    \end{tabular}
    \caption{Approximate execution times for the partial surface voxelization and boundary cell identification kernel variants applied to the BunnyL5 test on the 970M.}
    \label{tab:tests_voxelizer_vars}
\end{table}

We also varied the choice of floating-point precision to test performance with increased workloads. Floating-point precision strongly affects the triangle-AABB and partial surface voxelization kernels; both are computation-heavy, but the latter also increases total traffic due to the retrieval of larger vertex payloads. In fact, the partial surface voxelization kernels employ point-in-triangle tests based on the triangle-AABB overlap test to reduce ray cast intersection misses due to floating-point round-off. Table \ref{tab:tests_voxelization_precision} displays the execution times for the partial surface voxelization kernel in the DragonL7 and BunnyL7 tests on the H100. In both tests, switching from single to double precision results in an increase in execution time of $\sim 2.3\times$ on finer grids. On coarser grids, the workload remains too small to saturate bandwidth and the arithmetic units. The deviation from an ideal increase of $2\times$ indicates that the kernel is not fully memory- or compute-bound. Profiling via the NVIDIA Nsight Compute toolkit indicated that computation played a significant role, likely due to the use of triangle-AABB tests. The deviation from ideal is not severe, but exact floating-point arithmetic will be strongly considered in future work both to increase kernel efficiency and to eliminate false negatives in the point-in-triangle tests.
\begin{table}[h]
    \centering
    \small
    \begin{tabular}{c|rr|rr}
        \hline
        \multirow{3}{*}{Level} & \multicolumn{4}{c}{Execution Times ($\mu \text{s}$)} \\ \cline{2-5}
        & \multicolumn{2}{c}{DragonL7} & \multicolumn{2}{c}{BunnyL7} \\ \cline{2-5}
        & FP32 & FP64 & FP32 & FP64 \\
        \hline
        0 & $68\pm2$  &$96\pm2$  &$71\pm2$  &$108\pm4$  \\
        1 & $45\pm2$  &$74\pm2$  &$56\pm2$  &$102\pm2$  \\
        2 & $76\pm2$  &$150\pm2$  &$135\pm2$  &$285\pm2$  \\
        3 & $211\pm2$  &$466\pm2$  &$465\pm2$  &$1,054\pm2$  \\
        4 & $776\pm2$  &$1,801\pm2$  &$1,977\pm2$  &$4,587\pm2$  \\
        \rowcolor{gray!20}
        5 & $3,374\pm2$  &$7,933\pm2$  &$7,959\pm2$  &$18,566\pm2$  \\
        \rowcolor{gray!20}
        6 & $14,649\pm2$  &$34,535\pm4$  &$25,139\pm2$  &$57,119\pm4$  \\
        \hline
    \end{tabular}
    \caption{Execution times for the partial surface voxelization kernel invoked in the DragonL7 and BunnyL7 tests on the H100 in single- and double-precision.}
    \label{tab:tests_voxelization_precision}
\end{table}

%The partial surface voxelization kernel can be structured so that either 1) each cell-block loads vertex data component-by-component with broadcasts, or 2) threads all participate in a parallel read of the vertex data by exploiting the array of structures format. The latter approach requires broadcasts from shared memory, which is less expensive than broadcasts from global memory (even if the data is structured so that it is likely to be read from cached memory). These kernels can also be rewritten so that they take a purely warp-based approach. Some time can be saved by reading vertex data in parallel.

\subsection{Octree and Voxelization Visualizations}

\begin{figure}
    \centering
    \begin{subfigure}[b]{0.45\textwidth}
        \centering
        \includegraphics[width=1\linewidth]{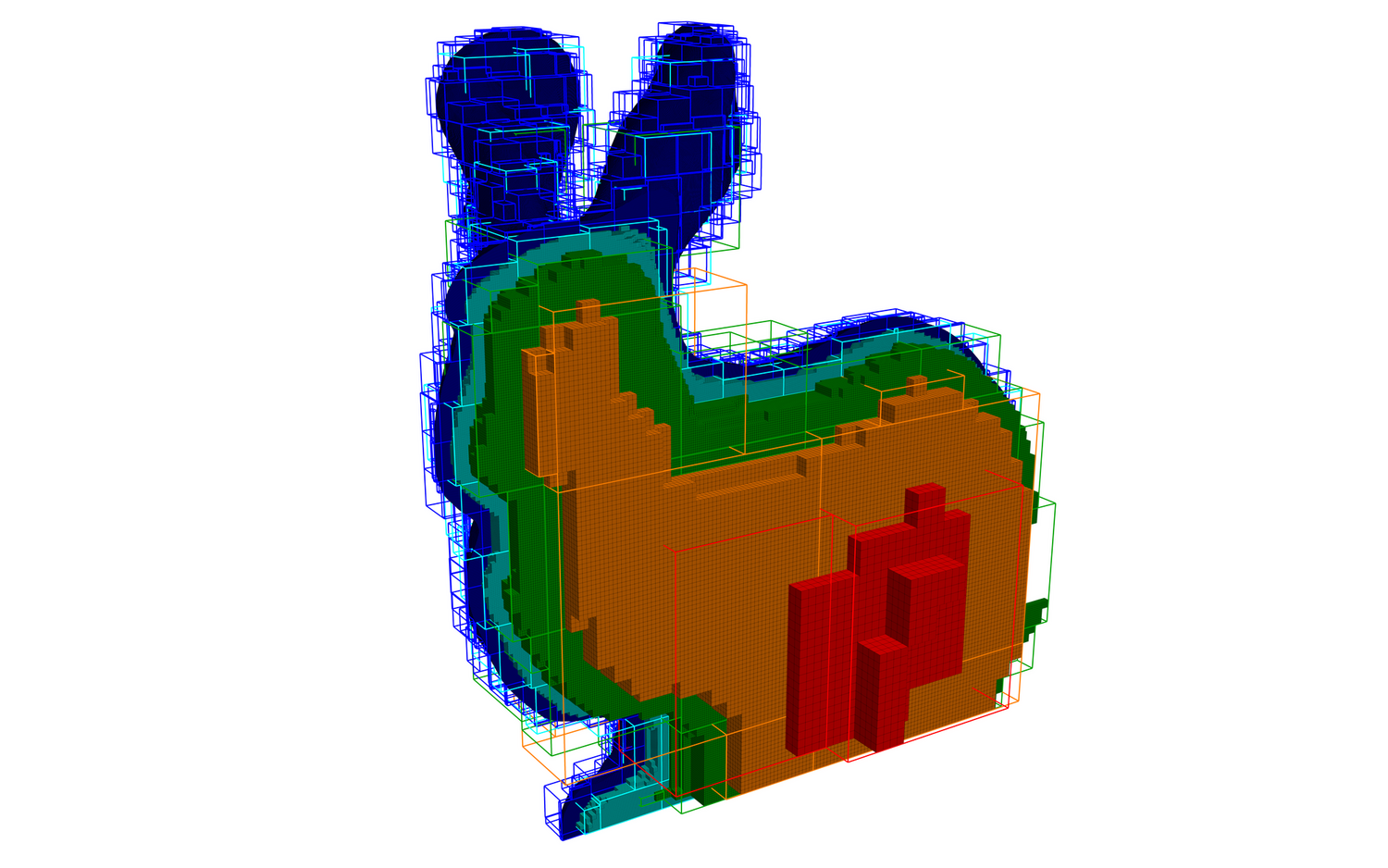}
        \caption{}
    \end{subfigure}
    \hfill
    \begin{subfigure}[b]{0.45\textwidth}
        \centering
        \includegraphics[width=1\linewidth]{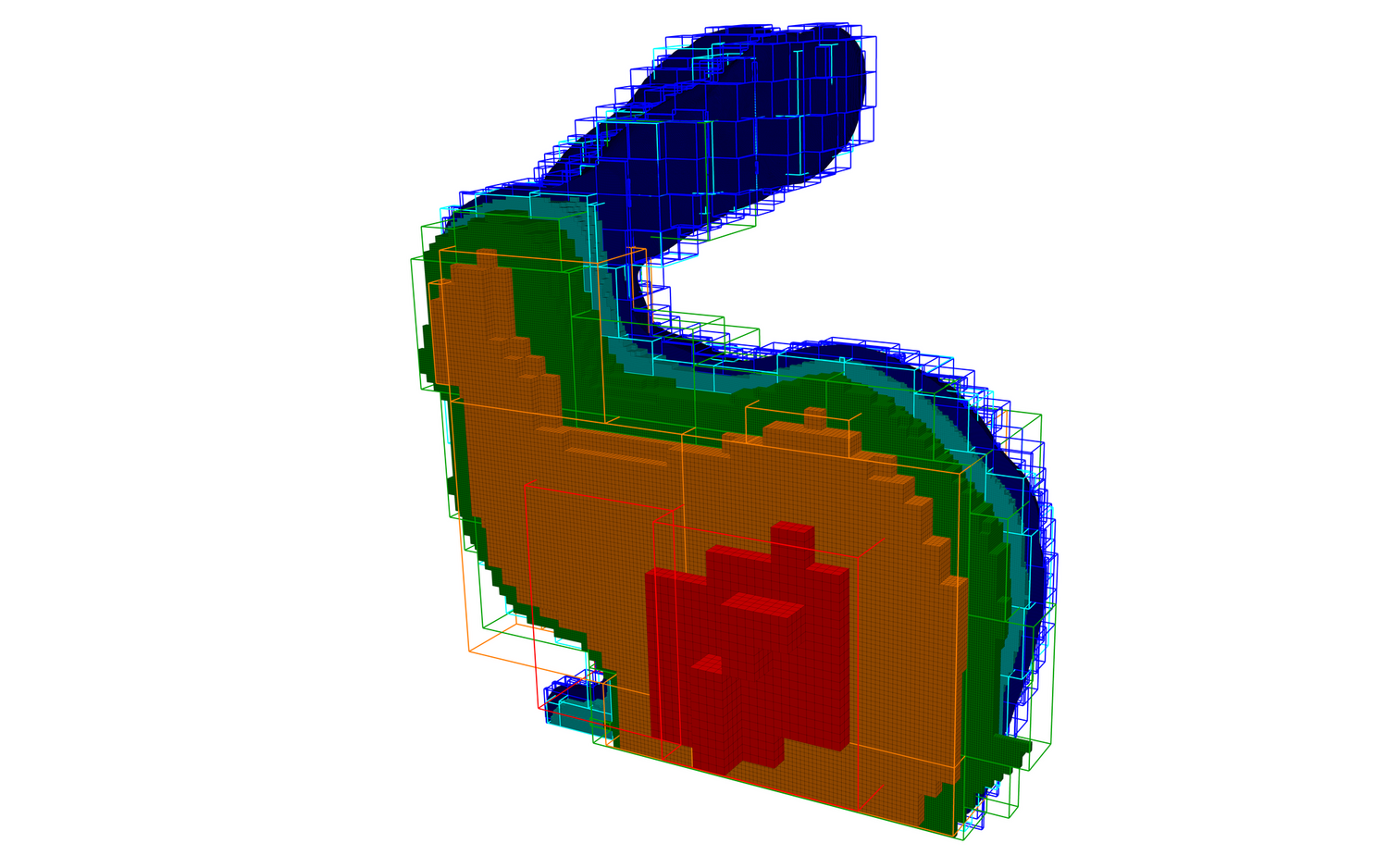}
        \caption{}
    \end{subfigure}
    \caption{Illustration of the block-based sparse voxel octree representation of the Stanford bunny. Levels $L < 2$ are omitted.}
    \label{fig:results_paraview_bunny_octree}
\end{figure}

\begin{figure}
    \centering
    \begin{subfigure}[b]{0.45\textwidth}
        \centering
        \reflectbox{\includegraphics[width=1\linewidth]{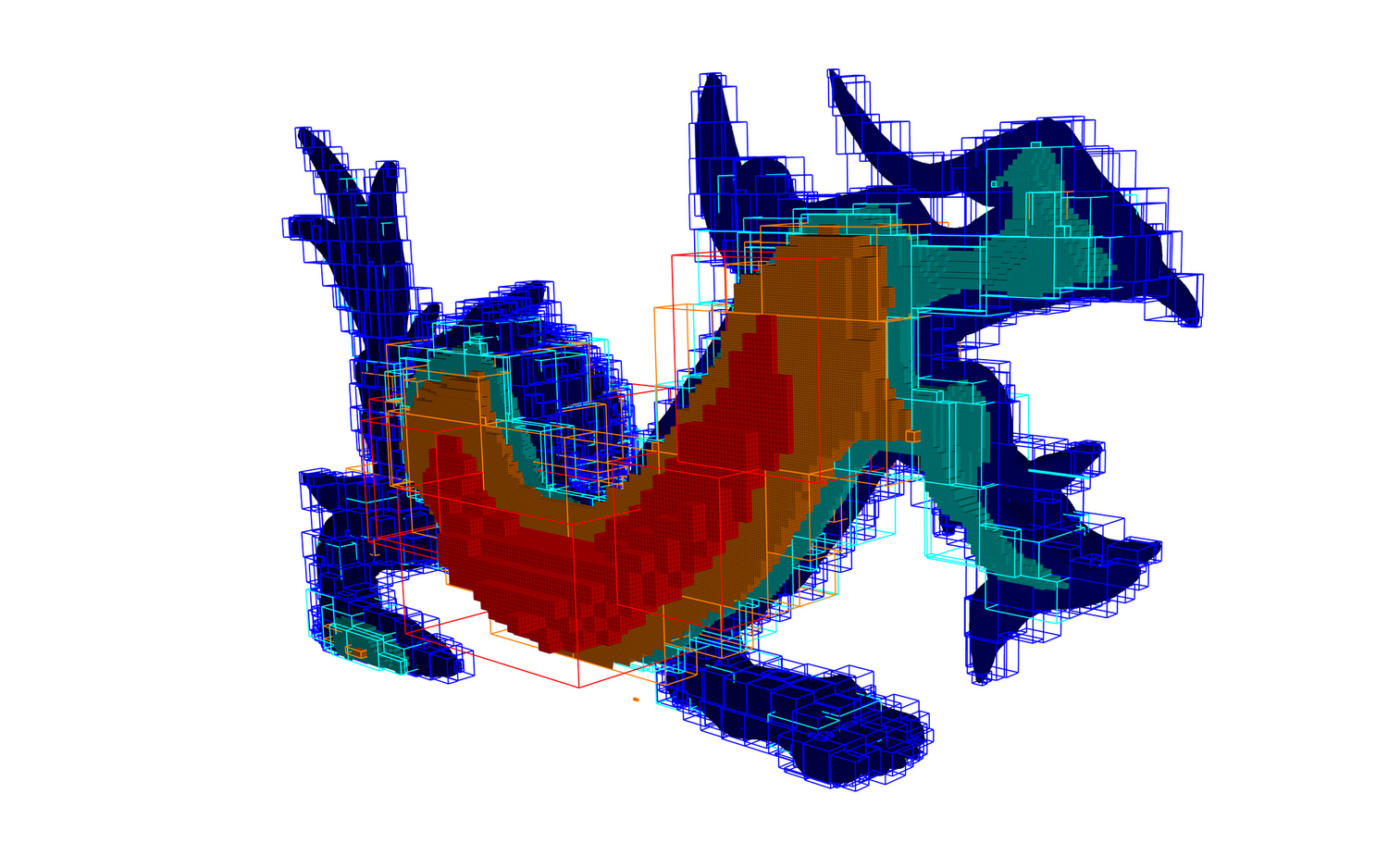}}
        \caption{}
    \end{subfigure}
    \hfill
    \begin{subfigure}[b]{0.45\textwidth}
        \centering
        \reflectbox{\includegraphics[width=1\linewidth]{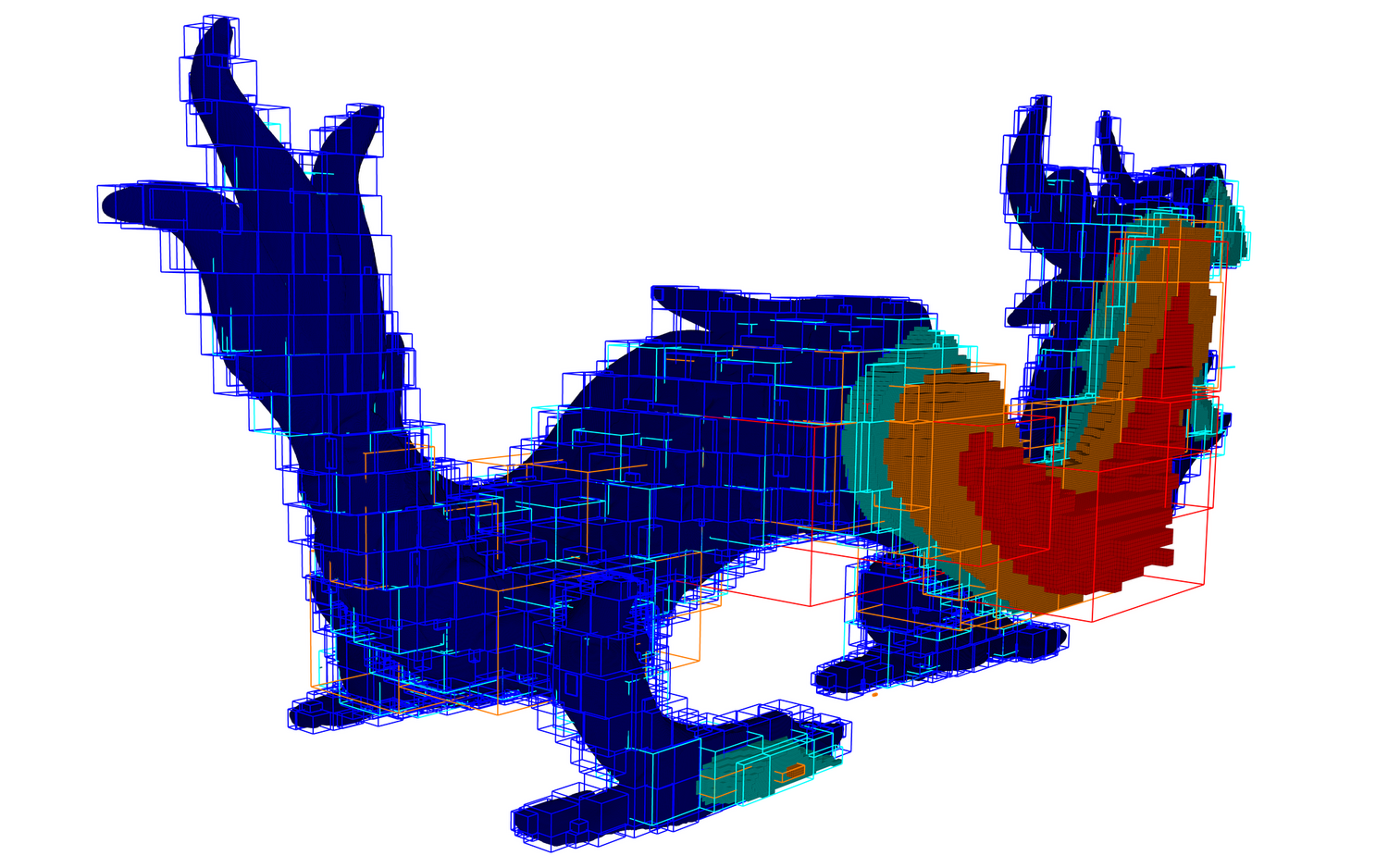}}
        \caption{}
    \end{subfigure}
    \caption{Illustration of the block-based sparse voxel octree representation of the XYZ RGB dragon (flipped). Levels $L < 3$ are omitted.}
    \label{fig:results_paraview_dragon_octree}
\end{figure}

\begin{figure}
    \centering
    \begin{subfigure}[b]{0.31\textwidth}
        \centering
        \includegraphics[width=1\linewidth]{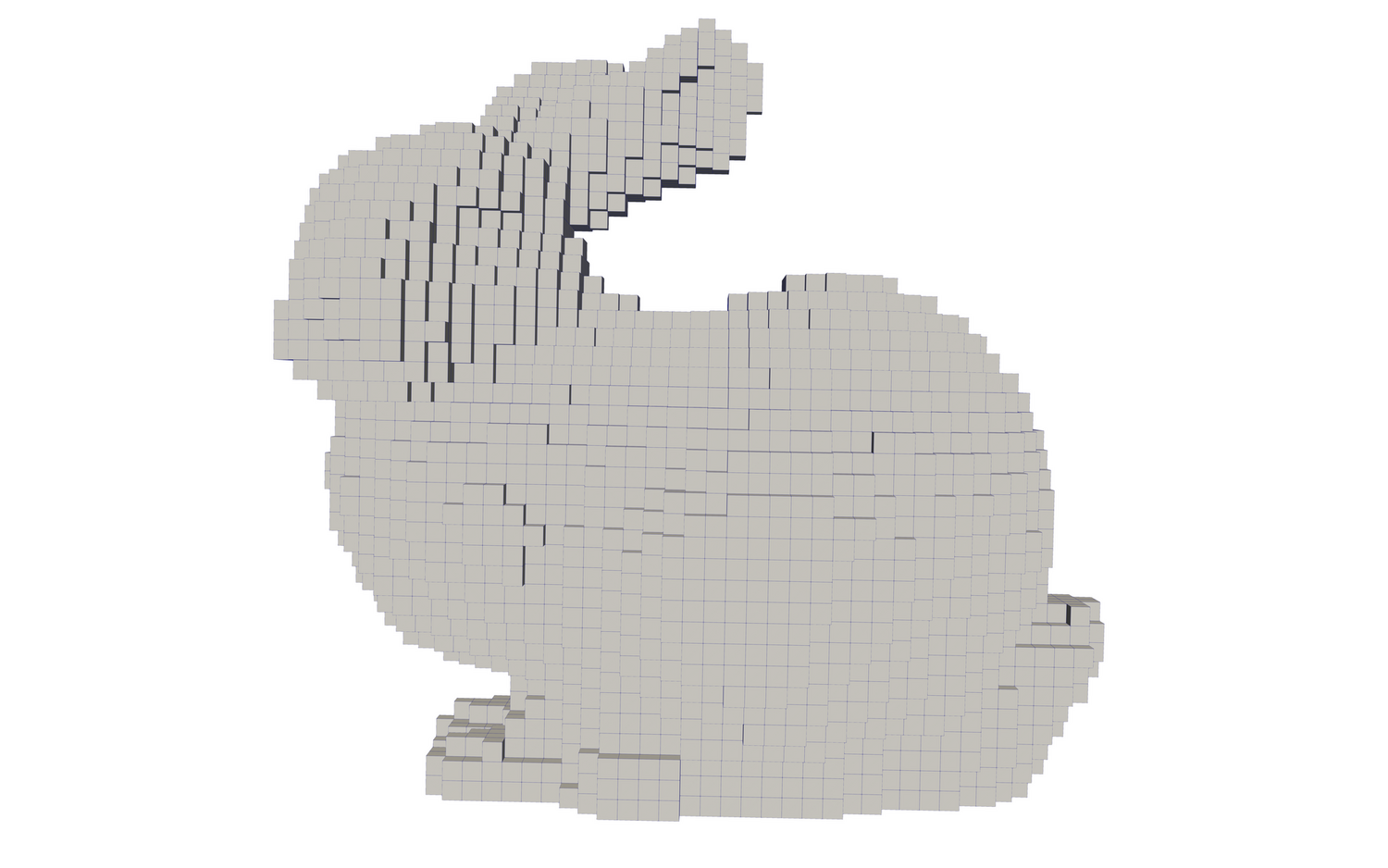}
        \caption{$L=0$}
    \end{subfigure}
    \hfill
    \begin{subfigure}[b]{0.31\textwidth}
        \centering
        \includegraphics[width=1\linewidth]{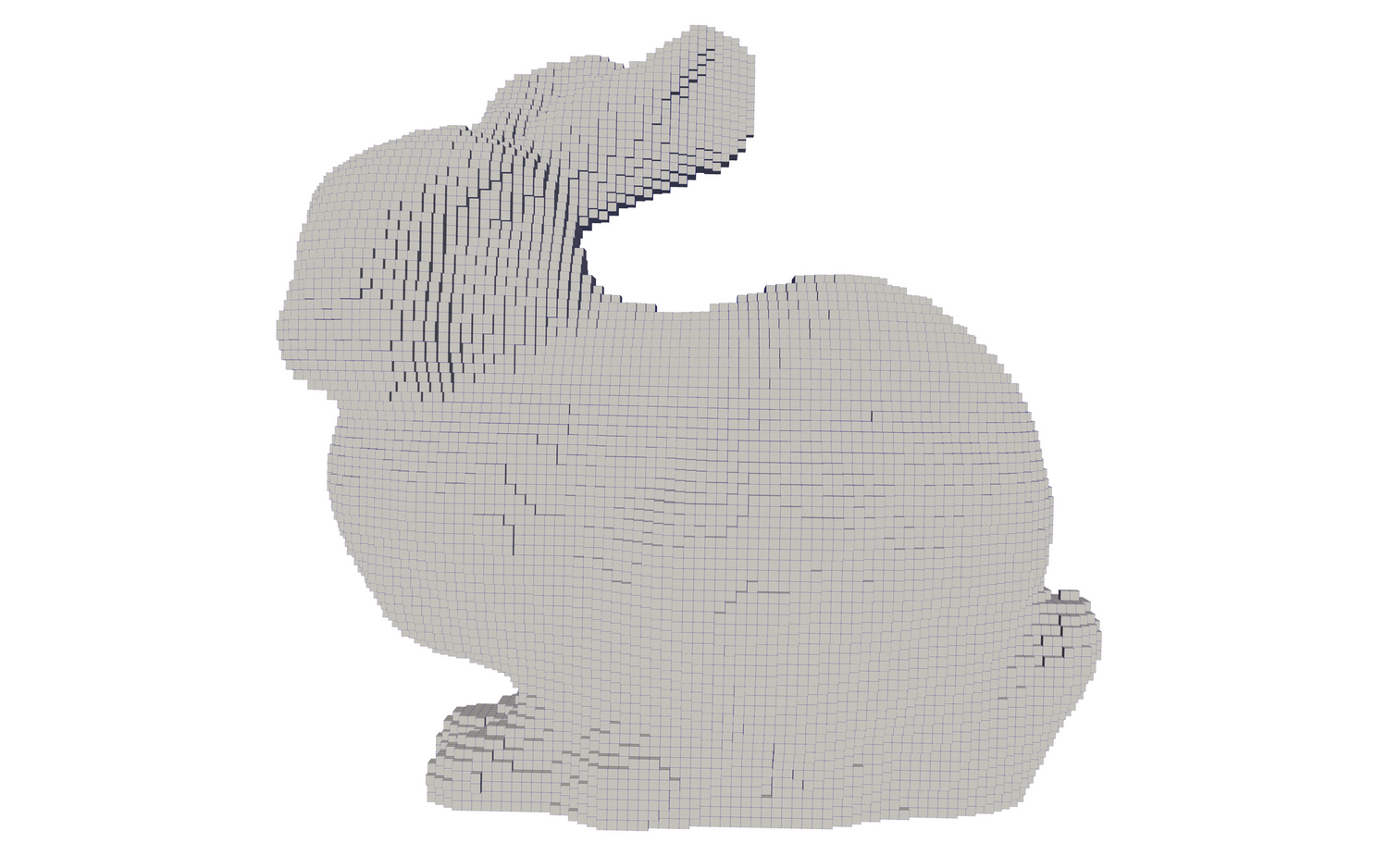}
        \caption{$L=1$}
    \end{subfigure}
    \hfill
    \begin{subfigure}[b]{0.31\textwidth}
        \centering
        \includegraphics[width=1\linewidth]{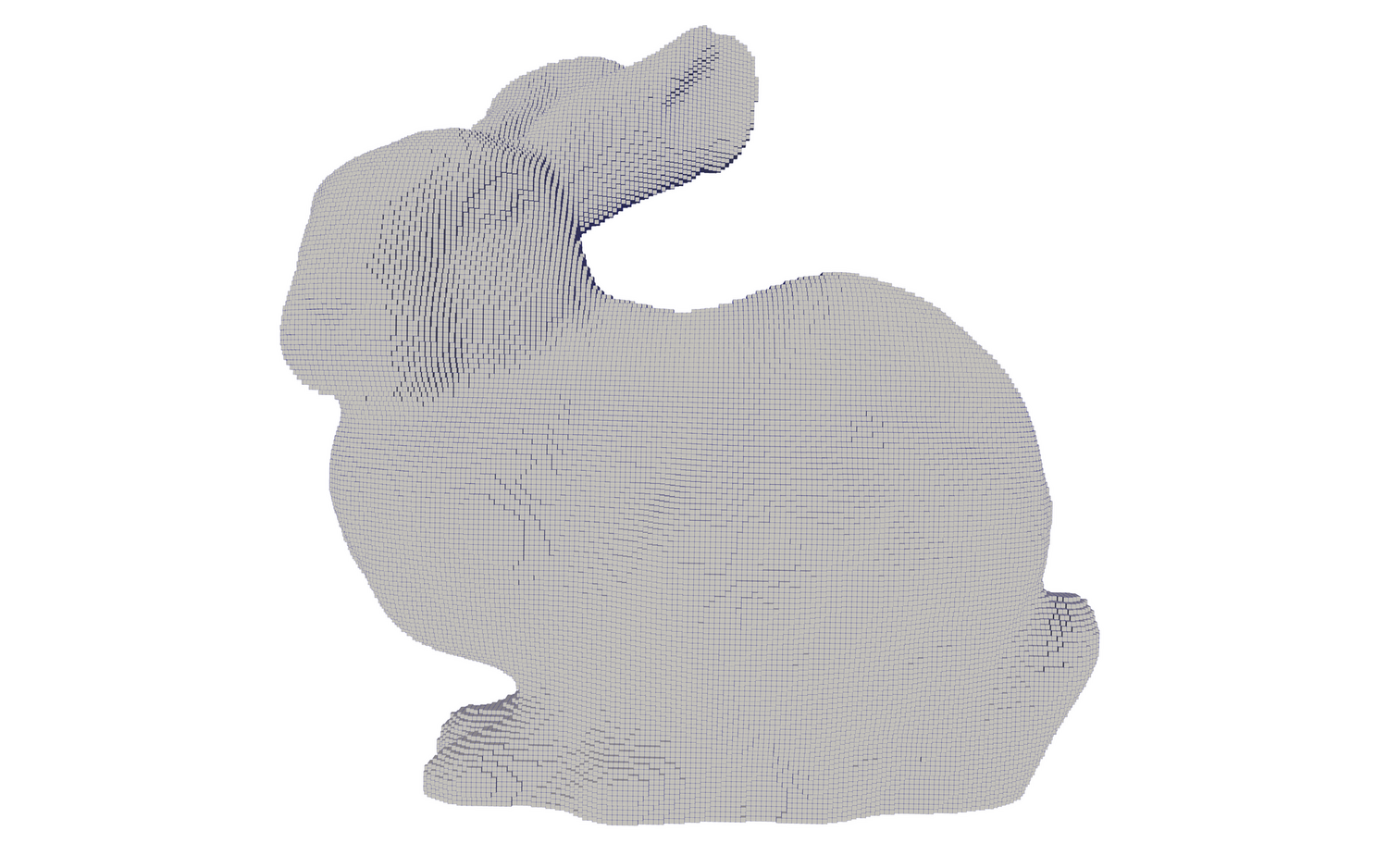}
        \caption{$L=2$}
    \end{subfigure}
    \begin{subfigure}[b]{0.31\textwidth}
        \centering
        \includegraphics[width=1\linewidth]{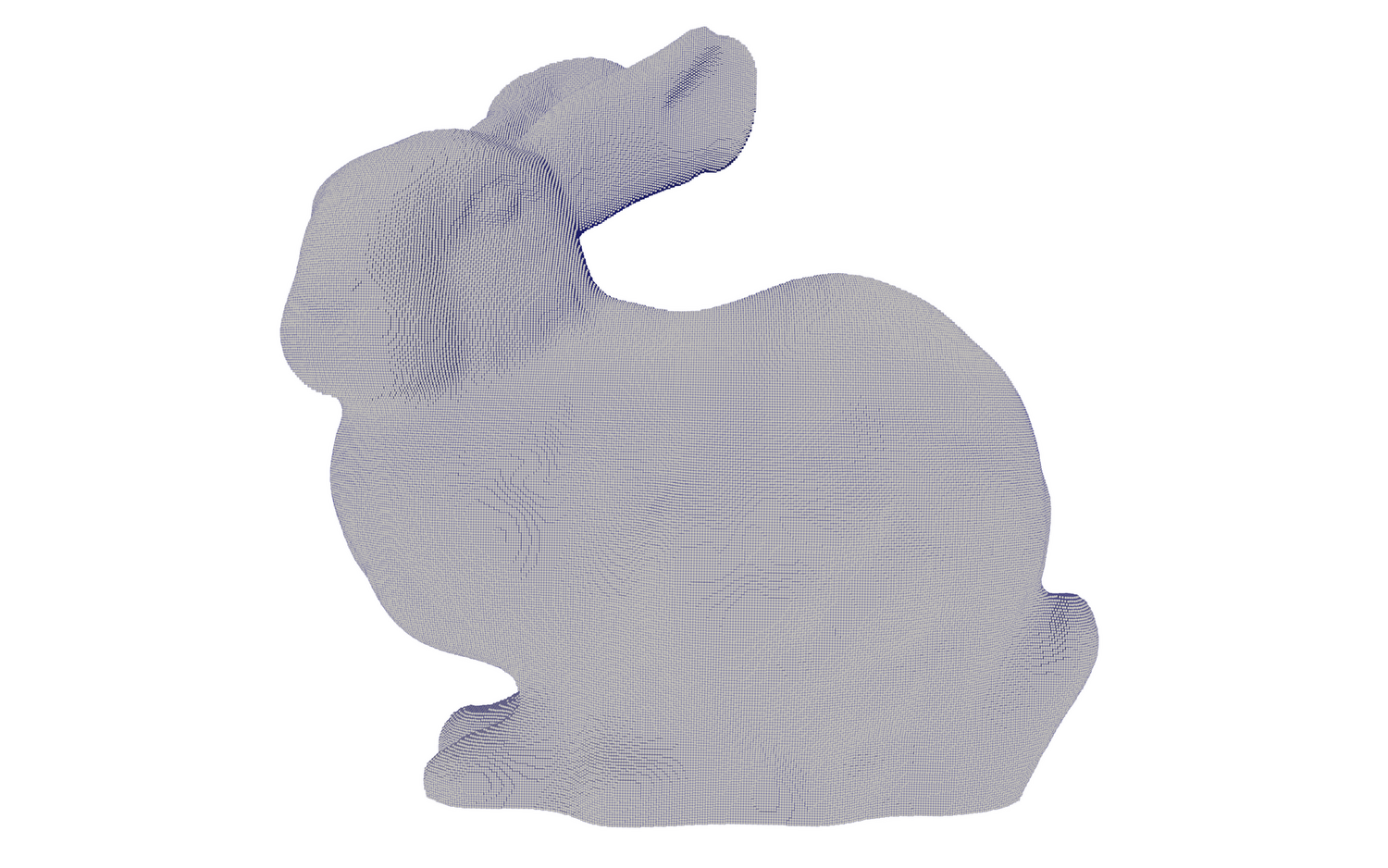}
        \caption{$L=3$}
    \end{subfigure}
    \hfill
    \begin{subfigure}[b]{0.31\textwidth}
        \centering
        \includegraphics[width=1\linewidth]{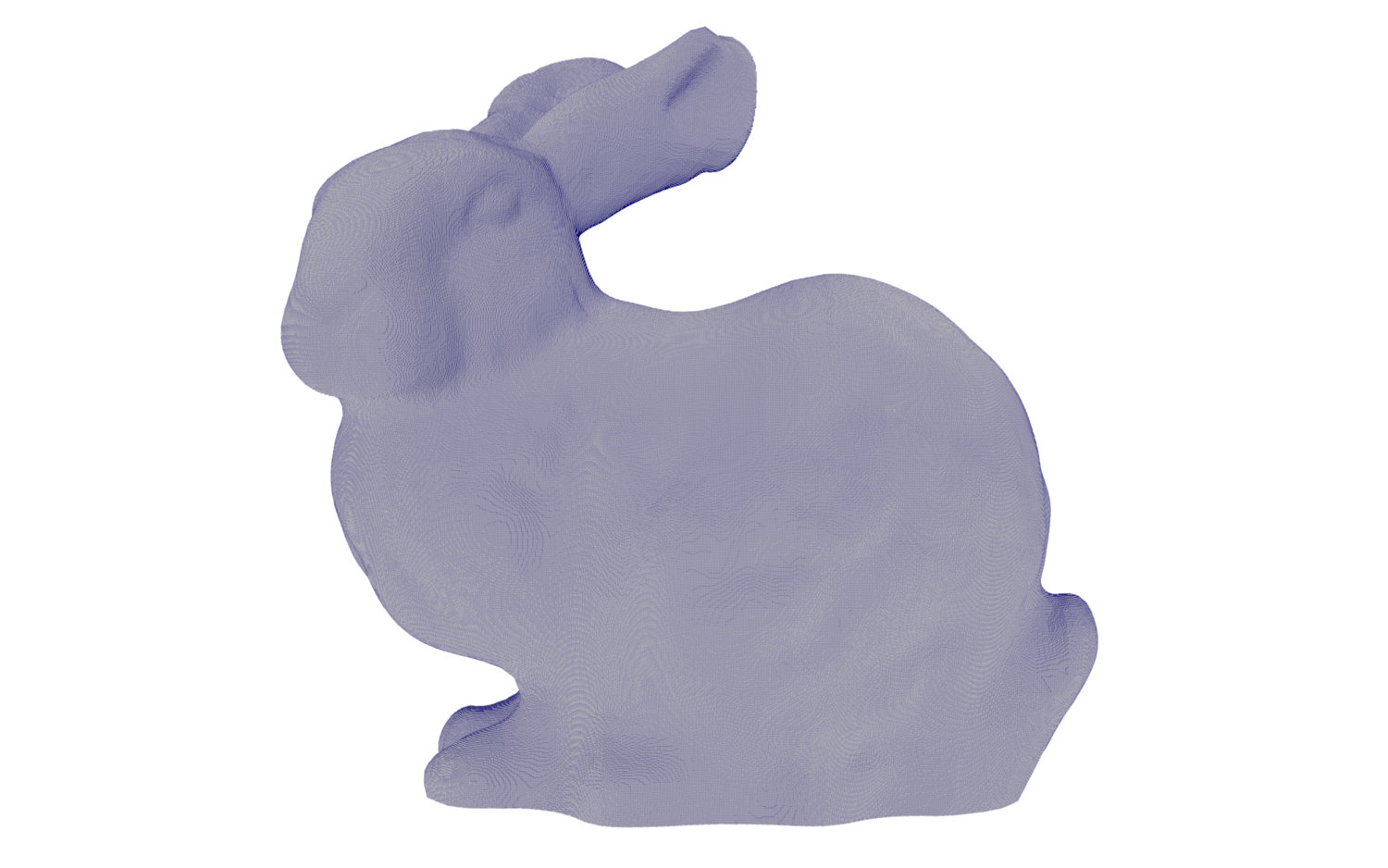}
        \caption{$L=4$}
    \end{subfigure}
    \hfill
    \begin{subfigure}[b]{0.31\textwidth}
        \centering
        \includegraphics[width=1\linewidth]{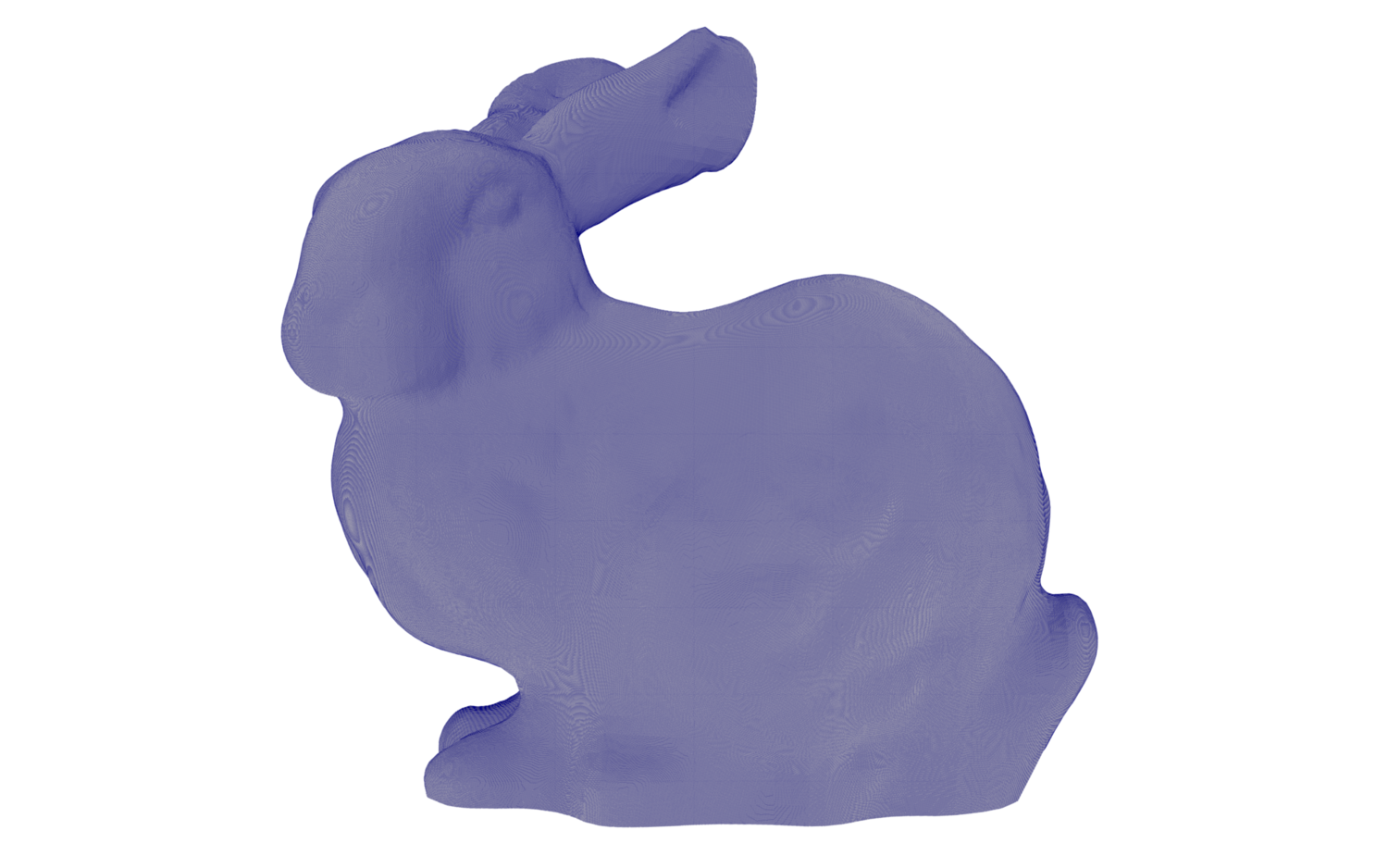}
        \caption{$L=5$}
    \end{subfigure}
    \caption{Voxelizations of the Stanford bunny extracted from a single forest-of-octrees grid.}
    \label{fig:results_paraview_bunny}
\end{figure}

\begin{figure}
    \centering
    \begin{subfigure}[b]{0.31\textwidth}
        \centering
        \includegraphics[width=1\linewidth]{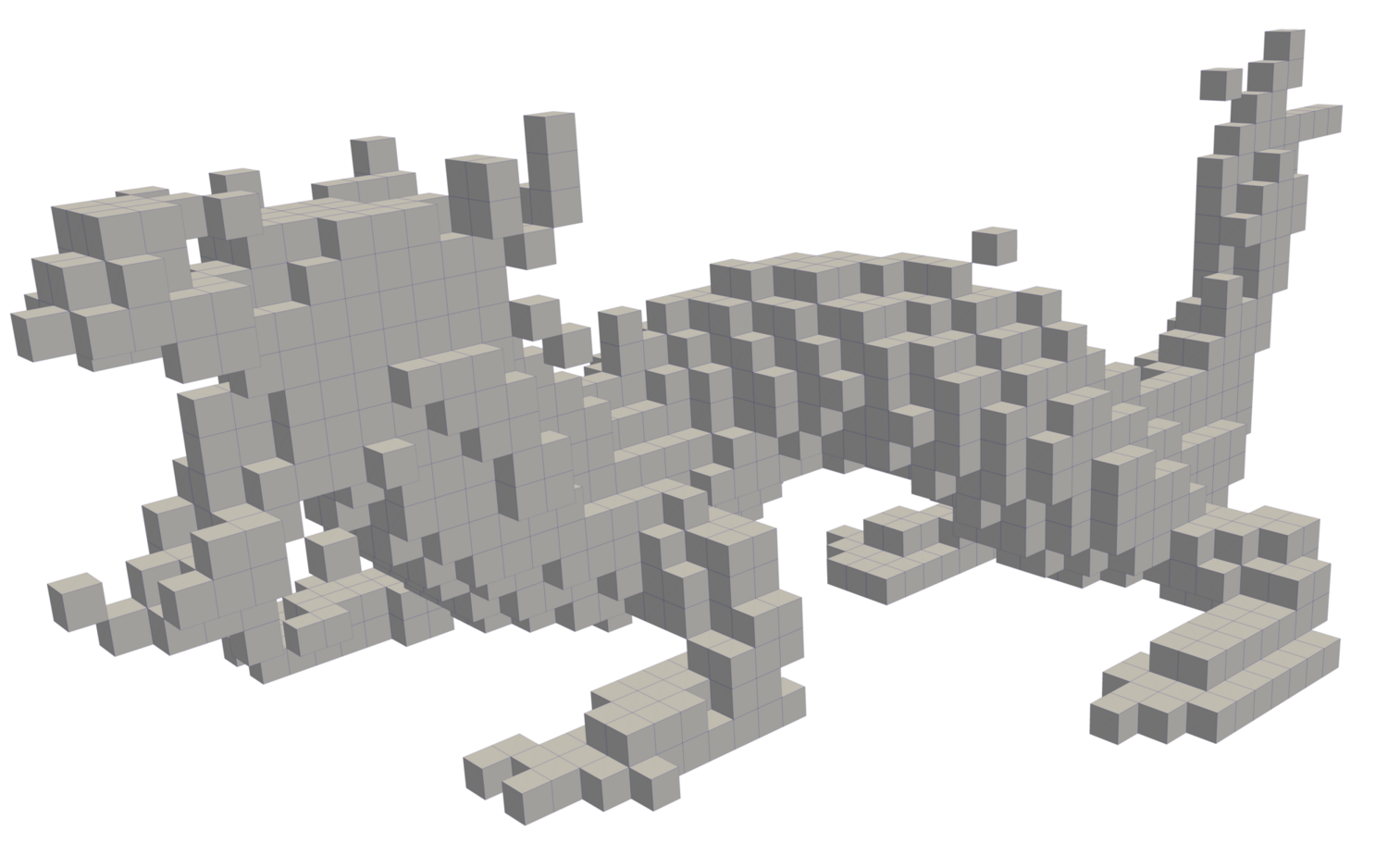}
        \caption{$L=0$}
    \end{subfigure}
    \hfill
    \begin{subfigure}[b]{0.31\textwidth}
        \centering
        \includegraphics[width=1\linewidth]{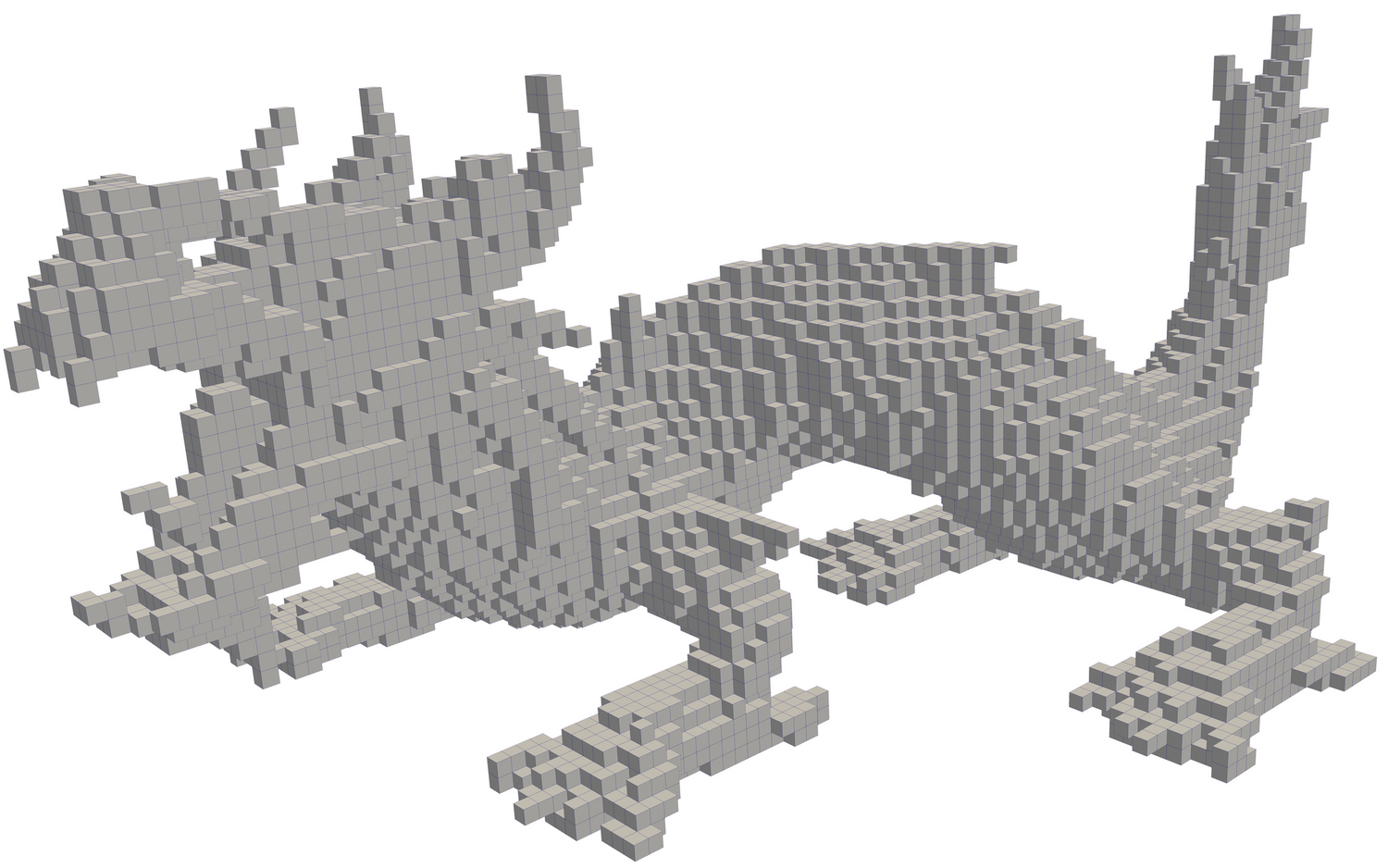}
        \caption{$L=1$}
    \end{subfigure}
    \hfill
    \begin{subfigure}[b]{0.31\textwidth}
        \centering
        \includegraphics[width=1\linewidth]{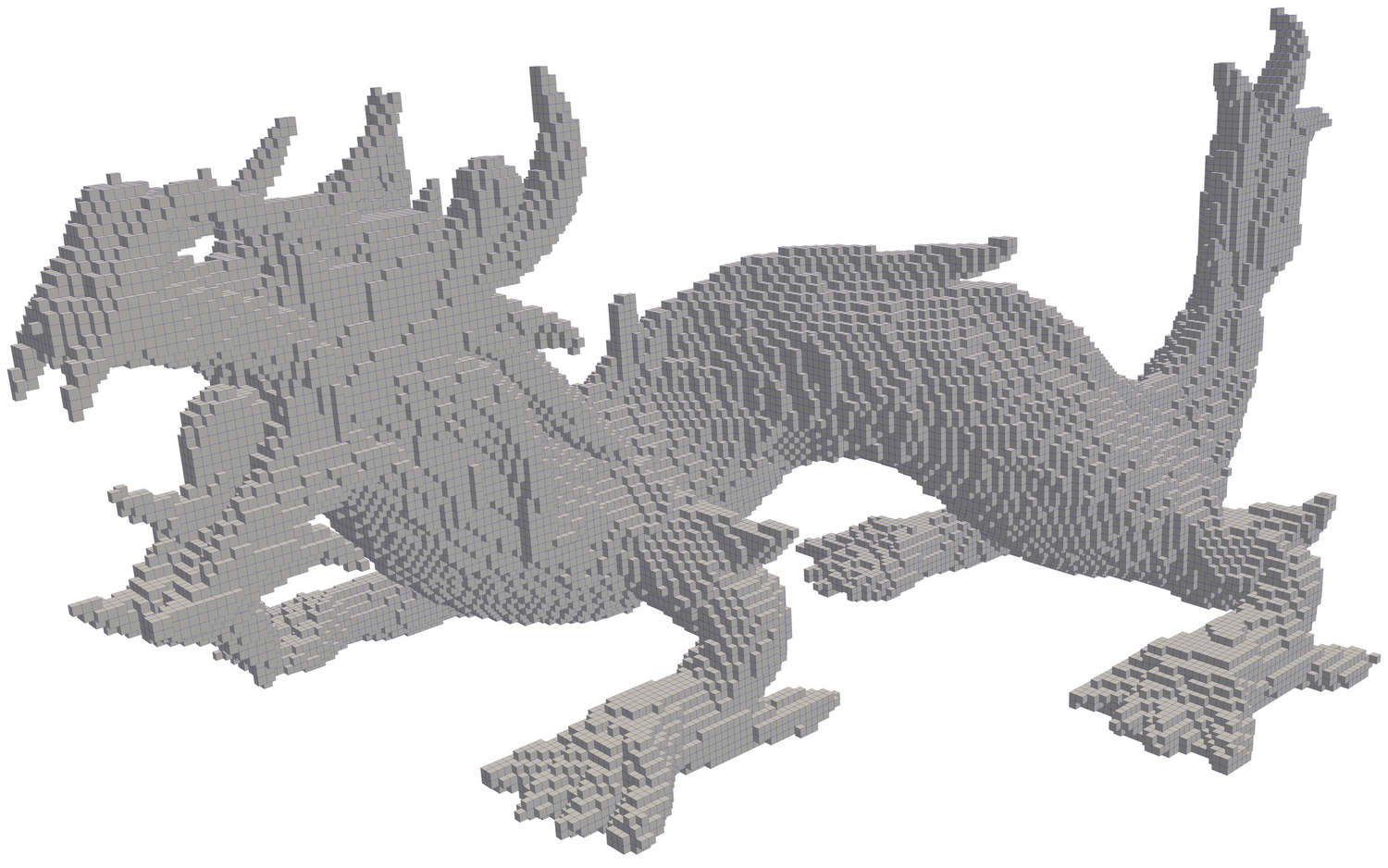}
        \caption{$L=2$}
    \end{subfigure}
    \begin{subfigure}[b]{0.31\textwidth}
        \centering
        \includegraphics[width=1\linewidth]{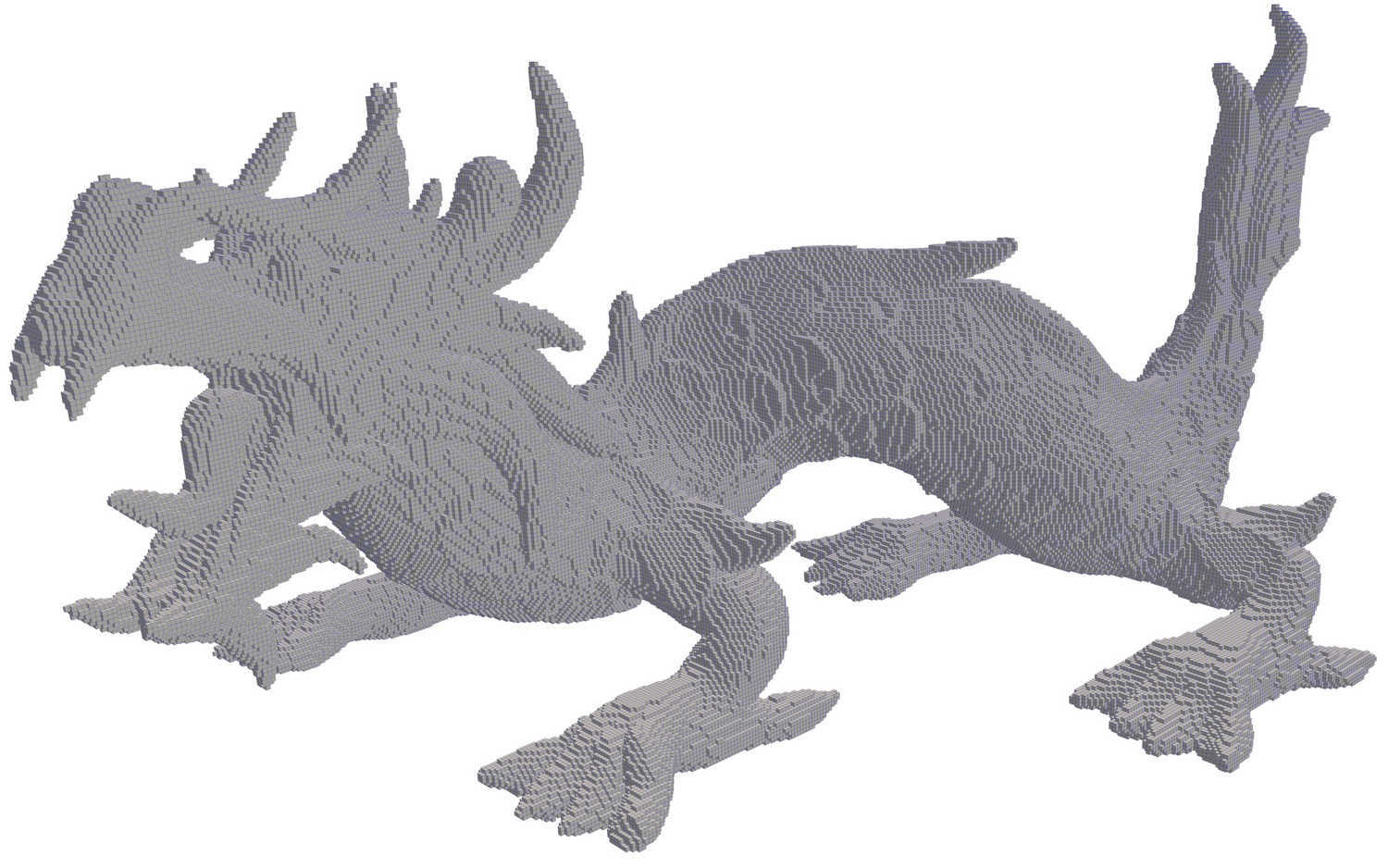}
        \caption{$L=3$}
    \end{subfigure}
    \hfill
    \begin{subfigure}[b]{0.31\textwidth}
        \centering
        \includegraphics[width=1\linewidth]{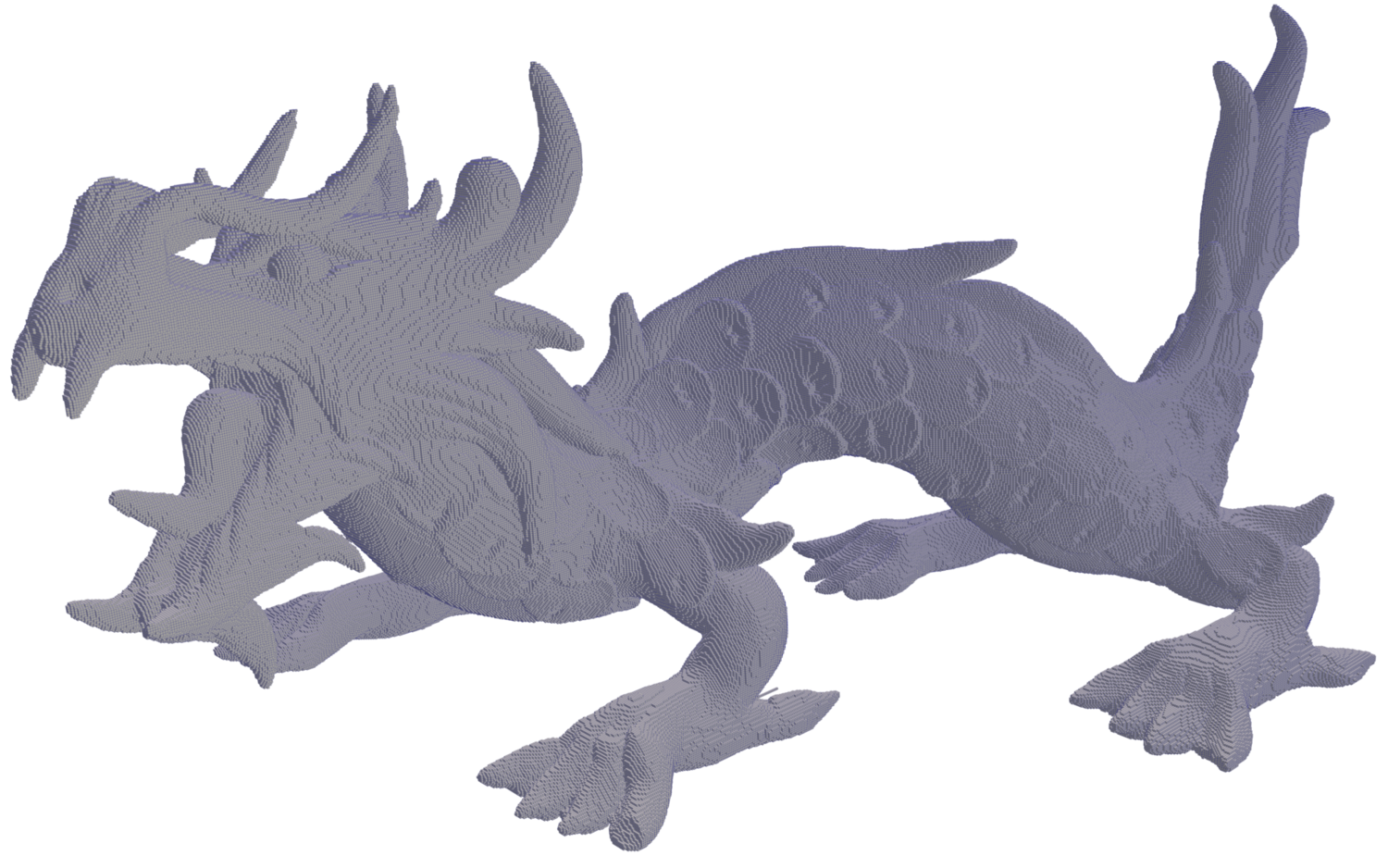}
        \caption{$L=4$}
    \end{subfigure}
    \hfill
    \begin{subfigure}[b]{0.31\textwidth}
        \centering
        \includegraphics[width=1\linewidth]{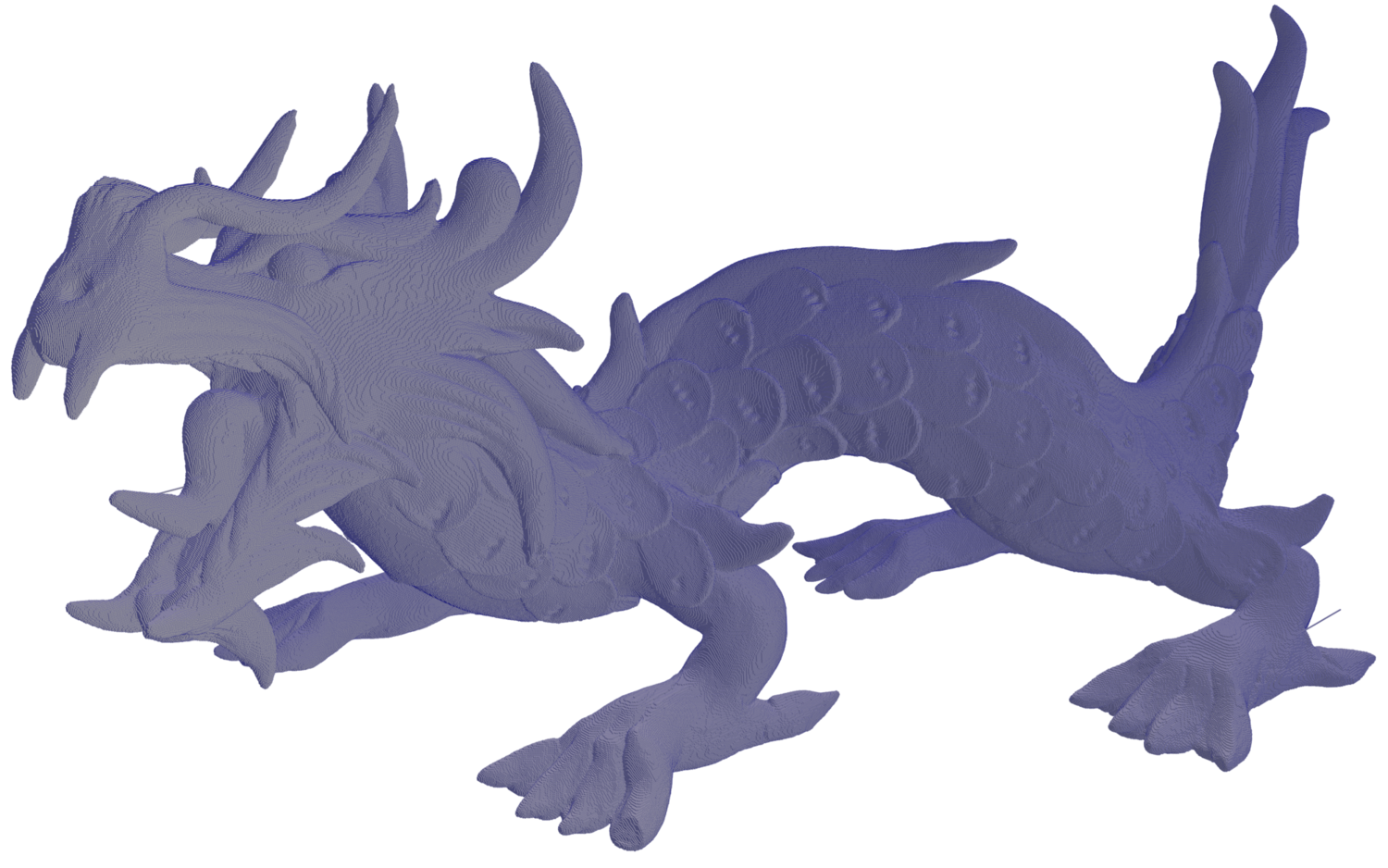}
        \caption{$L=5$}
    \end{subfigure}
    \caption{Voxelizations of the XYZ RGB dragon extracted from a single forest-of-octrees grid.}
    \label{fig:results_paraview_dragon}
\end{figure}

\begin{sidewaysfigure}
    \centering
    \includegraphics[width=\textheight]{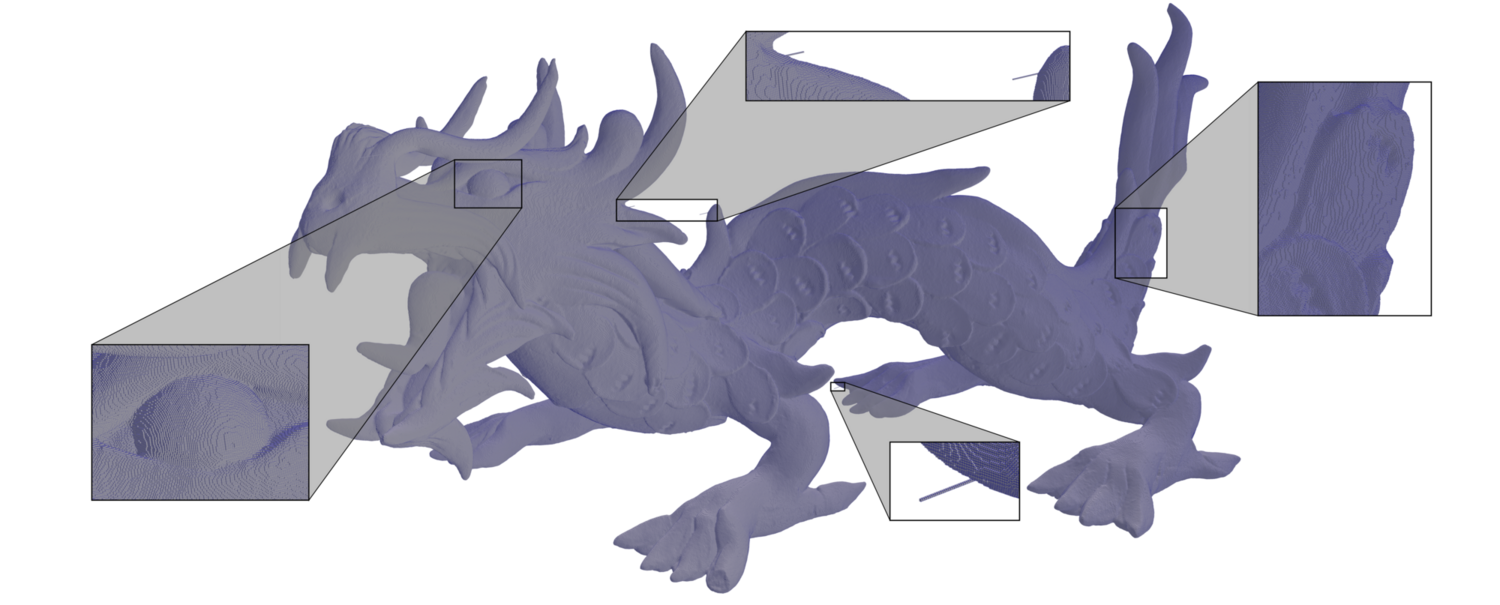}
    \caption{Voxelization of the XYZ RGB dragon on level $L=6$, with an effective grid resolution $64(2^6)=4096$. The algorithm is capable of resolving the complex features of the dragon. In the absence of exact floating-point precision, ray casts during partial surface voxelization will occasionally miss the geometry. Propagation without guard cells in these locations produce spikes. These misses are very rare and appear when then the effective resolution is high and round-off errors become significant.}
    \label{fig:results_bigone}
\end{sidewaysfigure}

\begin{figure}[h]
    \centering
    \begin{subfigure}[b]{0.85\textwidth}
        \centering
        \includegraphics[width=1\linewidth]{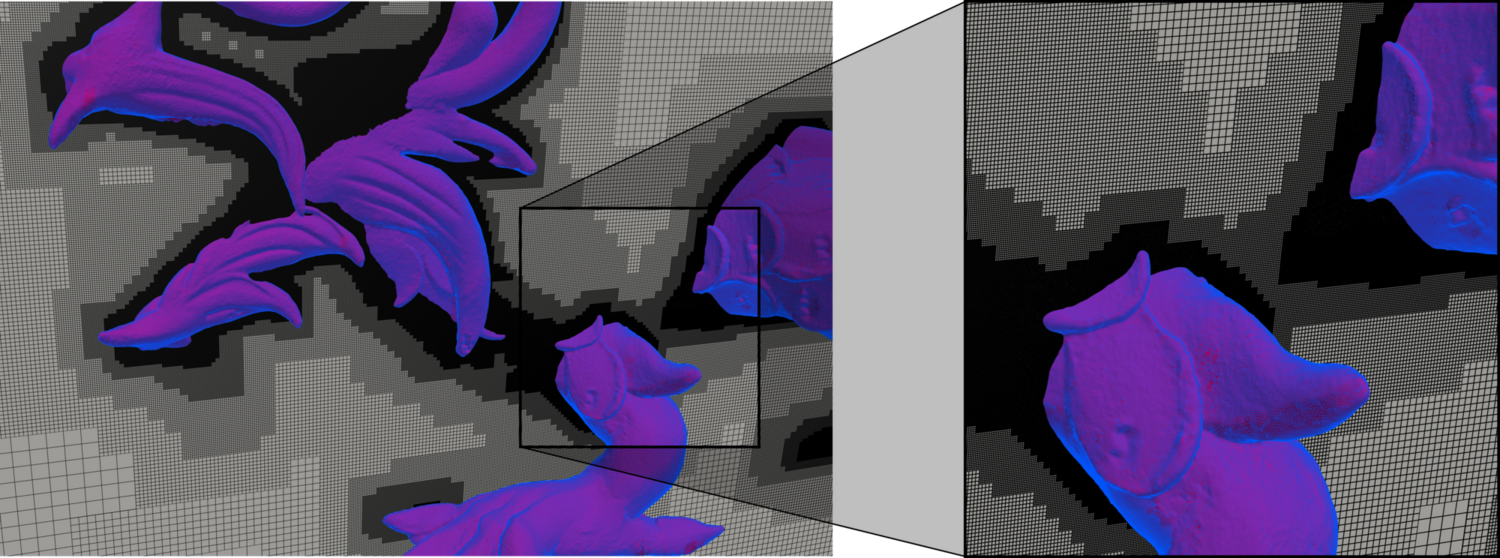}
        \caption{}
        \label{fig:results_paraview_balance_dragon}
    \end{subfigure}
    \begin{subfigure}[b]{0.85\textwidth}
        \centering
        \includegraphics[width=1\linewidth]{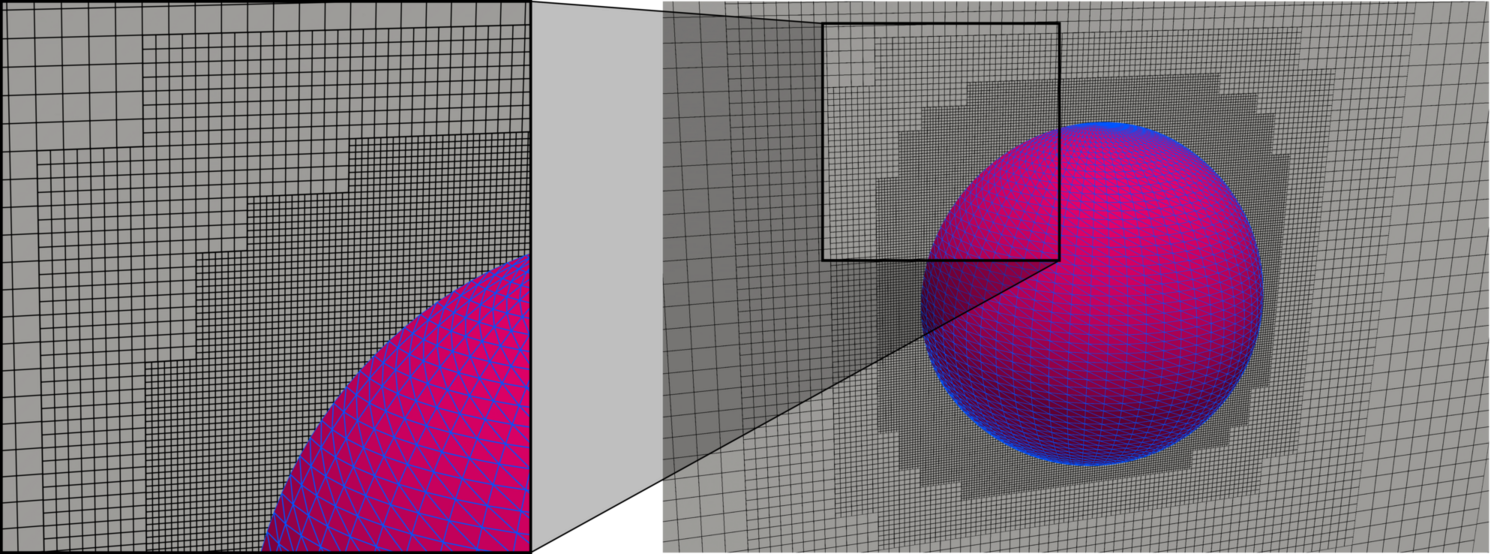}
        \caption{}
        \label{fig:results_paraview_balance_sphere}
    \end{subfigure}
    \caption{Illustrations of 2:1 balanced computational grids with an embedded a) XYZ RGB dragon, and b) sphere.}
    \label{fig:results_paraview_balance}
\end{figure}

The tree construction and voxelization results for the larger tests are illustrated with several figures. Figures \ref{fig:results_paraview_bunny_octree} and \ref{fig:results_paraview_dragon_octree} display clipped solid cells and cell-blocks by level for the BunnyL7 and DragonL7 tests, respectively. These blocks are coarser in the interior and gradually become finer towards the surface, eventually becoming fine-grained at the cell level at the boundary. Solid voxelizations that were extracted from each grid level are shown in Figures \ref{fig:results_paraview_bunny} and \ref{fig:results_paraview_dragon}. These voxelizations are available simultaneously for extraction at the end of the construction and voxelization. Figure \ref{fig:results_bigone} displays a solid voxelization extracted from the final level of the DragonL7 test, demonstrating that fine-scale features (such as the scales on dragon's body) are successfully captured. However, it also reveals the effects of floating-point round-off in a small set of `spikes', which result from incomplete partial surface voxelization in those regions prior to cell flag propagation. These spikes require exact floating-point arithmetic capabilities suitable for GPUs; this will be targeted in future work. Finally, Figures \ref{fig:results_paraview_balance_dragon} and \ref{fig:results_paraview_balance_sphere} display slices through the computational domains of embeddings of the Stanford dragon and a 3D sphere, respectively. The outlines of these slices illustrate that our tree construction produces 2:1 balanced forest-of-octrees grids.

%\footnote{The Stanford bunny is a well-known benchmark problem in the field of computer graphics. The original model \cite{Bunny} is in PLY format; we used a binary STL model obtained from Wikimedia Commons \cite{BunnyII} and transformed it to ASCII STL with the Blender software.}

\subsection{Validation}

\begin{sidewaysfigure}
    \centering
    \includegraphics[width=\textheight]{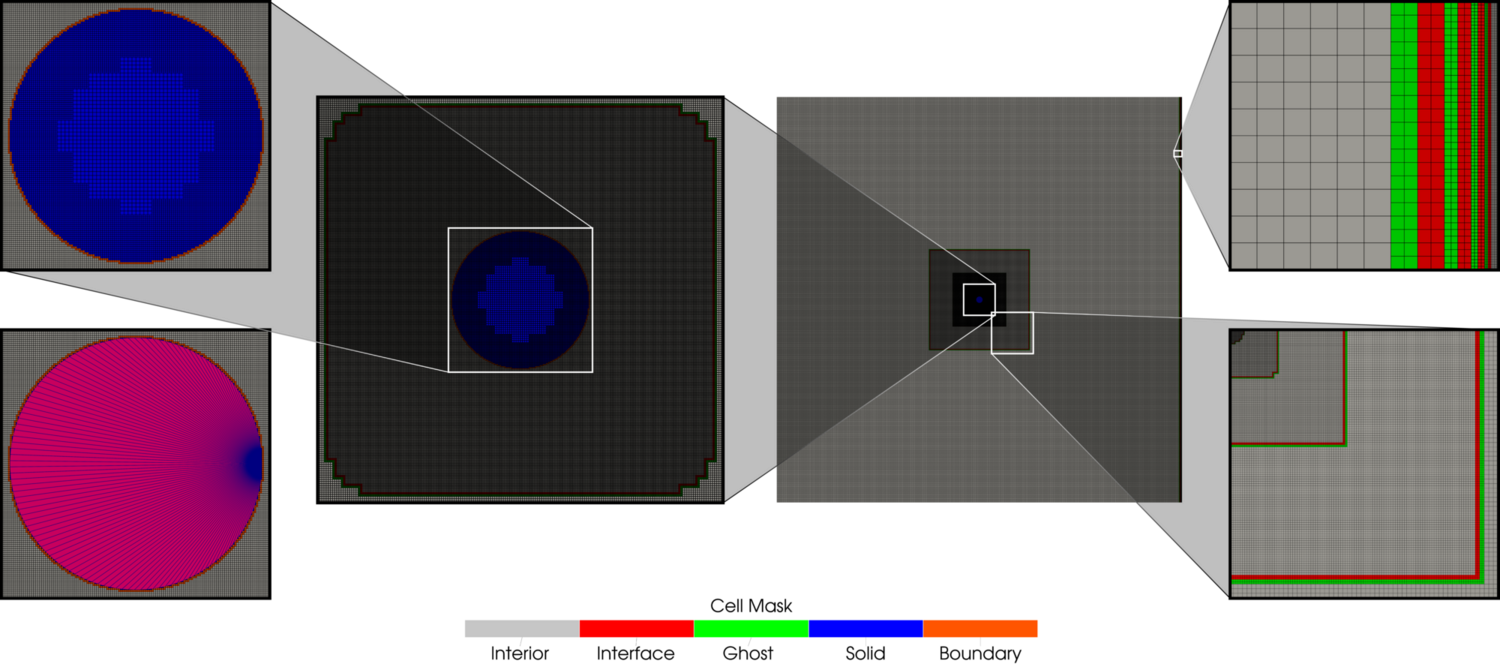}
    \caption{Illustration of the computational domain for the 2D flow past a circular cylinder with four levels of refinement (i.e., $L_{\text{max}}=5)$. From left to right: boundary and solid node classification within and around the circle, the finest grid in the hierarchy containing the embedding, the full computational domain, and visualizations of the refinement interface near the outlet (top) and the geometry (bottom).}
    \label{fig:results_bigone_domain}
\end{sidewaysfigure}
\begin{sidewaysfigure}
    \centering
    \includegraphics[width=\textheight]{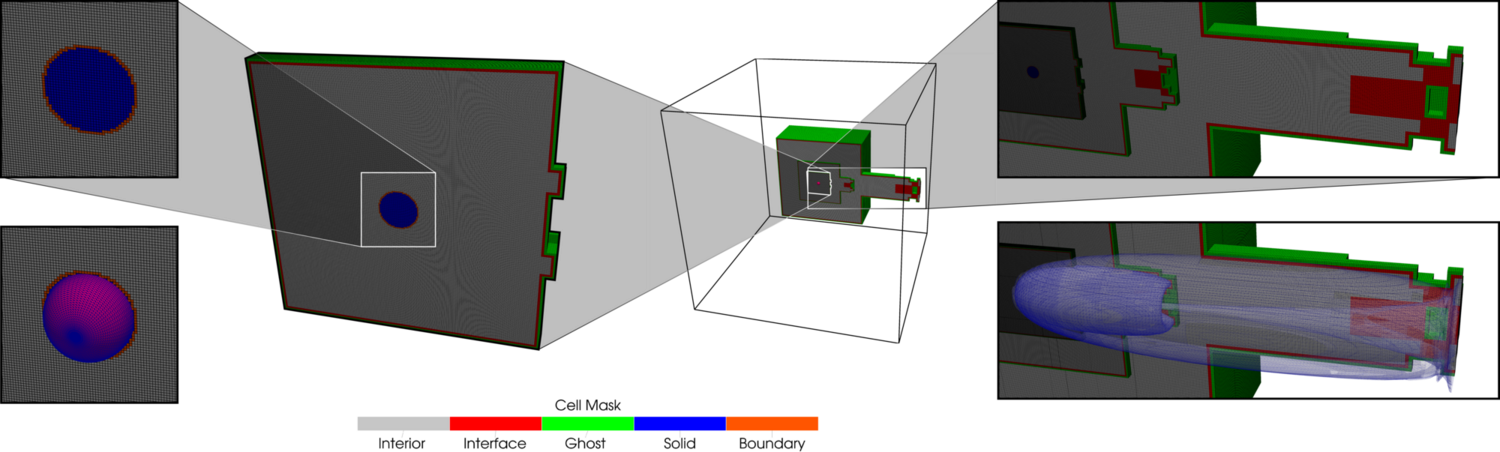}
    \caption{Illustration of the computational domain for the 3D flow past a sphere with three levels of refinement (i.e., $L_{\text{max}}=4)$. The root grid is hidden, but we have retained its solid outline. From left to right: boundary and solid node classification within and around the sphere, the finest grid in the hierarchy containing the embedding, the full computational domain, and visualization of refinement (top) based on vorticity magnitude isosurfaces (bottom).}
    \label{fig:results_bigone_domain_3d}
\end{sidewaysfigure}

The time-averaged drag coefficient $\overline{C_D}$, root-mean-square (RMS) of the lift coefficient $C_L$, and the Strouhal number ($\text{St}$) serve as validation metrics. These are given by
\begin{align}
    \overline{C_D}=\frac{1}{N_s}\sum_{i=1}^{N_s}\left(\frac{2F_x}{\rho u_c^2 A_c}\right)_i, \quad \text{RMS}(C_L)= \sqrt{\frac{1}{N_s}\sum_{i=1}^{N_s}\left(\frac{2F_h}{\rho u_c^2 A_c}\right)^2_i}, \quad \text{St}=\frac{f_s D_s}{u_c},
\end{align}

\noindent where $\textbf{F}=[F_x,F_y,F_z]$ is the force exerted by the fluid on the geometry, $F_h$ is the vertical component of the force, $f_s$ is the vortex shedding frequency, $N_s$ is the number of samples in the lift coefficient signal, and $u_c$, $D_s$, and $A_c$ are characteristic velocity, length, and surface area, respectively. In our framework, $F_h$ is given by $F_y$ and $F_z$ in 2D and 3D respectively, though only the former is considered in this paper. $f_s$ is estimated by fitting a sinusoidal profile through the lift coefficient signal and recovering the frequency.

% \subsubsection{The Lattice Boltzmann Method}

% We briefly re-state our adaptive mesh refinement scheme for the lattice Boltzmann method \cite{Kruger2017} and provide the equations for the simple and interpolated bounce-back conditions applied in the boundary. We refer the reader to \cite{Jaber2025} for further details. A time-step on grid level $L$ is given by:
% \begin{align}
%     f_p^*(t+\Delta t,\textbf{x}) &= f_p(t,\textbf{x}) - \frac{\Delta t}{\tau} \left(f_p - f_p^{\text{eq.}}\right)[t,\textbf{x}], \\
%     f_p(t + \Delta t, \textbf{x}+\Delta t \textbf{c}_p) &= f_p^*(t+\Delta t,\textbf{x}),
% \end{align}

% \noindent where $t$ and $\textbf{x}$ are time and space, $\Delta t_L$ and $\Delta x_L$ are the temporal and spatial step sizes on level $L$, $f_p$ are discrete density distribution functions (DDFs), $f_p^*$ are intermediate post-collision DDF values, $f_p^{\text{eq.}}$ are equilbiri and $\textbf{c}_p$

\subsubsection{Computational Setup}

A unit cube $[0,1]^D$ defines the computational domain. The root grid resolution $N_x$ is specified as $512^2$ in 2D and $256^3$ in 3D. Up to three additional refinements are applied for a total grid hierarchy size of $L_{\text{max}}=4$. The maximum effective resolution near the geometry, computed from $N_x(2^{L_{\text{max}}-1})$, is $4096$ in 2D and $2048$ in 3D. 2D simulations are performed for a total of $200N_x$ iterations, with force sampling beginning at $175N_x$ with a frequency of 1 sample per iteration. In 3D, the number of iterations is $100N_x$, and sampling begins at $75N_x$ at the same frequency. Figures \ref{fig:results_bigone_domain} and \ref{fig:results_bigone_domain_3d} display our computational grids for the flows past a circular cylinder and a sphere, including the output of cell classification in and around the geometry, as well as the refinement interface.

The 2D circular cylinder and 3D sphere (with diameter $D_s$) occupy the center. The 2D square (with dimensions $2D_s \times 2D_s $) is shifted slightly along $+x$ by $0.7/N_x$ to test the effect of interpolation. Placing the square in the center would result in identical results for the SBB and IBB simulations with the current  otherwise. We fix $D_s=1/64 \text{ m}$ so that sufficient spacing $31.5D$ is given between the cylinder/sphere and the inlet, outlet, and far-field. We apply a no-slip boundary condition on the surface of the geometry in the form of the bounce-back condition with optional linear interpolation according to the method of Bouzidi et al. \cite{Bouzidi2001}. The inlet, represented as a constant velocity profile $u_{\text{in}}=0.05 \text{ m/s}$ is also modeled using the bounce-back with a momentum correction term. Outflow is modeled with a specification of pressure (based on density) at the outlet. We implement this with the anti-bounce-back condition. An initial condition $\textbf{u}(t,\textbf{x})=u_{\text{in}}$ is set throughout the whole domain. Automated refinement around the geometry achieves a smooth transition from a coarse resolution at the domain boundaries to the finest one near the surface.
%All simulations are conducted in double precision to reduce the effects of error accumulation when $\textbf{F}$ is being calculated.

The solver offers the choices of linear and cubic interpolation at the refinement interface for coarse-to-fine data transfers to ghost cells. It has been shown that the order of accuracy of the chosen interpolation method at the interface has a significant impact on the accuracy of the generated pressure field. This, in turn, affects the values of the drag and lift coefficients. According to Lagrava et al. \cite{Lagrava2012}, the pressure field becomes discontinuous at the interface when linear interpolation is used due to an inconsistency with the order of accuracy of the numerical solution in the bulk of the flow. This result pertains to vertex-centered grids, however, in cell-centered grids, such as ours, the staggered pattern between the coarse and fine grids alleviates the discontinuity \cite{Palabos}. The latter grid arrangement still retains a non-negligible pressure error that requires multiple levels of refinement for an accurate computation of the drag and lift coefficients. We therefore employ cubic interpolation along the refinement interface in our validation.

\begin{figure}[]
    \centering
    \begin{subfigure}[b]{0.49\textwidth}
        \centering
        \includegraphics[width=1\linewidth]{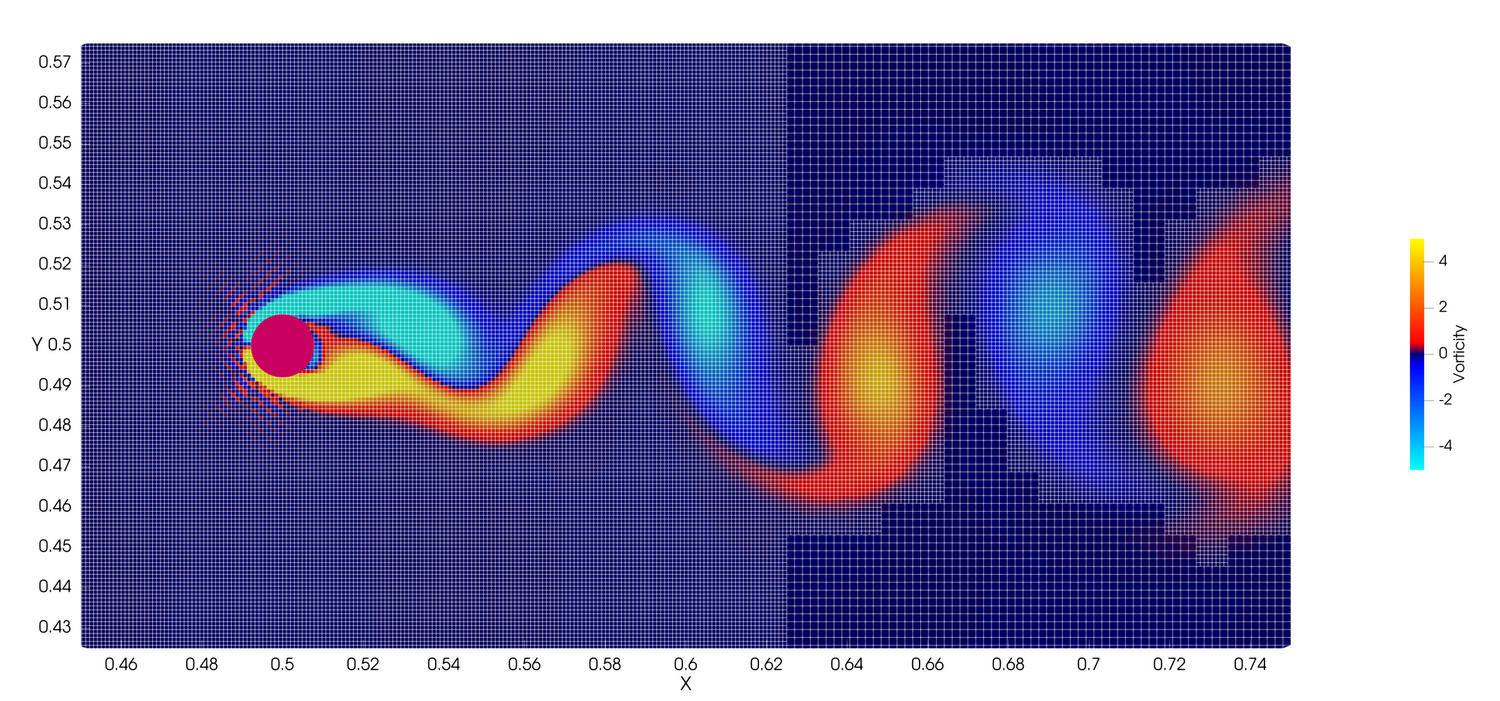}
        \caption{Circle, $L_{\text{max}}=2$.}
    \end{subfigure}
    \begin{subfigure}[b]{0.49\textwidth}
        \centering
        \includegraphics[width=1\linewidth]{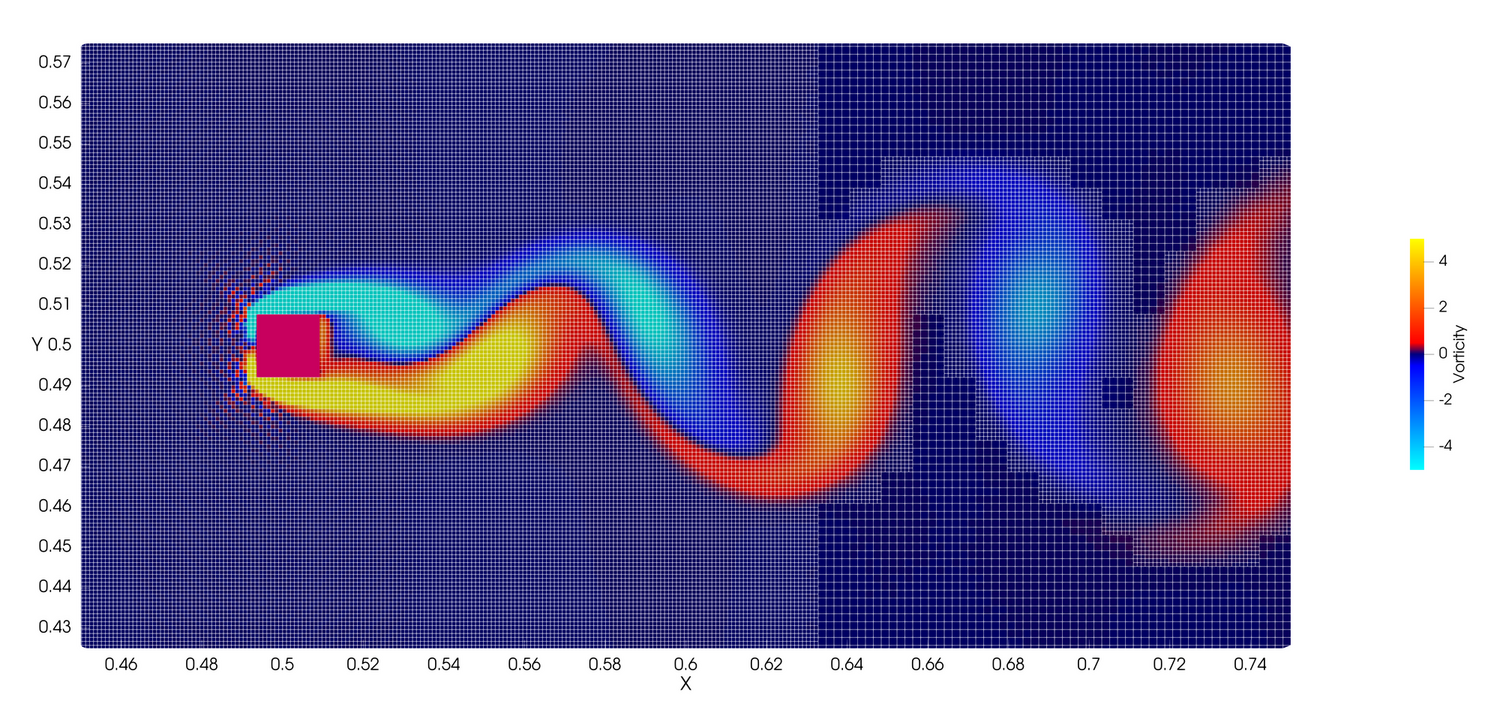}
        \caption{Square, $L_{\text{max}}=2$.}
    \end{subfigure}
    \begin{subfigure}[b]{0.49\textwidth}
        \centering
        \includegraphics[width=1\linewidth]{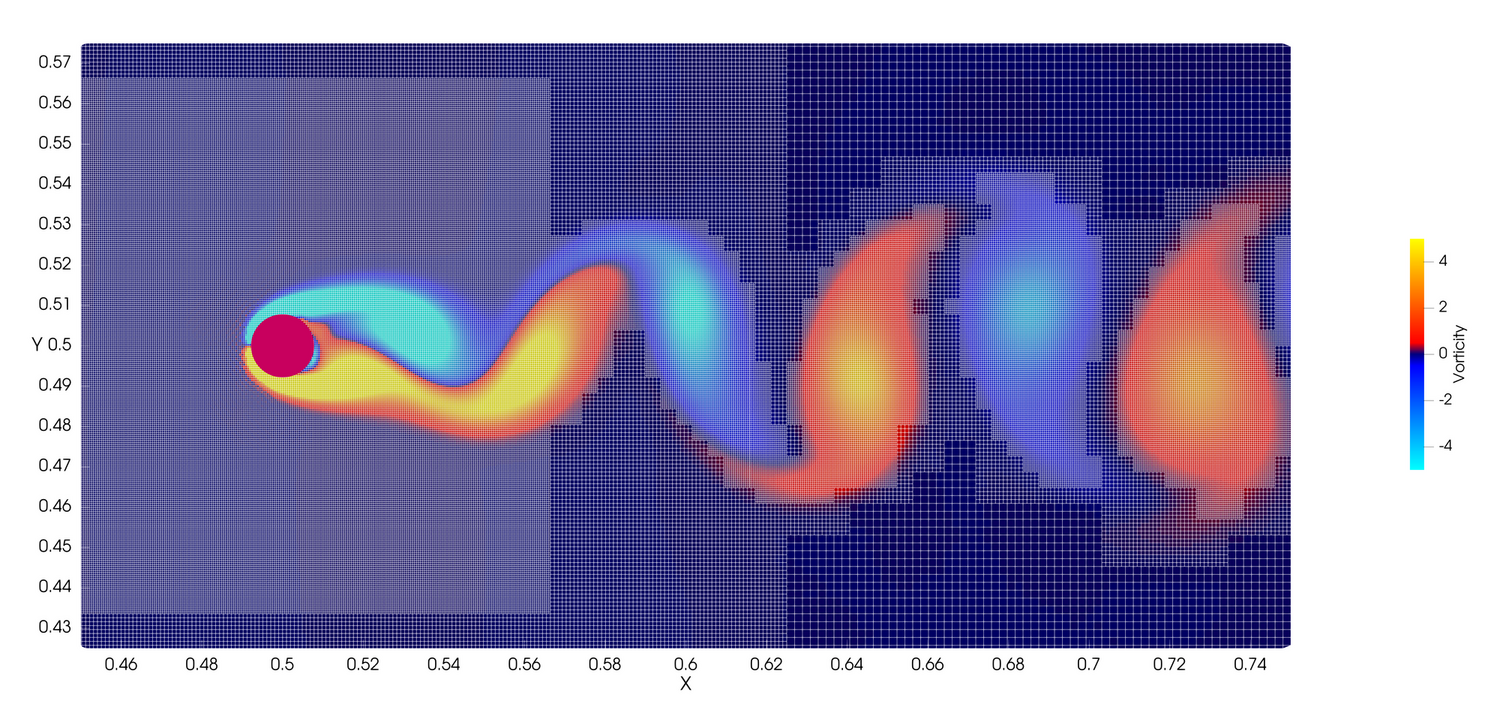}
        \caption{Circle, $L_{\text{max}}=3$.}
    \end{subfigure}
    \begin{subfigure}[b]{0.49\textwidth}
        \centering
        \includegraphics[width=1\linewidth]{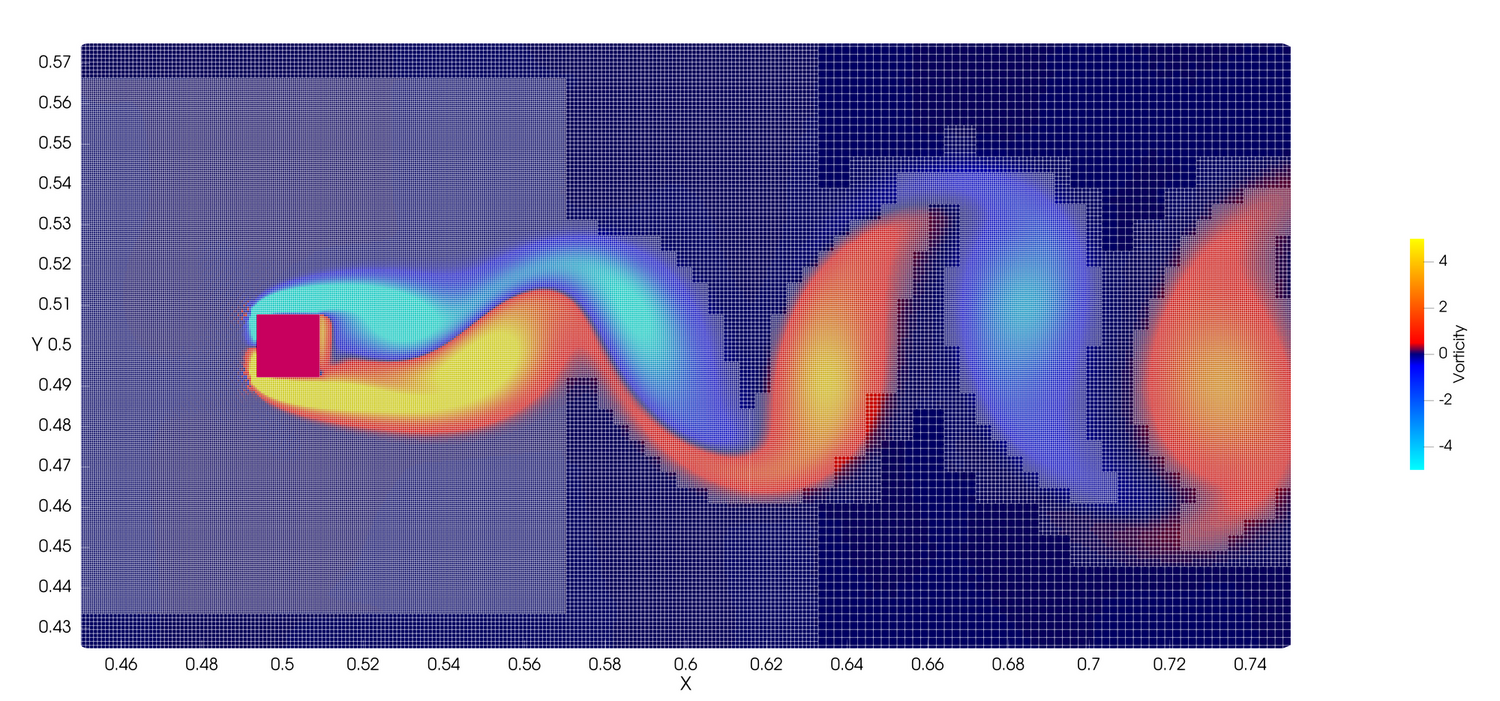}
        \caption{Square, $L_{\text{max}}=3$.}
    \end{subfigure}
    \begin{subfigure}[b]{0.49\textwidth}
        \centering
        \includegraphics[width=1\linewidth]{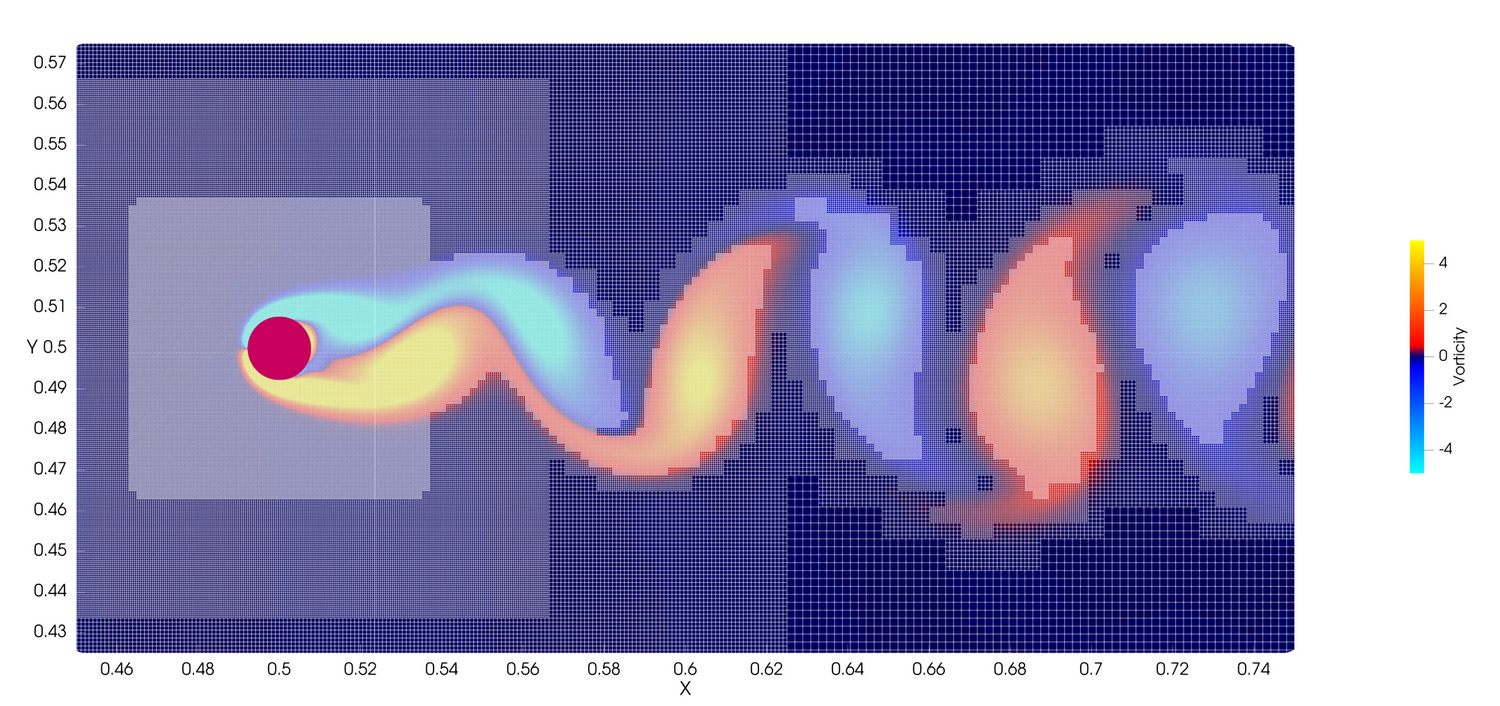}
        \caption{Circle, $L_{\text{max}}=4$.}
    \end{subfigure}
    \begin{subfigure}[b]{0.49\textwidth}
        \centering
        \includegraphics[width=1\linewidth]{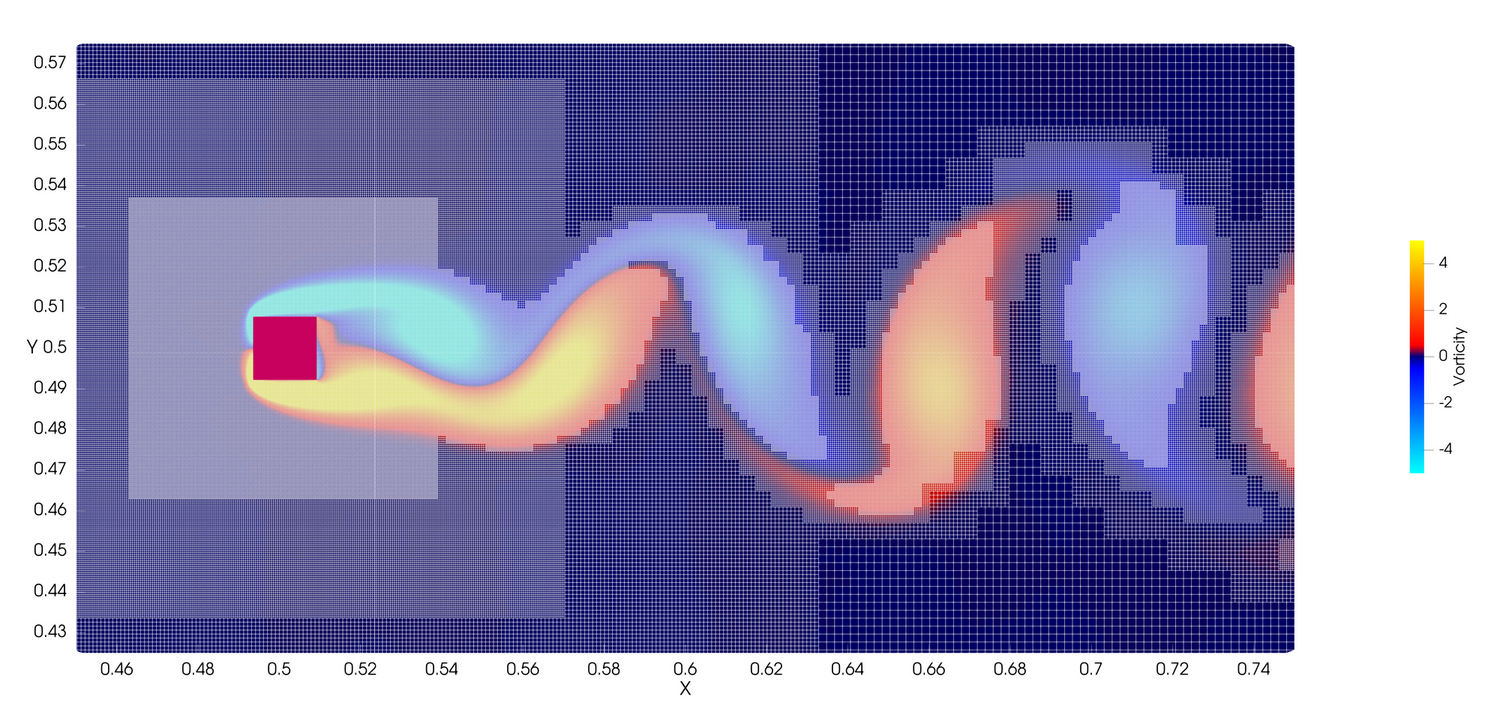}
        \caption{Square, $L_{\text{max}}=4$.}
    \end{subfigure}
    \begin{subfigure}[b]{0.49\textwidth}
        \centering
        \includegraphics[width=1\linewidth]{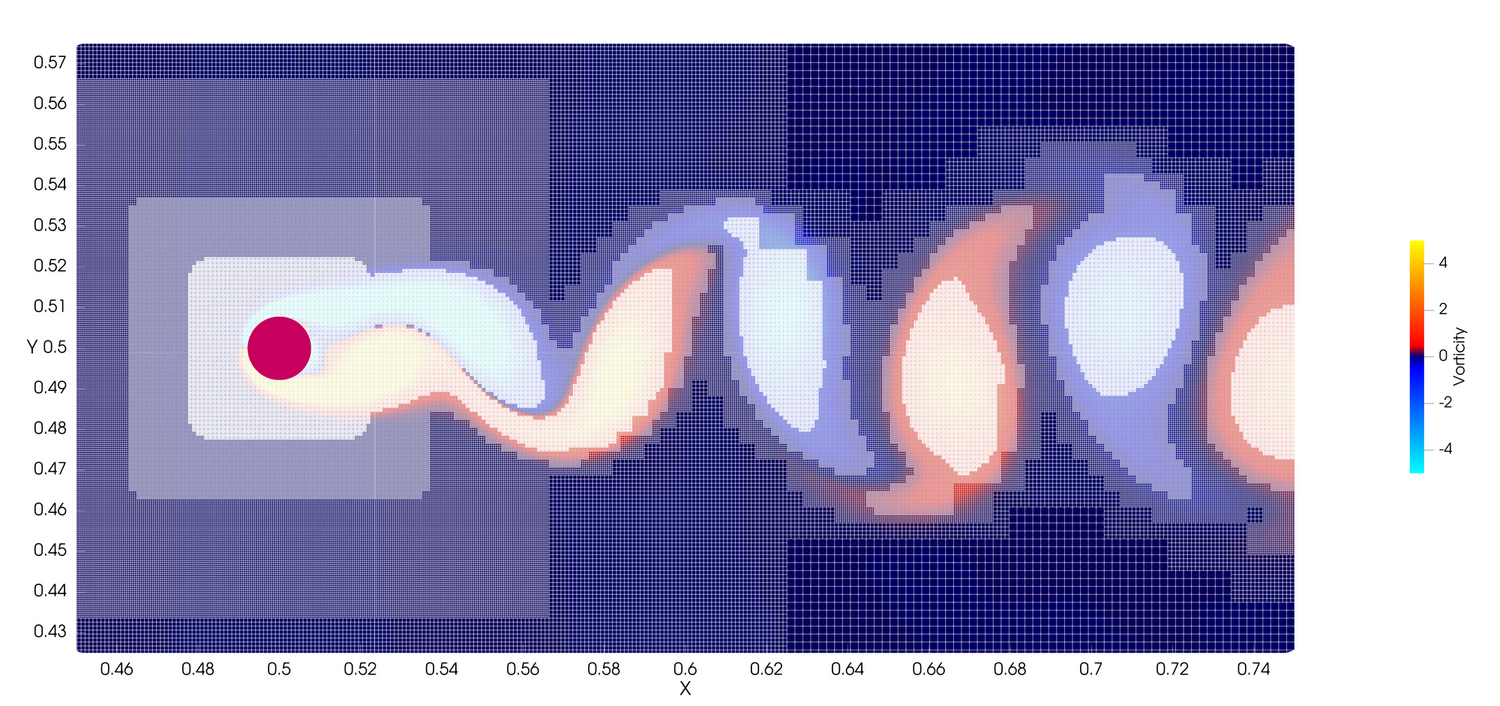}
        \caption{Circle, $L_{\text{max}}=5$.}
    \end{subfigure}
    \begin{subfigure}[b]{0.49\textwidth}
        \centering
        \includegraphics[width=1\linewidth]{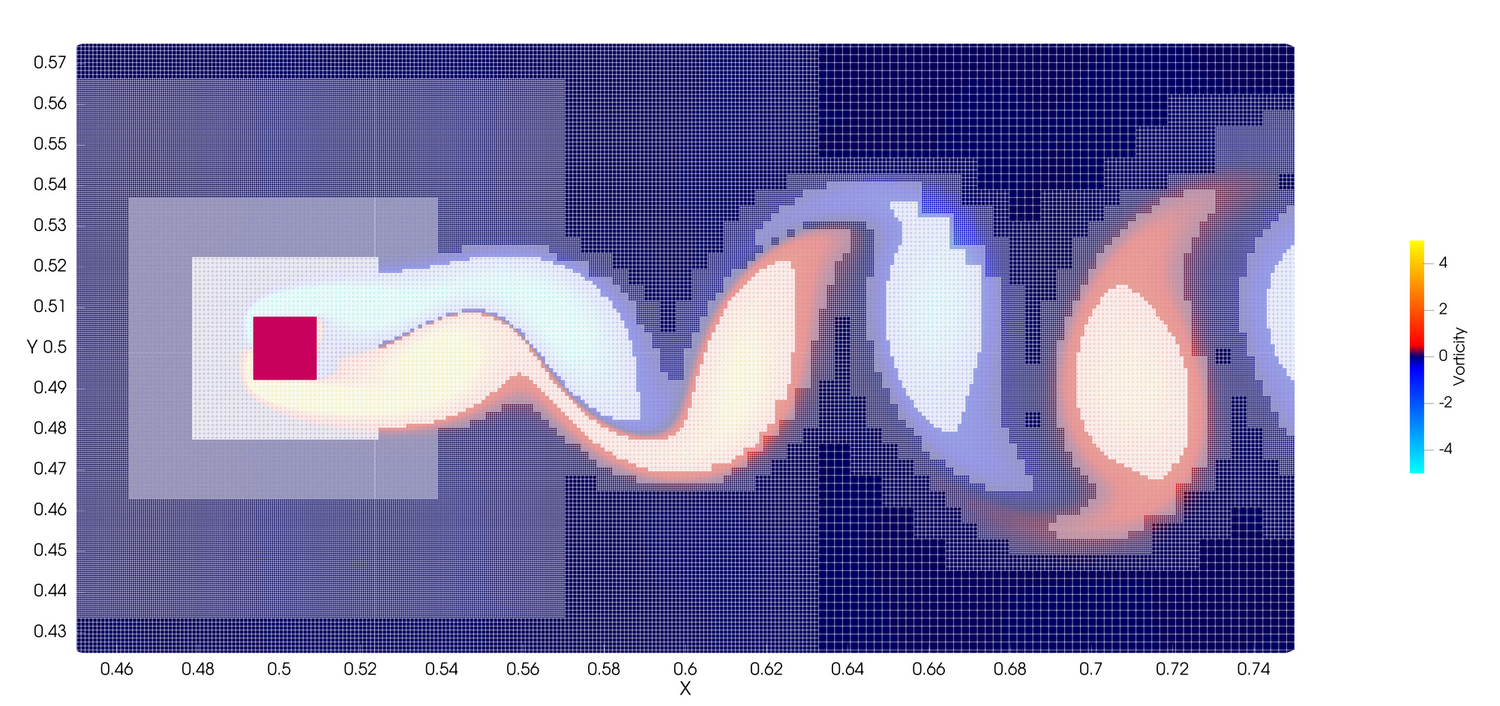}
        \caption{Square, $L_{\text{max}}=5$.}
    \end{subfigure}
    \caption{Vorticity plots with overlaid AMR grid outline extracted from the final time step.}
    \label{fig:tests_validation_vort}
\end{figure}

\begin{figure}[]
    \centering
    \begin{subfigure}[b]{0.49\textwidth}
        \centering
        \includegraphics[width=1\linewidth]{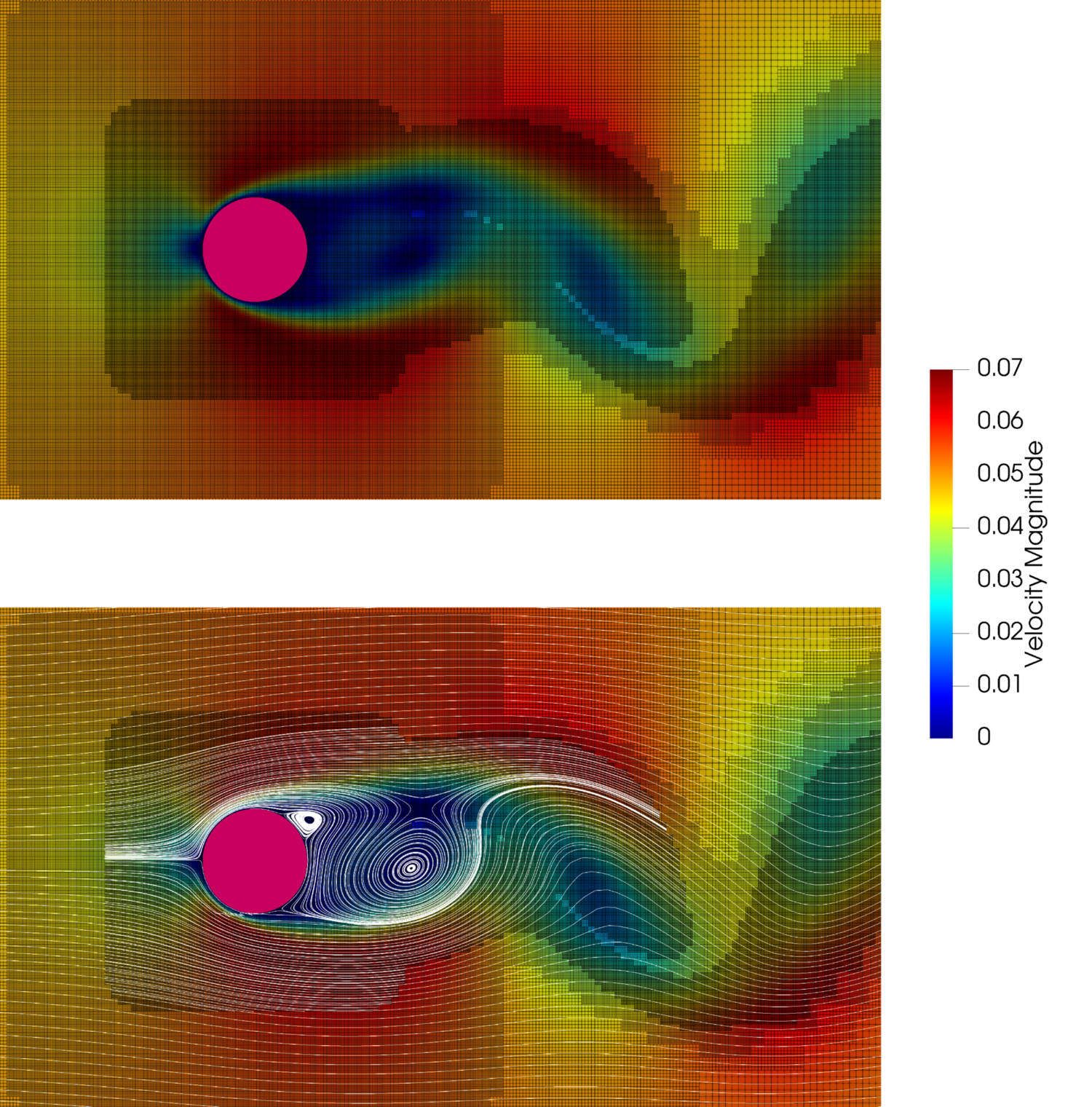}
        \caption{}
    \end{subfigure}
    \begin{subfigure}[b]{0.49\textwidth}
        \centering
        \includegraphics[width=1\linewidth]{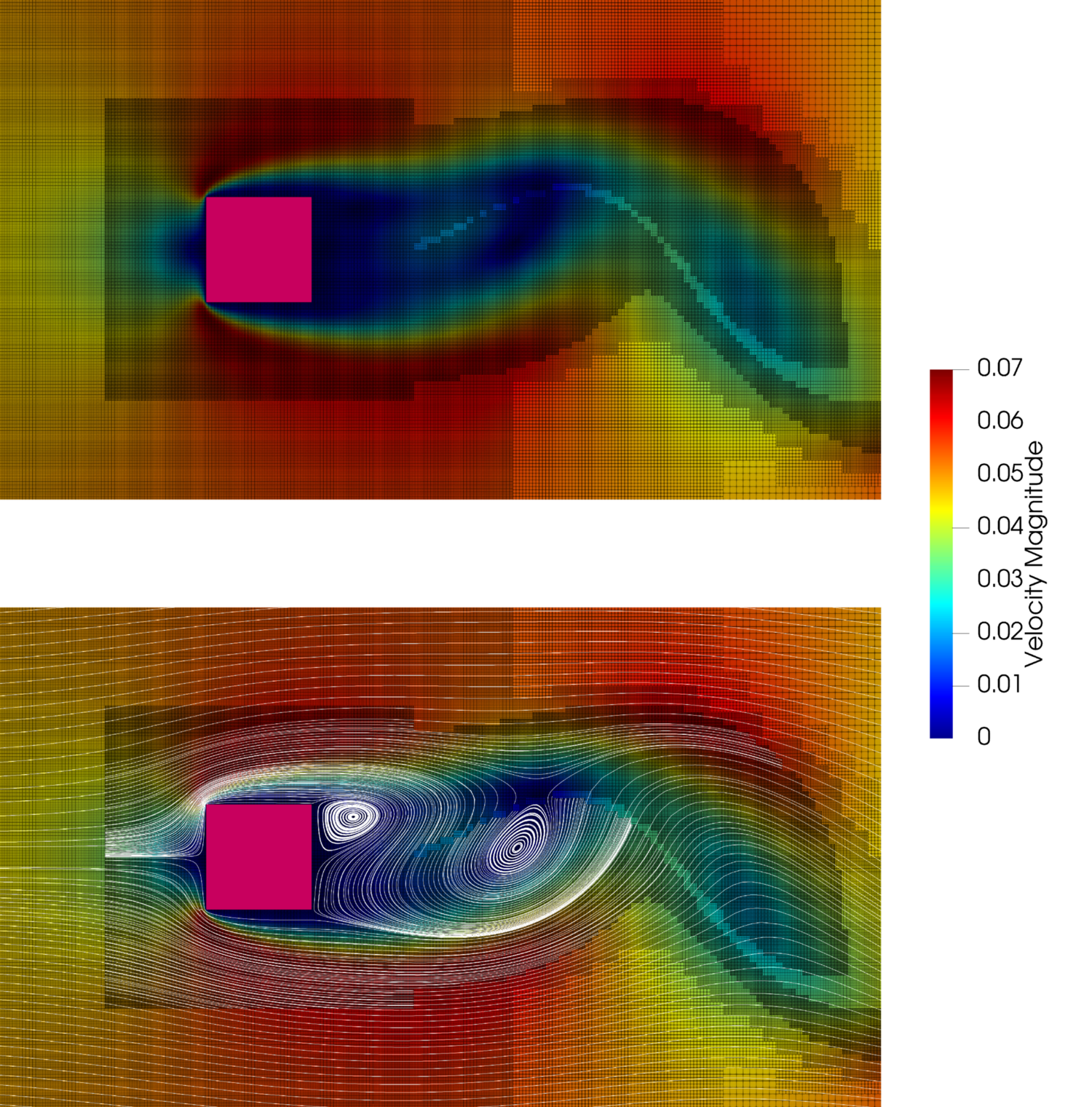}
        \caption{}
    \end{subfigure}
    \caption{Velocity magnitude plots (top) with overlaid streamlines (bottom) for the circular (left) and square (right) 2D cylinders with $L_{\text{max}}=5$.}
    \label{fig:tests_validation_vel}
\end{figure}

We set $\text{Re}=100$ for the 2D tests and use a range $\text{Re}\in \{10,15,20\}$ for the 3D tests. $\text{Re}$ is adjusted by setting the kinematic viscosity according to $\nu=uD_s/\text{Re}$. The relaxation rate $\tau$, used during the collision step, is automatically computed with the corresponding temporal time step (based on the grid resolution) and the kinematic viscosity. Plots of vorticity magnitude for the 2D cylinders are shown in Figure \ref{fig:tests_validation_vort} and overlaid with outlines of the AMR grids, illustrating that the implementation captures the coherent structures dynamically. Figure \ref{fig:tests_validation_vel} displays streamlines over velocity magnitude plots for the 2D AMR grids with $L_{\text{max}}=5$.
% D3Q19

\subsubsection{Flow Past 2D Circular Cylinder}

Tables \ref{tab:results_validation_2d_circle_theirs} and \ref{tab:results_validation_2d_circle_mine} display literature and current results for the flow past a circular cylinder, respectively. Our simulations converged towards values that agree well with the literature. With a single refinement of the root grid ($L_{\text{max}}=2$), we obtained values of $1.39$ (SBB) and $1.36$ (IBB) for the time-averaged drag coefficient. These values decrease by approximately $1\%$ with each successive increase in $L_{\text{max}}$ (beyond $L_{\text{max}}=3$) until both cases are identical up to two decimal places at $1.34$. The amplitude of the lift signal in the two cases differ by $0.031$ when $L_{\text{max}}=2$. This difference decreases to $0.003$ when $L_{\text{max}}=5$. The Strouhal number was found to be $0.164-0.165$ in most cases except for the case of $L_{\text{max}}=2$, where it was slightly smaller at $0.161-0.162$. 2D vorticity surface plots for these simulations indicate the presence of spurious oscillations near the cylinder which degrade the accuracy of the forces computed in the case $L_{\text{max}}=2$. For $L_{\text{max}}>2$, these oscillations disappear and this is reflected in the restored accuracy of the metrics in these simulations relative to the finest case with $L_{\text{max}}=5$.
\begin{table}[h]
    \centering
    
    \begin{subtable}[t]{0.45\textwidth}
        \centering
        \small
        \begin{tabular}{lc|ccc}
            \hline
            \multicolumn{2}{l}{Reference} & $\overline{C_D}$ & $\pm C_L$ & $\text{St}$ \\
            \hline
            Russel and Wang & \cite{Russell2003} & $1.34 \pm 0.007$ & $0.276$ & $0.165$ \\
            Calhoun & \cite{Calhoun2002} & $1.33 \pm 0.014$ & $0.298$ & $0.175$ \\
            Liu, Zheng and Sung & \cite{Liu1998} & $1.35 \pm 0.012$ & $0.339$ & $0.165$ \\
            Choi et al. & \cite{Choi2007} & $1.34 \pm 0.011$ & $0.315$ & $0.164$ \\
            Kim, Kim, and Choi & \cite{Kim2001} & $1.33$ & $0.320$ & $0.165$ \\
            \hline
        \end{tabular}
        \caption{Literature results.}
        \label{tab:results_validation_2d_circle_theirs}
    \end{subtable}
    \hfill
    \begin{subtable}[t]{0.45\textwidth}
        \centering
        \small
        \begin{tabular}{c|cc|cc|cc}
        \hline
        \multirow{2}{*}{$L_{\text{max}}$} & \multicolumn{2}{c}{$\overline{C_D}$} & \multicolumn{2}{c}{$\pm C_L$} & \multicolumn{2}{c}{St} \\ \cline{2-7}
        & SBB & IBB & SBB & IBB & SBB & IBB \\
        \hline
        2 & $1.39$ & $1.36$   & $0.345$ & $0.314$   & $0.161$ & $0.162$ \\
        3 & $1.36$ & $1.35$   & $0.340$ & $0.328$   & $0.164$ & $0.164$ \\
        4 & $1.35$ & $1.34$   & $0.334$ & $0.328$   & $0.164$ & $0.165$ \\
        \rowcolor{gray!40}
        5 & $1.34$ & $1.34$   & $0.330$ & $0.327$   & $0.165$ & $0.165$ \\
        \hline
        \end{tabular}
        \caption{Current results.}
        \label{tab:results_validation_2d_circle_mine}
    \end{subtable}

    \caption{Currents results for the flow past a 2D circular cylinder against the literature.}
    \label{tab:results_validation_2d_circle}
\end{table}

\subsubsection{Flow Past 2D Square Cylinder}

Tables \ref{tab:results_validation_2d_square_theirs} and \ref{tab:results_validation_2d_square_mine} display literature and current results in the case of the square cylinder. We converted RMS values reported in the literature for the sinusoidal lift signal into amplitudes for comparison. Our drag coefficient values range from $1.44-1.46$, which is up to $\sim 5\%$ smaller than the values reported in the literature. This includes our own previous simulations \cite{Jaber2025} in the case of axis-aligned geometry boundaries. We attribute this to the current size of the cylinder relative to the domain (which is half of what we previously used) rather than from AMR since the value of $\overline{C_D}$ obtained from high-resolution uniform grids was also $1.5$. A smaller cylinder decreases far-field effects on the force computation, which in LBM simulations is particularly sensitive to reflections at the outflow and transient effects from an inexact initialization of the pressure field. The SBB and IBB cases differ by $\sim 1\%$ across $L_{\text{max}} < 5$, but are identical to two decimal places when $L_{\text{max}}=5$. As with the circular cylinder simulations, the Strouhal number is largely the same in most cases, here in the range $0.142-0.144$. The values for $L_{\text{max}}=2,5$ are identical, indicating that differences may arise from factors such as the sinusoidal curve fitting used to extract the frequency of vortex shedding. Our current values differ from previous simulations \cite{Jaber2025} by about $3\%$. Relative to other works, this difference is $\sim 5\%$ and is at a maximum relative to Saha et al. \cite{Saha2000} at $\sim 10\%$ (though their value is higher than the others). The amplitude of the lift signal differs most when $L_{\text{max}}=3$ by about $0.007$, but is nearly identical otherwise. Values of the lift amplitude reported in the literature vary significantly, ranging in these selected works from $0.194$ to $0.277$, though our current values lie within this range. When compared with the LBM-AMR results of Fakhari and Lee (extracted from their curve plot), our converged value differs by $\sim 10\%$. Overall, our results are in general agreement with the literature.
\begin{table}[]
    \centering
    
    \begin{subtable}[t]{0.45\textwidth}
        \centering
        \small
        \begin{tabular}{lc|ccc}
            \hline
            \multicolumn{2}{l}{Reference} & $\overline{C_D}$ & $\pm C_L$ & $\text{St}$ \\
            \hline
            Fakhari and Lee & \cite{Fakhari2014} & $1.51$ & $0.277$ & $0.149$ \\
            Saha et al. & \cite{Saha2000} & $1.51$ & - & $0.159$ \\
            Singh et al. & \cite{Singh2009} & $1.52$ & $0.228$ & $0.150$ \\   
            Pavlov et al. & \cite{Pavlov2000} & $1.51$ & $0.194$ & $0.150$ \\
            Jaber et al. & \cite{Jaber2025} & $1.51$ & - & $0.147$ \\
            \hline
        \end{tabular}
        \caption{Literature results.}
        \label{tab:results_validation_2d_square_theirs}
    \end{subtable}
    \hfill
    \begin{subtable}[t]{0.45\textwidth}
        \centering
        \small
        \begin{tabular}{c|cc|cc|cc}
        \hline
        \multirow{2}{*}{$L_{\text{max}}$} & \multicolumn{2}{c}{$\overline{C_D}$} & \multicolumn{2}{c}{$\pm C_L$} & \multicolumn{2}{c}{St} \\ \cline{2-7}
        & SBB & IBB & SBB & IBB & SBB & IBB \\
        \hline
        2 & $1.45$ & $1.46$   & $0.262$ & $0.262$   & $0.143$ & $0.143$ \\
        3 & $1.44$ & $1.46$   & $0.255$ & $0.262$   & $0.144$ & $0.142$ \\
        4 & $1.44$ & $1.45$   & $0.255$ & $0.256$   & $0.143$ & $0.143$ \\
        \rowcolor{gray!40}
        5 & $1.44$ & $1.44$   & $0.252$ & $0.252$   & $0.143$ & $0.143$ \\
        \hline
        \end{tabular}
        \caption{Current results.}
        \label{tab:results_validation_2d_square_mine}
    \end{subtable}

    \caption{Current results for the flow past a 2D square cylinder against the literature.}
    \label{tab:results_validation_2d_square}
\end{table}

\subsubsection{Flow Past 3D Sphere}

\begin{figure}
    \centering
    \includegraphics[width=0.6\linewidth]{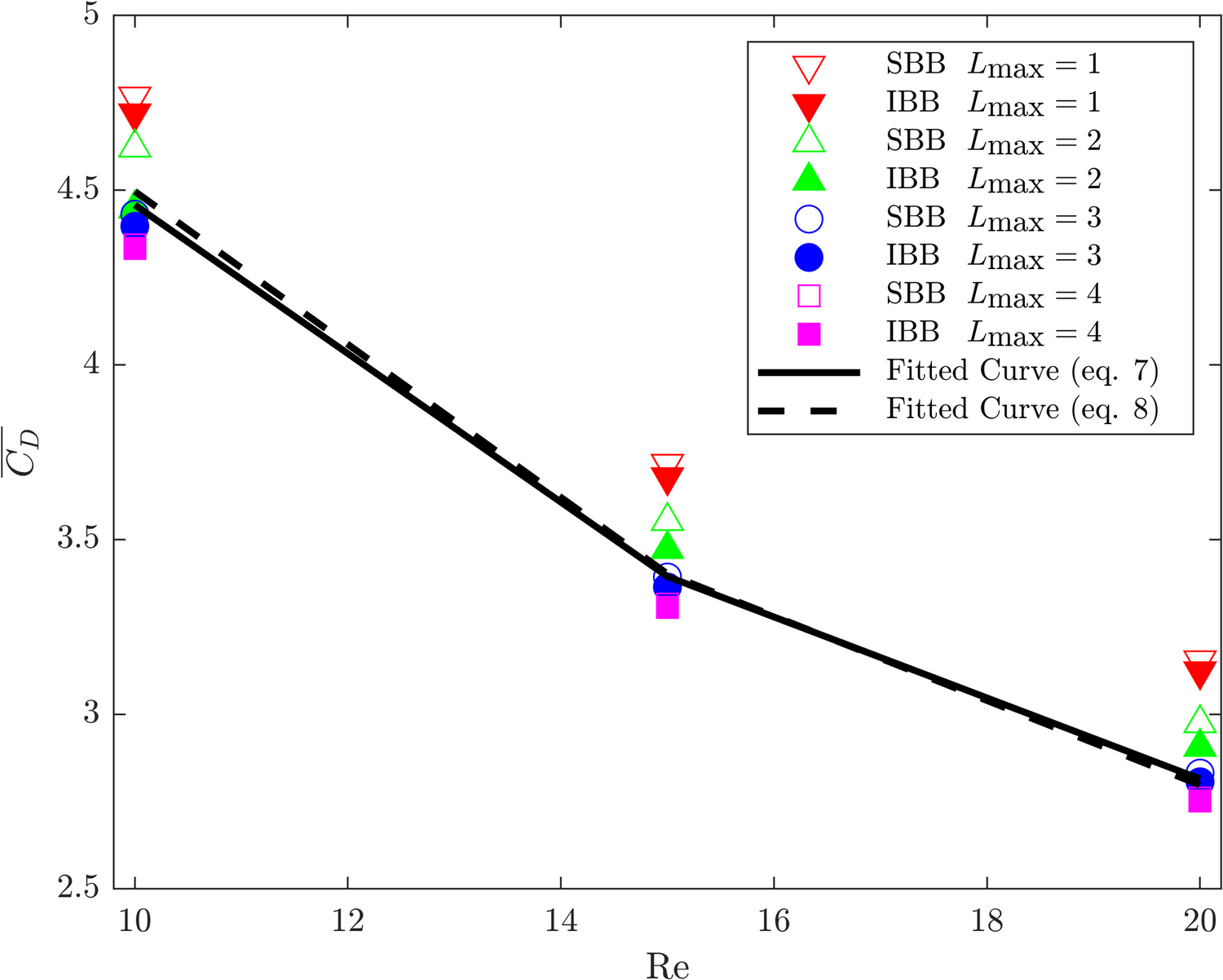}
    \caption{Time-averaged drag coefficients for flow past a sphere over the range $\text{Re}\in\{10,15,20\}$. Our results are compared with the curve fits of Mikhailov and Freire  \cite{Mikhailov2013} through the experimental data of Roos and Willlmarth \cite{Roos1971} and Brown and Lawler \cite{Brown2003}}
    \label{fig:results_forceplot_3d}
\end{figure}

Figure \ref{fig:results_forceplot_3d} displays our results for 3D flow past a sphere across $\text{Re}\in\{10,15,20\}$ compared with the experimental fits of Mikhailov and Freire \cite{Mikhailov2013}. Our results for the SBB and IBB generally converge to similar values as $L_{\text{max}}$ increases for all $\text{Re}$. The converged values at $L_{\text{max}}=4$ are slightly lower than those predicted by the fitted curves but deviate by no more than $\sim 4\%$ across all $\text{Re}$, indicating a general agreement.

As $L_{\text{max}}$ increases, the values of $\overline{C_D}$ corresponding to both the SBB and IBB cases decrease together. Simulations with the SBB consistently produce slightly larger values than those with the IBB. The percentage difference between the two cases depends strongly on the value of $L_{\text{max}}$ and takes on similar values across $\text{Re}$, ranging from a maximum of $2\%-4\%$ at $L_{\text{max}}=2$ to a minimum of $\sim 0.3\%$ at $L_{\text{max}}=4$. The difference between our results and those of the curve fit (eq. 7 of \cite{Mikhailov2013}) is at a minimum when $L_{\text{max}}=3$, ranging from $0.05\%-1.37\%$. However, this increases to $1.9\%-2.8\%$ when $L_{\text{max}}=4$. The case of $L_{\text{max}}=2$ and $\text{Re}=10$ with the IBB represents an outlier in that it differs from its SBB counterpart by $0.17$ (an increase by $\sim 4\%$ relative to the IBB value).

Eq. 8 of \cite{Mikhailov2013} produces a predicted value of the drag coefficient that is $\sim 1\%$ larger than that of eq. 7, which increases the relative error of our results at $L_{\text{max}}=4$ to $3.39\%-3.63\%$ (from $2.6\%-2.8\%$). At $\text{Re}=20$, the predicted value is $\sim 0.6\%$ smaller than that of eq. 7, which reduces the error of the converged SBB and IBB values to the range $1.3\%-1.6\%$ (from $1.9\%-2.2\%$).
% Numerical form of the data:
% L_max	Case	Re	Our Values	Curve Fit Values
% 1	SBB	10	4.7679	4.4578
% 1	SBB	15	3.7168	3.3943
% 1	SBB	20	3.1545	2.815
% 2	SBB	10	4.622	4.4578
% 2	SBB	15	3.553	3.3943
% 2	SBB	20	2.9742	2.815
% 3	SBB	10	4.4313	4.4578
% 3	SBB	15	3.3927	3.3943
% 3	SBB	20	2.8316	2.815
% 4	SBB	10	4.3442	4.4578
% 4	SBB	15	3.3149	3.3943
% 4	SBB	20	2.7605	2.815
% 1	IBB	10	4.7199	4.4578
% 1	IBB	15	3.6787	3.3943
% 1	IBB	20	3.1231	2.815
% 2	IBB	10	4.4474	4.4578
% 2	IBB	15	3.4726	3.3943
% 2	IBB	20	2.9063	2.815
% 3	IBB	10	4.3968	4.4578
% 3	IBB	15	3.3642	3.3943
% 3	IBB	20	2.8066	2.815
% 4	IBB	10	4.3333	4.4578
% 4	IBB	15	3.3058	3.3943
% 4	IBB	20	2.7524	2.815

%We validate the link-length computation kernel and the updated LBM solver using the 2D flow-past-a-cylinder and 3D flow-past-a-sphere benchmark tests.
%The same cylinder will be reused and simulated with the position of its center varied to show that the interpolation of DDFs using weights computed from the links recovers accuracy even when the cylinder edges are no longer aligned with the grid.

% Effective resolution of bunny (0.673, 0.549, 0.675):
% L0 64: 
% Effective resolution of dragon (0.729, 0.485, 0.327):
% L0 64: 
\section{Limitations and Future Work}

The two primary limitations in our implementation are that the geometry is stationary, and that the algorithm is designed for execution on a single GPU. These will need to be incorporated to enable more practical large-scale computational fluid dynamics simulations. Exact floating-point arithmetic will be pursued to make the voxelization robust. Lastly, we currently assume that no-slip conditions are imposed on the surface of embedded geometries. A future extension will enable the user to specify initial and boundary conditions that potentially vary in space and time.

The extension to moving geometries will likely require a mechanism for face re-binning, and further optimizations to the link-length computation routine to avoid a bottleneck in practical simulations. The current spatial binning and voxelization framework, which is fully implemented on the GPU, should be able to serve as a basis. For applicability of the proposed methodology to general immersed boundary simulations, the framework further requires a bilateral connection between the computational grid and the embedded geometry (in contrast to the current unilateral setup), and a consistent initialization approach for new fluid nodes spawned from previous solid nodes no longer covered by the geometry.

The multi-GPU extension is non-trivial, as tree-based AMR algorithms usually exploit the spatial locality and linearization properties of space-filling curves to achieve scalable communication and load balancing patterns. Load balancing will likely need to be incorporated within the level-by-level tree construction procedure outlined in the current methodology to ensure efficiency in even larger voxelizations.

\section{Conclusion}

We presented a methodology for embedding complex geometries in two- and three-dimensions into a hierarchical grid organized as a forest-of-octrees \cite{p4est}. The procedure is entirely GPU-native, requiring no communications with the device at any point in the process. Our C++/CUDA implementation accepts arbitrary geometries in ASCII STL format or generates primitive objects on the fly. Ray cast cell classification and flag propagation facilitate voxelization. A hierarchy of uniformly-distributed spatial bins restricts classification to cells in the vicinity of the geometry. Spatial binning also heavily filters out faces that never participate in ray cast computations in coarser grids, further accelerating voxelization. We implement optimizations such as stream compaction and index mapping to reduce the size of the binning problem as much as possible. Automated near-wall refinement is achieved with an initial refinement around the geometry surface that is followed by propagation of the refinement flags to a user-specified distance. The output of our implementation feeds into a pipeline for data-parallel forest-of-octrees refinement \cite{Jaber2025} to build and voxelize the grid level by level.

We compared execution times on consumer- and datacenter-grade GPUs with the sparse solid voxelization procedure of Schwarz and Seidel \cite{Schwarz2010} (developed for graphics processing). Our execution times were of comparable order of magnitude (albeit slower). Two major bottlenecks are the cell classification step within the tree construction procedure, and partial surface voxelization. Cell classification consistently makes up approximately half of the time taken during construction as the grid hierarchy increases in size, which could be optimized to accelerate the construction procedure. Partial surface voxelization currently relies on triangle-AABB overlap tests to reduce false negatives in point-in-triangle tests. Switching to standard half-plane tests with an offload to separate exact floating-point arithmetic routines for the few edge cases could further reduce total time taken.

A large scale test repeats the spatial binning and voxelization steps introduced in the methodology with different optimizations activated. The recorded execution times verify that filtering, index mapping, and stream compaction all contribute to reducing the overall time taken for both spatial binning and voxelization. The most consequential optimization is face filtering, which reduces the set of faces considered in both binning and partial surface voxelization. This affects time taken on the coarsest levels the most, where as many as $90\%$ of the faces are filtered out.

Finally, tests of variants introduced for two computationally expensive kernels in the voxelization pipeline (partial surface voxelization and boundary cell identification) determine the ideal underlying kernel structure and memory access patterns. We found that partial surface voxelization performs best when vertex data is arranged in an array of structures format and accessed via broadcast to threads component-by-component. Other variants implementing access of structure of arrays format, parallel reads to shared memory, or warp-level primitives for inter-block communication perform similarly but slightly slower.

The voxelization produces a stair-case approximation of the geometry when applied in computational fluid dynamics solvers. We implement a routine for computing the lengths of links between boundary and solid nodes which can be used to implement interpolation near the geometry (e.g., when imposing boundary conditions or computing finite difference stencils near the wall). We extend a previous lattice Boltzmann solver, previously limited to geometries whose surface aligns with Cartesian axes, with interpolated boundary conditions based on these link-lengths. Simulations of flow past 2D circular and square cylinders at $\text{Re}=100$, and a 3D sphere at $\text{Re}\in\{10,15,20\}$ validate the extended solver. The square cylinder test verifies that the output of the updated solver is accurate compared to our previous work \cite{Jaber2025}. We show that interpolated bounce-back conditions approach the grid-refined result faster than simple bounce-back in the 2D circular cylinder and 3D sphere tests.

In summary, the methodology presented in this paper enables embedding stationary complex geometries and applying them in computational fluid dynamics simulations. Solvers designed for axis-aligned multi-resolution grids like the lattice Boltzmann method can incorporate interpolation at the boundary to improve accuracy in external flows. Our work serves as a foundation for practical GPU-native meshing compatible with dynamic adaptive mesh refinement for computational fluid dynamics with complex geometries. Future work will improve robustness with exact floating-point precision, enable voxelizations of moving geometries, and extend the implementation for execution on multiple GPUs in distributed memory clusters.

\section{Acknowledgements}

%The authors gratefully acknowledge support from the Natural Sciences and Engineering Research Council of Canada, Canadian Microelectronics Corporation and the Digital Research Alliance of Canada. \hl{Mention OGS.}

The authors gratefully acknowledge support from the Natural Sciences and Engineering Research Council of Canada, Canadian Microelectronics Corporation, the Digital Research Alliance of Canada, and the Ontario Graduate Scholarship (OGS) funded by the Government of Ontario and the University of Toronto.

\section{Declaration of Generative AI Use}

%We used ChatGPT, Claude, and Deepseek to obtain feedback  on the logical flow, accuracy, and quality of the text in this manuscript. We applied this feedback manually to improve the text. ChatGPT was also used to write boilerplate code (such as expressions implementing Thrust's parallel algorithm routines). However, all custom CUDA kernels (and the host-side routines that call them) used in this work were hand-written. We did not use AI to generate new text, directly modify existing text, analyze our results, or develop the methodology.

We used ChatGPT, Claude, and Deepseek to provide feedback on the logical flow, accuracy, and clarity of the manuscript text, which we then applied manually during revision. ChatGPT was also used to generate boilerplate code, such as expressions implementing Thrust's parallel algorithm routines. All custom CUDA kernels and their associated host-side routines were authored directly without AI assistance. We did not use AI tools to generate new text, directly modify existing text, analyze results, or develop the methodology presented in this work.


\appendix

\section{Triangle-AABB Overlap}

Tests for overlap between triangles and axis-aligned bounding boxes are well-known in graphics processing. Akenine-M\"{o}ller \cite{Moller2001} developed one such test and provides a CPU-based implementation. However, this cannot be used directly in the current GPU-based framework. Our CUDA implementation (Algorithm \ref{alg:triangleAABB}) follows the description of Schwarz and Seidel \cite{Schwarz2010}. We implement the 2D cases of line-AABB overlap separately. Both implementations can be found in $\texttt{util\_tribin.h}$.

\setcounter{algocf}{0}
\renewcommand{\thealgocf}{\Alph{algocf}}
\begin{algorithm}[H]
\small
\caption{Triangle-AABB Intersection Test (3D)}
\label{alg:triangleAABB}
\SetAlgoLined
\DontPrintSemicolon

\SetKwProg{Fn}{Function}{}{}
\Fn{\texttt{TrianglePlaneCutsBox}($\textbf{v}_1, \textbf{v}_2, \textbf{v}_3, \textbf{v}_m, \textbf{v}_M$)}{
    Compute triangle normal $\textbf{n} = (\textbf{v}_2 - \textbf{v}_1) \times (\textbf{v}_3 - \textbf{v}_1)$\;
    Define $\textbf{c} = \left((\textbf{v}_{M,d} - \textbf{v}_{m,d}) \cdot (\textbf{n}_d > 0)\right)_{0 \leq d < D}^T$\;
    Compute $d = \textbf{n} \cdot \textbf{v}_m$\;
    Compute $d_1 = \textbf{n} \cdot (\textbf{c} - \textbf{v}_1)$\;
    Compute $d_2 = \textbf{n} \cdot ((\textbf{v}_M - \textbf{v}_m - \textbf{c}) - \textbf{v}_1)$\;
    \Return{$(d + d_1)(d + d_2) \leq 0$}\;
}

\vspace{1em}
\Fn{\texttt{TriangleBinOverlap2D\_AxisGap}($\textbf{Axis}, v_{1,x},v_{1,y},v_{2,x},v_{2,y},v_{3,x},v_{3,y},r_{m,x},r_{m,y},r_{M,x},r_{M,y} $)}{
    \textbf{Axis 0}: $\textbf{e} \leftarrow (1, 0)^T$\;
    \textbf{Axis 1}: $\textbf{e} \leftarrow (0, 1)^T$\;
    \textbf{Axis 2}: $\textbf{e} \leftarrow (v_{2,y} - v_{1,y}, v_{1,x} - v_{2,x})^T$\;
    \textbf{Axis 3}: $\textbf{e} \leftarrow (v_{3,y} - v_{2,y}, v_{2,x} - v_{3,x})^T$\;
    \textbf{Axis 4}: $\textbf{e} \leftarrow (v_{1,y} - v_{3,y}, v_{3,x} - v_{1,x})^T$\;

    Compute $t_{\min} = \min\left(\min(\textbf{v}_1 \cdot \textbf{e}, \textbf{v}_2 \cdot \textbf{e}), \textbf{v}_3 \cdot \textbf{e}\right)$\;
    Compute $t_{\max} = \max\left(\max(\textbf{v}_1 \cdot \textbf{e}, \textbf{v}_2 \cdot \textbf{e}), \textbf{v}_3 \cdot \textbf{e}\right)$\;

    Compute $r_{\min} = \min\Bigl(\min\bigl((r_{m,x}, r_{m,y})^T \cdot \textbf{e}, (r_{M,x}, r_{m,y})^T \cdot \textbf{e}\bigr), \min\bigl((r_{m,x}, r_{M,y})^T \cdot \textbf{e}, (r_{M,x}, r_{M,y})^T \cdot \textbf{e}\bigr)\Bigr)$\;

    Compute $r_{\max} = \max\Bigl(\max\bigl((r_{m,x}, r_{m,y})^T \cdot \textbf{e}, (r_{M,x}, r_{m,y})^T \cdot \textbf{e}\bigr), \max\bigl((r_{m,x}, r_{M,y})^T \cdot \textbf{e}, (r_{M,x}, r_{M,y})^T \cdot \textbf{e}\bigr)\Bigr)$\;

    \Return{$(t_{\max} < r_{\min})$ \textbf{or} $(r_{\max} < t_{\min})$}\;
}

\vspace{1em}
% Function 3
\Fn{\texttt{TriangleBinOverlap3D}($\textbf{v}_1, \textbf{v}_2, \textbf{v}_3, \textbf{v}_m, \textbf{v}_M$)}{
    \If{$\texttt{TrianglePlaneCutsBox}(\textbf{v}_1, \textbf{v}_2, \textbf{v}_3, \textbf{v}_m, \textbf{v}_M) = \text{false}$}{
        \Return{\textbf{false}}\;
    }
    \For{$k \leftarrow 0$ \KwTo $4$}{
        \If{\texttt{TriangleBinOverlap2D\_AxisGap}$(k, v_{1,x}, v_{1,y}, v_{2,x}, v_{2,y}, v_{3,x}, v_{3,y}, v_{m,x}, v_{m,y}, v_{M,x}, v_{M,y})$}{
            \Return{\textbf{false}}\;
        }
    }
    \For{$k \leftarrow 0$ \KwTo $4$}{
        \If{\texttt{TriangleBinOverlap2D\_AxisGap}$(k, v_{1,y}, v_{1,z}, v_{2,y}, v_{2,z}, v_{3,y}, v_{3,z}, v_{m,y}, v_{m,z}, v_{M,y}, v_{M,z})$}{
            \Return{\textbf{false}}\;
        }
    }
    \For{$k \leftarrow 0$ \KwTo $4$}{
        \If{\texttt{TriangleBinOverlap2D\_AxisGap}$(k, v_{1,z}, v_{1,x}, v_{2,z}, v_{2,x}, v_{3,z}, v_{3,x}, v_{m,z}, v_{m,x}, v_{M,z}, v_{M,x})$}{
            \Return{\textbf{false}}\;
        }
    }
    \Return{\textbf{true}}\;
}

\end{algorithm}

\nomenclature{$B$}{Bin density}
\nomenclature{$C_L$}{Lift coefficient}
\nomenclature{$\overline{C_D}$}{Time-averaged drag coefficient}
\nomenclature{$D$}{Number of dimensions}
\nomenclature{$D_s$}{Diameter}
\nomenclature{$d$}{Distance}
\nomenclature{$\Delta t_L$}{Temporal step size on grid level $L$}
\nomenclature{$\Delta x_L$}{Spatial step size on grid level $L$}
\nomenclature{$\Delta x_{b,L}$}{Cell-block length along one axis}
\nomenclature{$L$}{Grid level}
\nomenclature{$L_{\text{max}}$}{Grid hierarchy size}
\nomenclature{$L_{\text{max},B}$}{Number of bin levels}
\nomenclature{$M_b$}{Cell-block size}
\nomenclature{$M_t$}{Thread-block size (primary mode)}
\nomenclature{$\overline{M_t}$}{Thread-block size (secondary mode)}
\nomenclature{$\mathcal{T}$}{Geometry face (line/triangle in 2D/3D)}
\nomenclature{$\textbf{I}$}{Local cell indices $(I,J,K)$}
\nomenclature{$\textbf{e}_k$}{Unit vector along axis $k$}
\nomenclature{$\textbf{n}$}{Normal}
\nomenclature{$\textbf{v}$}{Vertex}
\nomenclature{$t$}{Time}
\nomenclature{$x$}{Space}
\nomenclature{$l_{x/y/z}$}{Domain length along $x/y/z$}
\nomenclature{$\text{St}$}{Strouhal number}

%\printnomenclature

\bibliographystyle{elsarticle-num} 
\bibliography{refs}


%\begin{thebibliography}{00}



%\end{thebibliography}
\end{document}